\documentclass{LMCS}

\usepackage{amsmath}
\usepackage{amssymb}
\usepackage{stmaryrd}
\usepackage{subfigure}
\usepackage{gastex,amsfonts,amssymb,latexsym}
\usepackage[all]{xy}
\usepackage{arrows}
\usepackage{enumerate,hyperref}

\renewcommand{\emptyset}{\varnothing}

\newcommand{\mc}[1]{\mathcal{#1}}

\newcommand{\mb}{\mathbb}

\newcommand{\<}{\langle}
\renewcommand{\>}{\rangle}
\newcommand{\les}{\leqslant}
\newcommand{\ges}{\geqslant}

\newcommand{\mX}{\mathcal{X}}

\newcommand{\mG}{\mathcal{G}}
\newcommand{\mD}{\mathcal{D}}
\newcommand{\mC}{\mathcal{C}}

\newcommand{\mB}{\mathcal{B}}
\newcommand{\mA}{\mathcal{A}}

\newcommand{\mF}{\mathcal{F}}

\newcommand{\mV}{\mathcal{V}}

\newcommand{\BSCC}{\textsc{BSCC}}

\newcommand{\CTL}{\textsc{CTL}}

\newcommand{\CSLTA}{\CSL$^{\!\textrm{\uppercase{TA}}}$}

\newcommand{\CTMC}{\textsc{{CTMC}}}

\newcommand{\DTMP}{\textsc{DTMP}}
\newcommand{\DTA}{\textsc{DTA}}
\newcommand{\DTAr}{\DTA$^{\!\Ever}$}
\newcommand{\DTAo}{\DTA$^{\omega}$}
\newcommand{\DMTAr}{\DMTA$^{\!\Ever}$}
\newcommand{\DMTAo}{\DMTA$^{\omega}$}

\newcommand{\DTMC}{\textsc{DTMC}}

\newcommand{\CSL}{\textsc{CSL}}

\newcommand{\DMTA}{\textsc{DMTA}}

\newcommand{\PDP}{\textsc{PDP}}
\newcommand{\ODE}{\textsc{ODE}}
\newcommand{\PDE}{\textsc{PDE}}

\newcommand{\mv}[1]{\singlearrow{#1}}

\newcommand{\emb}{\textsl{emb}}

\newcommand{\bdD}{{\bf D}}
\newcommand{\bdE}{{\bf E}}
\newcommand{\bdF}{{\bf F}}
\newcommand{\bdB}{{\bf B}}

\newcommand{\bdM}{{\bf M}}

\newcommand{\bdPi}{\mathbf{\Pi}}

\renewcommand{\P}{{\bf P}}

\newcommand{\Q}{{\bf Q}}

\newcommand{\AP}{\textsc{AP}}

\newcommand{\Distr}{{\it Distr}}

\newcommand{\Paths}{{\it Paths}}

\newcommand{\Path}{{\it Paths}}

\newcommand{\AccPaths}{{\it AccPaths}}

\newcommand{\Prob}{{\it Prob}}

\newcommand{\Pro}{{\mathbb{P}}}

\newcommand{\W}{\mathbin{\mathsf{W}}}   

\newcommand{\F}{\mathop{\diamondsuit}}

\newcommand{\Inv}{\mathit{Inv}}

\newcommand{\Nats}{\mathbb{N}}
\newcommand{\Reals}{\mathbb{R}}

\newcommand{\Rationals}{\mathbb{Q}}

\newcommand{\Ever}{\F}

\newcommand{\updownmapsto}[4]{\xymatrix{#1\;\ar @{|->}[r]^{#2}_{#3}&#4}}

\newcommand{\updownsquigarrow}[4]{\xymatrix{#1\;\ar @{~>}[r]^{#2}_{#3}&#4}}

\def\topbotatom#1{\hbox{\hbox to 0pt{$#1\bot$\hss}$#1\top$}}

\makeatletter
\newcommand{\be}{%
\begin{group}
\eqnarray%
\@ifstar{\nonumber}{}%
}

\makeatother

\def\doi{7 (1:12) 2011}
\lmcsheading%
{\doi}
{1--34}
{}
{}
{Dec.~25, 2009}
{Mar.~29, 2011}
{}

\begin{document}

\title[Model Checking of CTMCs Against Timed Automata]{%
Model Checking of Continuous-Time Markov Chains \\
Against Timed Automata Specifications}

\author[T.~Chen]{Taolue Chen\rsuper a} 
\address{{\lsuper a}Formal Methods and Tools,
University of Twente, The Netherlands}  
\email{chent@ewi.utwente.nl}  
\thanks{{\lsuper a}This research is funded by the DFG
research training group 1295 AlgoSyn, the SRO DSN project of CTIT, University of Twente, the EU FP7 project QUASIMODO
and the DFG-NWO ROCKS project}  

\author[T.~Han]{Tingting Han\rsuper b} 
\address{{\lsuper{b,d}}Software Modelling and
Verification, RWTH Aachen University, Germany
}  
\email{\{tingting.han,mereacre\}@cs.rwth-aachen.de}  
\thanks{}   

\author[J.-P.~Katoen]{Joost-Pieter Katoen\rsuper c}  
\address{{\lsuper c}Software Modelling and
Verification, RWTH Aachen University, Germany;\newline
Formal Methods and Tools, University of Twente, The Netherlands}    
\email{katoen@cs.rwth-aachen.de}  
\thanks{}   

\author[A.~Mereacre]{Alexandru Mereacre\rsuper d}   
\address{\vskip-6 pt}  
\thanks{}

\keywords{continuous-time Markov chains, deterministic timed automata,
linear-time specification, model checking, piecewise-deterministic Markov processes}
\subjclass{D.2.4}

\begin{abstract}
\noindent
We study the verification of a finite continuous-time Markov chain $($\CTMC$)$
$\mc{C}$ against a linear real-time specification given as a deterministic
timed automaton $($\DTA$)$ $\mc{A}$ with finite or Muller acceptance
conditions.
The central question that we address is: what is the probability of the set of
paths of $\mc{C}$ that are accepted by $\mc{A}$, i.e., the likelihood that
$\mC$ satisfies $\mA$?
It is shown that under finite acceptance criteria this equals the reachability
probability in a finite piecewise deterministic Markov process $($\PDP$)$, whereas
for Muller acceptance criteria it coincides with the reachability probability
of terminal strongly connected components in such a PDP.
Qualitative verification is shown to amount to a graph analysis of the PDP.
Reachability probabilities in our PDPs are then characterized as the least solution
of a system of Volterra integral equations of the second type and are shown
to be approximated by the solution of a system of partial differential equations.
For single-clock \DTA, this integral equation system can be transformed into
a system of linear equations where the coefficients are solutions of ordinary
differential equations.
As the coefficients are in fact transient probabilities in CTMCs, this result implies that
standard algorithms for CTMC analysis suffice to verify single-clock DTA specifications.
\end{abstract}

\maketitle
%
\vfill

\section{Introduction}

Continuous-time Markov chains (\CTMC s) are one of the most prominent models
in performance and dependability analysis.
They are exploited in a broad range of applications, and constitute the underlying
semantical model of a plethora of modeling formalisms for real-time probabilistic
systems such as Markovian queueing networks, stochastic Petri nets, stochastic
variants of process algebras, and calculi for systems biology.
\CTMC\ model checking has been mainly focused on the branching-time temporal
logic CSL (Continuous Stochastic Logic \cite{ASSB00,BHHK03}), a variant of timed
\CTL\ where the \CTL\ universal and existential path quantifiers are replaced by a
probabilistic operator.
Like \CTL\ model checking, CSL model checking of finite CTMCs proceeds by a
recursive descent over the parse tree of the CSL formula.
One of the key ingredients is that time-bounded reachability probabilities can be
approximated arbitrarily closely by a reduction to transient analysis in \CTMC s
\cite{BHHK03}.
This results in an efficient polynomial-time algorithm that has been realized in
model-checking tools such as PRISM \cite{HKNP06} and MRMC \cite{KHHJZ09} and
has been successfully applied to various case studies from diverse application areas.

Verifying a finite CTMC $\mc{C}$ against linear-time (but untimed) specifications
in the form of a regular or $\omega$-regular language is rather straightforward
and boils down to computing reachability probabilities in discrete-time Markov
chains (DTMCs).
This can be seen as follows.
Assume that the specification is provided as a deterministic automaton $\mc{A}$
on finite words, or alternatively as a deterministic Muller automaton $\mc{A}$.
The underlying idea is that the evolution of a CTMC is ``synchronized'' with an
accepting run of $\mc{A}$ by considering the state labels in a CTMC, i.e., atomic
propositions, as letters read by $\mc{A}$.
As $\mc{A}$ does not constrain the timing of events in the CTMC $\mc{C}$, it
suffices to take a synchronous product of $\mc{A}$ and $\mc{C}$'s embedded
DTMC, denoted $\emb(\mc{C})$, which is obtained by just ignoring the random
state residence times in $\mc{C}$ while keeping all other ingredients, in particular
the transition probabilities and state labels.
For finite acceptance criteria, the probability that $\mc{C} \models \mc{A}$, i.e., the
probability of the set of paths in $\mc{C}$ that are accepted by $\mc{A}$,
$\Pr(\mc{C} \models \mc{A})$ for short, is obtained as the reachability probability in
the product $\emb(\mc{C}) \otimes \mc{A}$ of the final states in $\mc{A}$.
Since $\mA$ is deterministic, $\emb(\mC)\otimes\mA$ is a DTMC.
In case of Muller acceptance criteria, $\Pr(\mc{C} \models \mc{A})$ corresponds
to the reachability probability of accepting terminal strongly connected components
in $\emb(\mc{C}) \otimes \mc{A}$.  This follows directly from results in~\cite{CY95}.
The reachability probabilities in a DTMC can be obtained by solving a system of linear
equations whose size is linear in the size of the DTMC, see, e.g., \cite{HJ94}.

In this paper, we consider the verification of CTMCs against linear \emph{real-time}
specifications that are given as deterministic \emph{timed} automata (DTA)~\cite{AD94}.
That is to say, we explore the following problem: given a \CTMC\ $\mc{C}$, and a
linear real-time specification provided as a \emph{deterministic timed automaton}
$\mc{A}$, what is the probability of the set of paths of $\mc{C}$ that are accepted
by $\mc{A}$, i.e., what is $\Pr(\mc{C} \models \mc{A})$?

\begin{exa}
Let us illustrate the usage of \DTA\ specifications by means of a small example.
Consider a robot randomly moving in some area. It starts in some zone ($A$, say)
and has to reach zone $B$ within 10 time units, cf.\ Figure\,\ref{fig:Robot_CTMC}.
(For simplicity, all zones on the map are equally-sized, but this is not a restriction.)
The robot randomly moves through the zones, and resides in a zone for an
exponentially distributed amount of time.
The robot may pass through all zones to reach $B$, but should not stay longer
than 2 time units in any gray zone.
The specification ``reach $B$ from $A$ within 10 time units while residing in any
gray zone for at most 2 time units'' is modeled by a simple \DTA\ which accepts
once location $q_2$ is reached, cf.\ Figure\,\ref{fig:Robot_DTA}.
Clock $x$ controls the timing constraint on the residence times of the gray zones
(assumed to be labeled with $g$), while clock $y$ controls the global time constraint
to reach zone $B$.
In state $q_0$, the robot traverses non-gray zones, in $q_1$ gray zones, and in
$q_2$ it has reached the goal zone $B$.
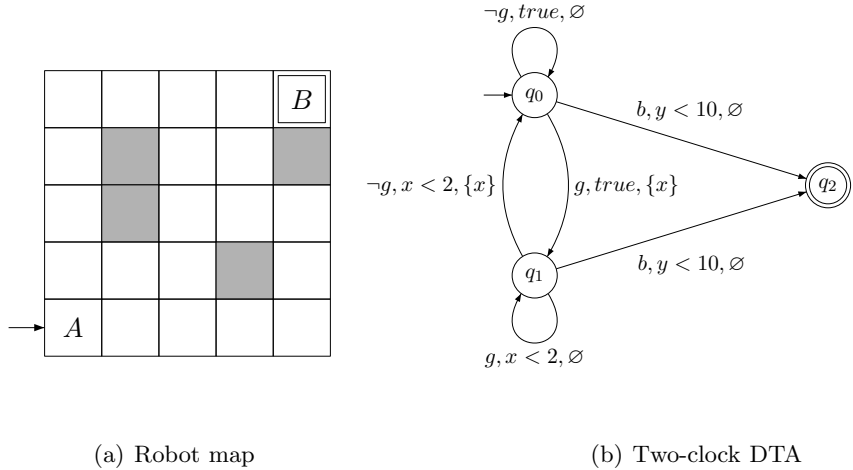
\begin{figure}[h]
\begin{center}
\hspace{-0.7cm}
\subfigure[Robot map]{\scalebox{0.95}{ \begin{picture}(59,62)(0,-62)
\put(0,-62){}
\node[Nmr=0.0](n1)(16.0,-16.0){}

\node[Nmr=0.0](n2)(24.0,-16.0){}

\node[Nmr=0.0](n3)(32.0,-16.0){}

\node[Nmr=0.0](n4)(40.0,-16.0){}

\node[NLangle=0.0,Nmarks=r,Nmr=0.0](n5)(48.0,-16.0){$B$}

\node[Nmr=0.0](n6)(16.0,-24.0){}

\node[Nfill=y,fillgray=0.7,Nmr=0.0](n7)(24.0,-24.0){}

\node[Nmr=0.0](n8)(32.0,-24.0){}

\node[Nmr=0.0](n9)(40.0,-24.0){}

\node[Nfill=y,fillgray=0.7,Nmr=0.0](n10)(48.0,-24.0){}

\node[Nmr=0.0](n11)(16.0,-32.0){}

\node[Nfill=y,fillgray=0.7,Nmr=0.0](n12)(24.0,-32.0){}

\node[Nmr=0.0](n13)(32.0,-32.0){}

\node[Nmr=0.0](n14)(40.0,-32.0){}

\node[Nmr=0.0](n15)(48.0,-32.0){}

\node[Nmr=0.0](n16)(16.0,-40.0){}

\node[Nmr=0.0](n17)(24.0,-40.0){}

\node[Nmr=0.0](n18)(32.0,-40.0){}

\node[Nfill=y,fillgray=0.7,Nmr=0.0](n19)(40.0,-40.0){}

\node[Nmr=0.0](n20)(48.0,-40.0){}

\node[NLangle=0.0,Nmarks=i,Nmr=0.0](n21)(16.0,-48.0){$A$}

\node[Nmr=0.0](n22)(24.0,-48.0){}

\node[Nmr=0.0](n23)(32.0,-48.0){}

\node[Nmr=0.0](n24)(40.0,-48.0){}

\node[Nmr=0.0](n25)(48.0,-48.0){}

\end{picture}\label{fig:Robot_CTMC}}}\hspace{0.6cm}
\subfigure[Two-clock DTA]{\scalebox{0.75}{\begin{picture}(88,75)(0,-75)
\put(0,-75){}
\node[NLangle=0.0,Nmarks=i](n26)(16.0,-16.0){$q_0$}

\node[NLangle=0.0](n27)(16.0,-48.0){$q_1$}

\node[NLangle=0.0,Nmarks=r](n28)(68.0,-32.0){$q_2$}

\drawedge(n26,n28){$b,y<10,\emptyset$}

\drawedge[ELside=r,ELdist=1.83](n27,n28){$b,y<10,\emptyset$}

\drawloop(n26){$\neg g,true,\emptyset$}%

\drawloop[loopangle=-90.0](n27){$g,x<2,\emptyset$}

\drawedge[curvedepth=6.0](n26,n27){$g,true,\{x\}$}

\drawedge[curvedepth=5.64](n27,n26){$\neg g,x<2,\{x\}$}

\end{picture}\label{fig:Robot_DTA}}}
\end{center}
\caption{A robot example\label{fig:robot}}
\end{figure}
\end{exa}

Like in the untimed setting discussed before, we consider two variants: DTA that accept
finite timed words, and DTA that accept infinite timed words according to a Muller
acceptance condition.
(Note that DTA with Muller acceptance condition are strictly more expressive than
DTA with B\"uchi acceptance conditions~\cite{AD94}.)
The considered verification problem is substantially harder than the case for untimed
linear specifications, e.g., as the DTA may constrain the timing of events in $\mc{C}$, it
does not suffice to take the embedded DTMC $\emb(\mc{C})$ as starting-point.
In addition, the product of a CTMC and a DTA is neither a CTMC nor a DTA, and has an
infinite state space.
It is unclear which (and whether a) stochastic process is obtained from such infinite
product, and if so, how to analyze it.

We tackle the verification of a finite CTMC against a DTA specification as follows:
\begin{enumerate}[(1)]
\item
We first show that the problem $\mc{C} \models \mc{A}$ is well-defined in the
sense that the set of paths of $\mc{C}$ that are accepted by $\mc{A}$ is measurable.
\item
We define the product $\mc{C} \otimes \mc{A}$ for \CTMC\ $\mathcal{C}$ and \DTA\
$\mathcal{A}$ as a variant of \DTA\ in which, besides the usual ingredients of timed
automata like guards and clock resets, the location residence time is exponentially
distributed, and define a probability space over sets of timed paths in this model.
In particular, we show that the probability of $\mc{C} \models \mc{A}$ coincides with
the reachability probability of accepting paths in $\mc{C} \otimes \mc{A}$.
\item
We adapt the standard region construction for timed automata \cite{AD94} to this
variant of \DTA, and show that the thus obtained region automata are in fact \emph{%
piecewise deterministic Markov processes} (\PDP s) \cite{Dav93}, a model that is frequently
used in, e.g., stochastic control theory and financial mathematics.
The characterization of region automata as PDPs sets the ground for obtaining the
following results concerning qualitative and quantitative verification of CTMCs against
DTA.
\item
For finite acceptance criteria, we show that $\Pr(\mc{C} \models \mc{A})$ equals the
reachability probability in the embedded PDP of $\mc{C} \otimes \mc{A}$.
Under Muller acceptance criteria, $\Pr(\mc{C} \models \mc{A})$ equals the reachability
probability of accepting terminal strongly connected components in this embedded
PDP.
In case of qualitative verification ---does CTMC $\mc{C}$ satisfy $\mc{A}$ with probability
larger than zero, or equal to one?--- a graph traversal of the (embedded) PDP suffices.
\item
We then show that reachability probabilities in our PDPs can be characterized as the
least solution of a system of \emph{Volterra integral equations} of the second type
\cite{AWW95}.
This probability is shown to be approximated by the solution of a system of partial
differential equations (PDEs).
\item For the case of single-clock DTA, we show that the system of integral equations can be
transformed into a system of \emph{linear equations}, whose coefficients are solutions
of some ordinary differential equations (ODEs).
For these coefficients either an analytical solution (for small state space) can be obtained
or an arbitrarily closely approximated solution can be determined efficiently.
\end{enumerate}

\subsection*{Related work}
Model checking CTMCs against linear real-time specifications has received scant attention
so far.
To our knowledge, this issue has only been (partly) addressed in \cite{DHS09,BCHKS07}.
Baier et al.~\cite{BCHKS07} define the logic asCSL where path properties are characterized
by (time-bounded) regular expressions over actions and state formulas.
The truth value of path formulas depends not only on the available actions in a given time
interval, but also on the validity of certain state formulas in intermediate states.
asCSL is strictly more expressive than CSL~\cite{BCHKS07}.
Model checking asCSL is performed by representing the regular expressions as finite-state
automata, followed by computing time-bounded reachability probabilities in the product
of CTMC $\mc{C}$ and this automaton.
In \CSLTA~\cite{DHS09}, time constraints of until modalities are specified by single-clock
\DTA; the resulting logic is at least as expressive as asCSL~\cite{DHS09}.
The combined behavior of $\mC$ and DTA $\mA$ is interpreted as a Markov renewal process
and model checking \CSLTA\ is reduced to computing reachability probabilities in a \DTMC\
whose transition probabilities are given by subordinate \CTMC s.
This paper takes a completely different approach.
The technique of \cite{DHS09} cannot be generalized to multiple clocks, whereas our approach
does not restrict the number of clocks and thus supports more specifications than \CSLTA.
The DTA specification of our robot example, for instance, can neither be expressed in \CSLTA\
nor in asCSL.
For the single-clock case, our approach produces the same result as~\cite{DHS09}, but
yields a (in our opinion) conceptually simpler formulation whose correctness can be
derived by simplifying the system of integral equations obtained for the general case.
Moreover, measurability has not been addressed in \cite{DHS09}.
Other related work~\cite{BBBBG07,BBBBG08,BBBM08} provides a quantitative interpretation
to timed automata where delays and discrete choices are interpreted probabilistically.
In this approach, delays of unbounded clocks are governed by exponential distributions
like in \CTMC s.
Decidability results have been obtained for almost-sure properties~\cite{BBBBG08} and
quantitative verification~\cite{BBBM08} for (a subclass of) single-clock timed automata.

\subsection*{Organization of the paper.}
Section~\ref{sec:prelim} defines the three models that are central to this paper: CTMCs,
DTA, and PDPs.
Section~\ref{sec:DTA} shows that the set of paths in CTMC $\mc{C}$ accepted by DTA
$\mc{A}$ is measurable and coincides with reachability probabilities in the product
$\mc{C} \otimes \mc{A}$.
It also shows that the underlying region graph of $\mc{C} \otimes \mc{A}$ is a (simple
instance of a) PDP.
Section~\ref{sec:finite} constitutes the main part of the paper and deals with the verification
of DTA with finite acceptance conditions, and analyzes the quantitative reachability problem
in our PDPs, for both the general case and single-clock DTA.
Section~\ref{sec:infinite} considers DTA with Muller acceptance criteria, as well as qualitative
verification.
Finally, section~\ref{sec:concl} concludes.

This paper extends the conference paper~\cite{CHKM09_LICS} with complete proofs,
illustrative examples, and by considering Muller acceptance criteria.

\section{Preliminaries}\label{sec:prelim}

Given a set $H$, let $\Pr:\mc{F}(H)\to[0,1]$ be a probabi\-li\-ty
measure on the measurable space $(H,\mc{F}(H))$, where $\mc{F}(H)$
is a $\sigma$-algebra over $H$. Let $\Distr(H)$ denote the set of
probability measures on this measurable space.

\subsection{Continuous-time Markov chains}
\begin{defi}[\CTMC] A (labeled) \emph{continuous-time Markov chain} (\CTMC) is a
tuple $\mc{C}=(S,\AP,L,\alpha,\P,E)$ where $S$ is a \emph{finite}
set of \emph{states}; \ \ $\AP$ is a finite set of \emph{atomic
propositions}; \ \ $L:S\to 2^{\AP}$ is the \emph{labeling
function};\ \ $\alpha\in \Distr(S)$ is the \emph{initial
distribution};\ \ $\P:S\times S\to [0,1]$ is a stochastic
\emph{transition probability matrix}; \ and $E:S\to \Reals_{\ges
0}$ is the \emph{exit rate function}.
\end{defi}
The probability to exit state $s$ in $t$ time units is given by $\int_{0}^{t}
E(s){\cdot} e^{-E(s)\tau} d\tau$; the probability to take the transition
$s\to s'$ in $t$ time units equals $\P(s,s'){\cdot}\! \int_{0}^{t}E(s)
e^{-E(s){\cdot}\tau} d\tau$.
A state $s$ is \emph{absorbing} if ${\P(s,s)=1}$.
The \emph{embedded} discrete-time Markov chain (\DTMC) of \CTMC\
$\mc{C}$ is obtained by deleting the exit rate function $E$, i.e.,
$\emb(\mc{C})=(S,\AP,L,\alpha,\P)$.

\begin{defi}[Timed paths]
Let $\mc{C}$ be a \CTMC.
$\Paths^{\mc{C}}_n:=S\times{\left(\Reals_{> 0}\times S\right)}^n$
is the set of paths of length $n$ in $\mc{C}$; the set of finite
paths in $\mc{C}$ is defined by
$\Paths^{\mc{C}}_{\star}=\bigcup_{n\in\Nats}\Paths^{\mc{C}}_n$
and $\Paths^{\mc{C}}_{\omega}:={\left(S\times\Reals_{>
0}\right)}^\omega$ is the set of infinite paths in $\mc{C}$.
$\Paths^{\mc{C}}=\Paths^{\mc{C}}_{\star}\cup\Paths^{\mc{C}}_{\omega}$
denotes the set of all paths in $\mc{C}$.
\end{defi}

We denote a path $\rho\in\Paths^{\mc{C}}(s_0)$
($\rho\in\Paths(s_0)$ for short) as the sequence
$\rho=s_0\mv{t_0}s_1\mv{t_1}s_2 \cdots$ starting in state $s_0$
such that for $n\leqslant|\rho|$ ($|\rho|$ is the number of
transitions in $\rho$ if $\rho$ is finite); $\rho[n]:=s_n$ is the
$n$-th state of $\rho$ and $\rho\<n\>:=t_n$ is the time spent in
state $s_n$. Let $\rho@t$ be the state occupied in $\rho$ at
time $t\in\Reals_{\geqslant 0}$, i.e. $\rho@t:=\rho[n]$ where $n$ is
the smallest index such that $\sum_{i=0}^{n}\rho\<i\>
> t$. We assume w.l.o.g.\ $t_i>0$ for any $i$.

The definition of a Borel space on paths through \CTMC s follows
\cite{Var85,BHHK03}. A \CTMC\ $\mc{C}$ yields a probability measure ${\Pr}^\mc{C}$ on paths as
follows.  Let $s_0, \ldots\!, s_k\in S$ with $\P(s_i, s_{i+1})>0$
for $0\leqslant i<k$ and $I_0, \ldots\!, I_{k-1}$ nonempty
intervals in $\Reals_{\ges 0}$.
Let $C(s_0, I_0, \ldots\!, I_{k-1}, s_k)$ denote the \emph{cylinder
set} consisting of all paths $\rho\in\Paths(s_0)$ such that
$\rho[i]=s_i$ ($i\leqslant k$), and $\rho\<i\>\in I_i$ ($i<k$).
$\mc{F}(\Paths(s_0))$ is the smallest $\sigma$-algebra on
$\Paths(s_0)$ which contains all sets $C(s_0, I_0, \ldots\!, I_{k-1},
s_k)$ for all state sequences $(s_0,\ldots\!,s_k)\in S^{k+1}$ with
$\P(s_i, s_{i+1})>0$ ($0\leqslant i<k$) and $I_0, \ldots\!, I_{k-1}$
range over all sequences of nonempty intervals in $\Reals_{\ges 0}$.
The probability measure ${\Pr}^\mc{C}$ on $\mc{F}(\Paths(s_0))$
is the unique measure defined by induction on $k$ by
${\Pr}^{\mc{C}}(C(s_0))=\alpha(s_0)$ and for $k> 0$:
\begin{align}\label{eqn:ctmc}
&{\Pr}^{\mc{C}}\big(C(s_0, I_0, \ldots\!, I_{k-1}, s_k)\big)=
{\Pr}^{\mc{C}}\big(C(s_0, I_0, \ldots\!, I_{k-2}, s_{k-1})\big)\nonumber\\
& \quad
\cdot\int_{I_{k-1}}\P(s_{k-1},s_k)E(s_{k-1}){\cdot}e^{-E(s_{k-1})\tau}d\tau.
\end{align}

\begin{figure}
 \scalebox{0.85}{\begin{picture}(73,36)(0,-36)
\put(0,-36){}
\node[Nmarks=i](n0)(8.32,-16.3){$s_0$}

\node(n1)(32.51,-16.3){$s_1$}

\drawedge[curvedepth=3.0](n0,n1){1}

\drawedge[curvedepth=3.0](n1,n0){$0.5$}

\node(n2)(52.48,-10.19){$s_2$}

\node(n3)(52.48,-22.19){$s_3$}

\drawedge(n1,n2){0.2}

\drawedge[ELside=r,ELdist=1.31](n1,n3){0.3}

\drawloop[loopdiam=6.0,loopangle=0.0](n2){1}

\drawloop[loopdiam=6.0,loopangle=0.0](n3){1}

\node[Nframe=n](n4)(8.48,-10.16){$\{a\}$}

\node[Nframe=n](n5)(32.48,-10.16){$\{a\}$}

\node[Nframe=n](n6)(56.16,-4.16){$\{b\}$}

\node[Nframe=n](n7)(56.16,-28.16){$\{c\}$}

\node[Nframe=n](n8)(49.16,-28.16){$r_3$}

\node[Nframe=n](n9)(49.16,-4.16){$r_2$}

\node[Nframe=n](n10)(32.16,-22.16){$r_1$}

\node[Nframe=n](n11)(8.16,-22.16){$r_0$}

\end{picture}}\vspace{-0.6cm}\caption{An example CTMC\label{fig:CTMC_1c_1}}
\end{figure}
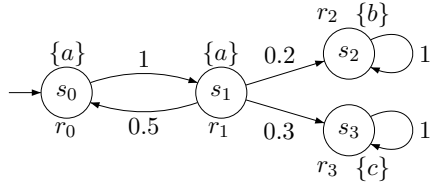

\begin{exa}
An example \CTMC\ is illustrated in Figure~\ref{fig:CTMC_1c_1}, where $\AP=\{a,b,c\}$ and $s_0$ is
the initial state, i.e., $\alpha(s_0)=1$ and $\alpha(s)=0$ for any
$s\neq s_0$. The exit rates are indicated at the states, whereas the
transition probabilities are attached to the transitions. An example
timed path is $\rho = s_0 \mv{2.5} s_1 \mv{1.4} s_0 \mv{2} s_1
\mv{2\pi} s_2 \cdots$ with $\rho[2] = s_0$ and $\rho@6 = \rho[3] =
s_1$.
\end{exa}

\subsection{Deterministic timed automata}


\newcommand{\mCC}{\mathcal{C}\mathcal{C}}
\newcommand{\mRe}{\mathcal{R}e}

Let $\mX = \{x_1,\ldots\!, x_n\}$ be a set of \emph{nonnegative} real-valued variables,
called \emph{clocks}.
An $\mc{X}$-valuation is a function $\eta:\mX\to \Reals_{\ges 0}$ assigning
to each variable $x$ a nonnegative real value $\eta(x)$.
Let $\mc{V}(\mc{X})$ denote the set of all valuations over $\mc{X}$.
A \emph{clock constraint} on $\mc{X}$, denoted by $g$, is a conjunction of expressions
of the form $x \bowtie c$ for clock $x \in \mc{X}$, comparison operator $\bowtie\ \in
\{<, \leqslant, >, \geqslant\}$ and $c \in \Nats$.
Let $\mCC(\mc{X})$ denote the set of clock constraints over $\mc{X}$.
An $\mc{X}$-valuation $\eta$ \emph{satisfies} constraint $x \bowtie c$, denoted
$\eta \models x \bowtie c$, if and only if $\eta(x) \bowtie c$; it satisfies a conjunction
of such expressions if and only if $\eta$ satisfies all of them.
Let $\vec{0}$ denote the valuation that assigns 0 to all clocks.
For a subset $X\subseteq \mc{X}$, the reset of $X$, denoted $\eta[X\!:=0]$, is the
valuation $\eta'$ such that $\forall x\in X.\ \eta'(x)\!:=0$ and
$\forall x\notin X.\ \eta'(x)\!:=\eta(x)$.
For $\delta\in \Reals_{\geqslant 0}$ and $\mc{X}$-valuation $\eta$, $\eta{+}\delta$
is the $\mc{X}$-valuation $\eta''$ such that $\forall x\in \mc{X}$. $\eta''(x)\!:=\eta(x)
{+}\delta$, which implies that all clocks proceed at the same speed.

\begin{defi}[\DTA]\ A \emph{deterministic timed automaton} (or
  \DTA\ for short)
is a tuple $\mc{A}=(\Sigma,\mc{X},Q,q_0,$ $Q_\mathbf{F},\rightarrow)$ where
$\Sigma$ is a finite \emph{alphabet}; $\mc{X}$ is a finite set of \emph{clocks};
$Q$ is a nonempty, finite set of \emph{locations} with \emph{initial location}
$q_0\in Q$; $Q_\mathbf{F}$ is the \emph{acceptance condition}, which is either:
\begin{enumerate}[$\bullet$]
\item
$Q_F\subseteq Q$, a set of \emph{accepting locations} (reachability or finite acceptance), or
\item
$Q_\mF\subseteq 2^Q$, an \emph{acceptance family} (Muller acceptance).
\end{enumerate}
The relation $\rightarrow\ \subseteq Q\times \Sigma\times \mCC(\mc{X})\times
2^{\mc{X}}\times Q$ is the \emph{edge relation} satisfying:
$$
\big(q\mv{a,g,X}q' \mbox{ and } q\mv{a,g',X'}q'' \mbox{ with } g
\neq g'\big) \quad \mbox{implies} \quad g \cap g' = \emptyset.
$$
\end{defi}

We refer to $q\mv{a,g,X}q'$ as an \emph{edge}, where $a\in \Sigma$ is an input symbol,
the \emph{guard} $g$ is a clock constraint on the clocks of $\mc{A}$, $X$ is the set of
clocks that are to be reset and $q'$ is the successor location.
Intuitively, the edge $q\mv{a,g,X}q'$ asserts that the \DTA\ $\mc{A}$ can move from
location $q$ to $q'$ when the input symbol is $a$ and the guard $g$ holds, while the
clocks in $X$ should be reset when entering $q'$.
DTA are deterministic as they have a single initial location, and outgoing edges of a
location labeled with the same input symbol are required to have disjoint guards.
In this way, the next location is uniquely determined for a given location and a given
clock valuation.
In case no guard is satisfied in a location for a given clock valuation, time can progress.
If the advance of time will never reach a situation in which a guard holds, the DTA will
stay in that location ad infinitum.
Note that DTA do not have location invariants, as in safety timed automata.
For the sake of simplicity, diagonal constraints like $x-y\bowtie c$ are not considered.
This restriction does, however, not harm the expressiveness~\cite{BPDG98}.

An (infinite) \emph{timed path} of \DTA\ $\mc{A}$ is of the form $\theta = q_0 \mv{a_0,
t_0} q_1\mv{a_1, t_1}\cdots$ such that $\eta_0 = \vec{0}$, and for all $j \ges 0$, it holds
$t_j > 0$, $\eta_j{+}t_j \models g_j$, $\eta_{j+1}=(\eta_j{+}t_j)[X_j:=0]$, where $\eta_j$
is the clock evaluation when \emph{entering} $q_j$.
The definitions on timed paths (such as $\theta[i]$, $\theta@t$, and so forth) for CTMCs
can readily be adapted for DTA.
We consider DTA with two types of acceptance criteria.
Let \DTAr\ and \DTAo\ denote the set of \DTA\ with reachability and Muller acceptance
conditions, respectively.
\DTA\ denotes the general case covering both \DTAr\ and \DTAo.

\begin{defi}[DTA accepting paths]
An infinite timed path $\theta$ is \emph{accepted} by a \DTAr\ if $\theta[i] \in Q_F$ for
some $i\ges 0$; ${\theta}$ is accepted by a \DTAo\ if ${in\!f}({\theta})\in Q_\mF$, where $in\!f({\theta})$ is the set of states $q \in Q$ such that $q=q_i$ for infinitely many $i \ges
0$.
\end{defi}

The timed path $\theta$ is accepted according to a reachability criterion if it reaches
some final location, whereas it is accepted according to a Muller acceptance condition
if the set of infinitely visited locations equals some  set in $Q_\mF$.
As a convention, we assume each location $q \in Q_F$ in \DTAr\ to be a sink.

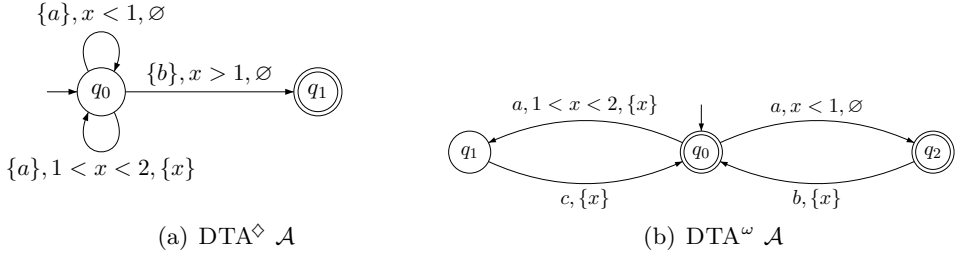
\begin{figure}\begin{center}
\subfigure[\DTAr\
 $\mc{A}$]{\scalebox{0.8}{\begin{picture}(63,32)(0,-32)
\put(0,-32){}
\node[Nmarks=i](n24)(12.16,-12.32){$q_0$}

\node[Nmarks=r](n25)(48.16,-12.32){$q_1$}

\drawloop[loopdiam=6.0](n24){$\{a\},x<1,\emptyset$}

\drawloop[loopdiam=6.0,loopangle=270](n24){$\{a\},1<x<2,\{x\}$}

\drawedge(n24,n25){$\{b\},x>1,\emptyset$}

\end{picture}\label{fig:DTA_1c_ss}}
 }\subfigure[\DTAo\
 $\mc{A}$]{\hspace{-0.8cm}\scalebox{0.7}{\begin{picture}(118,31)(0,-31)
\put(0,-31){}
\node[iangle=90.0,Nmarks=ir](n91)(64.0,-20.0){$q_0$}

\node[Nmarks=r](n92)(108.0,-20.0){$q_2$}

\node(n93)(20.0,-20.0){$q_1$}

\drawedge[curvedepth=6](n91,n92){$a,x<1,\emptyset$}

\drawedge[curvedepth=6](n92,n91){$b,\{x\}$}

\drawedge[ELside=r,curvedepth=-6](n91,n93){$a,1<x<2,\{x\}$}

\drawedge[ELside=r,curvedepth=-6](n93,n91){$c,\{x\}$}

\end{picture}\label{fig:DTMA_Muller_s}}
 }
\end{center}\caption{DTA with (a) reachability and (b) Muller acceptance conditions \label{fig:DTMA_Muller_sss}}
\end{figure}

\begin{exa}
Figure~\ref{fig:DTA_1c_ss} depicts an example \DTAr\ over the alphabet $\{a,b\}$
with initial location $q_0$.
The timed automaton is deterministic as $q_0$ is the only initial location and both
$a$-labeled edges have disjoint guards.
Any timed path ending in $Q_F = \{q_1\}$ is accepting.

Figure~\ref{fig:DTMA_Muller_s} depicts an example \DTAo\ over the alphabet $\{a,b,c\}$.
Its initial location is $q_0$; its Muller acceptance family equals $Q_\mF = \big\{\{ q_0,
q_2\}\big\}$.
Any accepting path should cycle between the locations $q_0$ and $q_1$ \emph{finitely}
often, and between $q_0$ and $q_2$ \emph{infinitely} often.
\end{exa}

\begin{rem}$[$Expressive power of \DTAo$]$
\DTAo \ is the set of (deterministic) timed Muller automata, \emph{(D)MTA}, for short.
A (deterministic) timed B\"uchi automaton, \emph{(D)TBA} for short, has a set $Q_F$ of
accepting locations, and accepts an infinite timed path $\theta$ if $\theta$ visits
some location in $Q_F$ infinitely often, i.e., $\inf(\theta) \, \cap \, Q_F\neq\emptyset$.
The expressive power of \emph{(D)TMA} and \emph{(D)TBA} is related as follows~\cite{AD94}:
\[ \emph{TMA} = \emph{TBA} > \emph{DTMA} > \emph{DTBA}.\]
Note that in nondeterministic \emph{TMA} and \emph{TBA}, guards on edges emanating from a
location may overlap.
\emph{DTMA} are closed under all Boolean operators $($union, intersection,
and complement$)$, while \emph{DTBA} are \emph{not} closed under complement.
\end{rem}

\begin{rem}$[$Successor location$]$
Since DTA are deterministic, the edge relation $\rightarrow$ can be replaced
by a (partial) function $succ: Q\times \Sigma\times \mCC(\mc{X}) \mapsto 2^{\mc{X}}
\times Q$.
If only the successor location is of interest, we simpy use the function
$\widetilde{succ}: Q \times \Sigma \times \mCC(\mc{X}) \mapsto Q$, i.e., $q' =
\widetilde{succ}(q,a,g)$.
\end{rem}

\subsection{Piecewise-deterministic Markov processes}\label{sec:PDP}

PDPs~\cite{Dav84} constitute a general model for stochastic systems without
diffusions~\cite{Dav93} and has been applied to a variety of problems in
engineering, operations research, management science, and economics.
Powerful analysis and control techniques for PDPs have been developed%
~\cite{Len85,Len91,Cos88}.
A \PDP\ is a hybrid stochastic process involving discrete control (i.e.,
locations) and continuous variables.

Let us introduce some auxiliary notions.
Let $\mX = \{x_1, \ldots, x_n\}$ be a set of variables in $\Reals$.
Note that clock variables are a special case of these variables.
A \emph{constraint} over $\mX$, denoted by $g$, is a subset of $\Reals^n$.
Let $\mB(\mX)$ denote the set of constraints over $\mX$.
An $\mX$-valuation $\eta$ satisfies constraint $g$, denoted $\eta \models
g$, if and only if $(\eta(x_1),...,\eta(x_n)) \in g$.
For $g \in \mc{B}(\mc{X})$, a constraint over $\mc{X} = \{ \, x_1, \ldots, x_n
\, \}$, let $\overline{g}$ be the closure of $g$, $\mathring{g}$ the interior
of $g$, and $\partial g = \overline{g} \setminus \mathring{g}$ the boundary
of $g$.
For instance, for $g = {x_1^2 - 2x_2 \les 1.5} \wedge {x_3> 2}$, we have
$\mathring{g} = {x_1^2-2x_2< 1.5 \wedge x_3>2}$, $\overline{g} = x_1^2
- 2x_2 \leqslant 1.5 \wedge x_3 \geqslant2$, and $\partial g$ equals $x_1^2
- 2x_2 = 1.5\wedge x_3 = 2$.

To each control location $z$ of a PDP, an \emph{invariant} $\Inv(z)$ is associated,
a constraint over $\mc{X}$ which constrains the variable values in $z$.
The state of a PDP is a pair $(z, \eta)$ with control location $z$ and $\eta$
a variable valuation.
Let $\mb{S} = \left\{ \, (z,\eta) \mid z\in Z, \eta \models \Inv(z) \, \right\}$,
where $Z$ is the set of locations.
The notions of closure, interior and boundary can be lifted to $\mb{S}$ in
a straightforward manner, e.g., $\partial\mb{S} = \bigcup_{z\in Z} \{z\}
\times \partial \Inv(z)$ is the boundary of $\mb{S}$; $\mathring{\mb{S}}$
and $\overline{\mb{S}}$ are defined in a similar way.
\begin{defi}[\PDP\,\cite{Dav93}]
A \emph{piecewise-deterministic $($Markov$)$ process} (PDP) is a tuple $\mc{Z}
= \left(Z,\mc{X},\Inv,\phi,\Lambda,\mu\right)$ where $Z$ is a finite set of
\emph{locations}, $\mc{X}$ is a finite set of \emph{variables}, $\Inv: Z \to
\mc{B}(\mc{X})$ is an \emph{invariant function}, and
\begin{enumerate}[$\bullet$]
\item
$\phi: Z \times \mc{V}(\mc{X}) \times \Reals \to \mc{V}(\mc{X})$ is a \emph{flow
function}, which is the solution of a system of ODEs with a Lipschitz continuous vector
field,
\item
$\Lambda: \mb{S} \to \Reals_{\ges 0}$ is an \emph{exit rate function}
satisfying for any $\xi\in\mb{S}$:
$$
\ \  \hspace{2.5cm} \exists \epsilon(\xi)>0.\, \mbox{function } t\mapsto \Lambda(\xi\oplus t)
\mbox{ is integrable on } [0,\epsilon(\xi)),\hspace{2.5cm} (\triangle)
$$
where $(z,\eta)\oplus t =\big(z,\phi(z,\eta,t)\big)$,  and
\item $\mu:
\bar{\mb{S}}\rightarrow\Distr(\mb{S})$
is the \emph{transition probability function} satisfying:
$$
\mu(\xi,\{\xi\}) = 0 \quad \mbox{and} \quad
\xi\mapsto \mu(\xi,A) \mbox{ is measurable for any } A\in\mc{F}(\mb{S}),
$$
where $\mu(\xi,A)$ denotes $(\mu(\xi))(A)$, $\mc{F}(\mb{S})$ is a $\sigma$-algebra
generated by 
$\bigcup_{z\in Z}\{z\}\times A_z$ with $A_z \subseteq \mc{F}(\Inv(z))$, and
$\mc{F}(\Inv(z))$ is a $\sigma$-algebra generated by $\Inv(z)$.
\end{enumerate}
\end{defi}



\noindent Let us explain the behavior of a PDP.
A PDP can reside in a state $\xi = (z,\eta) \in \mathring{\mb{S}}$ as long as
$\Inv(z)$ holds.
In state $\xi = (z,\eta)$, the \PDP\ can either \emph{delay} or take a
\emph{Markovian jump}.
Delaying by $t$ time units yields the next state $\xi' = \xi \oplus t$, i.e., the
PDP remains in location $z$ while all its continuous variables are updated
according to $\phi(z,\eta,t)$.
The flow function $\phi$ defines the time-dependent behavior in a single
location, in particular, it specifies how the variable valuations change when
time elapses.
In case of a Markovian jump in state $\xi$, the next state $\xi'' = (z'',\eta'')
\in \mb{S}$ is reached with probability $\mu(\xi,\{\xi''\})$.
The residence time of a state is exponentially distributed; this is defined
by the function $\Lambda$.
A third possibility for a PDP to evolve is by taking \emph{forced transitions}.
When the variable valuation $\eta$ satisfies the boundary of the invariant, i.e.,
$\eta \models \partial\Inv(z)$, the \PDP\ is forced to take a boundary jump,
i.e., it has to leave state $\xi$.
With probability $\mu(\xi,\{\xi''\})$ it then moves to state $\xi''$.
For any $T \in \Reals_{\ges 0}$, the function $\Lambda$ is integrable as the
interval $[0,T]$ can be divided into finitely many small intervals, on which
by equation $(\triangle)$, the function $\Lambda$ is integrable.

A \PDP\ is named piecewise-deterministic because in each location (one piece)
the behavior is deterministically determined by the flow function $\phi$.
The PDP is Markovian as the current state contains all the information to
determine the future progress of the PDP.

\subsection{Embedded PDP}
The embedded  \emph{discrete-time Markov process} (\DTMP) $\emb(\mc{Z})$
of the \PDP\ $\mc{Z}$ has the same state space $\mb{S}$ as $\mc{Z}$ and is
equipped with a transition probability function $\hat{\mu}$.
The one-jump \emph{transition probability} from a state $\xi$ to a set $A
\subseteq \mb{S}$ of states (with different location as $\xi$), denoted
$\hat\mu(\xi, A)$, is given by~\cite{Dav93}:
\begin{eqnarray}
\hat\mu(\xi,A)&=&\int_0^{{\flat}(\xi)}\hspace{-0.1cm}(\mc{Q}\mathbf{1}_{A})(\xi\oplus
t){\cdot}\Lambda\left(\xi\oplus t\right)
e^{-\int_0^t\Lambda\left(\xi\oplus \tau\right)d\tau}\ dt \label{eq:embedded}\\
&+&(\mc{Q}\mathbf{1}_{A})(\xi\oplus\flat(\xi))\cdot
e^{-\int_0^{\flat(\xi)}\Lambda\left(\xi\oplus\tau\right)d\tau} \qquad\qquad\quad
\label{eq:embedded2}
\end{eqnarray}
where $\flat(\xi) = \inf\{{t>0}\!\mid\!{\xi \oplus t \in \partial\mb{S}}\}$ is the
minimal time to hit the boundary if such time exists; $\flat(\xi)= \infty$ otherwise.
$(\mc{Q}\mathbf{1}_{A})(\xi)=\int_{\mb{S}}\mathbf{1}_{A}(\xi')\mu(\xi,d\xi')$
is the accumulative (one-jump) transition probabi\-li\-ty from $\xi$ to $A$ and
$\mathbf{1}_{A}(\xi)$ is the characteristic function such that $\mathbf{1}_{A}(\xi)
= 1$ when $\xi \in A$ and $\mathbf{1}_{A}(\xi) = 0$ otherwise.
Term~\eqref{eq:embedded} specifies the probability to delay to state ${\xi\oplus t}$
(on the same location) and take a Markovian jump from ${\xi\oplus t}$ to $A$.
Note the delay $t$ can take a value from $[0,\flat(\xi))$.
Term~\eqref{eq:embedded2} is the probability to stay in the same location for
$\flat(\xi)$ time units and then it is forced to take a boundary jump from $\xi \oplus
\flat(\xi)$ to $A$ since $\Inv(z)$ will be by any delay invalid.

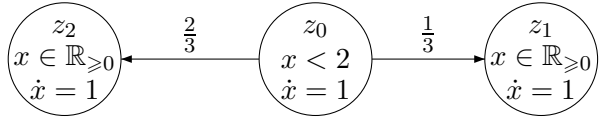
\begin{figure} \begin{center}\scalebox{1}{
\begin{picture}(85,20)(0,-20)
\node[Nw=15.0,Nh=15.0,Nmr=9.0](n0)(43.61,-8.31){}

\node[Nframe=n](n1)(43.61,-4.31){$z_0$}

\node[Nframe=n](n2)(43.61,-8.31){$x<2$}

\node[Nframe=n](n3)(43.61,-12.31){$\dot{x}=1$}

\node[Nw=15.0,Nh=15.0,Nmr=9.0](n4)(73.61,-8.31){}

\drawedge(n0,n4){$\frac{1}{3}$}

\node[Nframe=n](n5)(73.61,-4.31){$z_1$}

\node[Nframe=n](n6)(73.61,-8.31){$x\in \Reals_{\ges 0}$}

\node[Nframe=n](n7)(73.61,-12.31){$\dot{x}=1$}

\node[Nw=15.0,Nh=15.0,Nmr=9.0](n9)(10.25,-8.31){}

\node[Nframe=n](n10)(10.25,-4.31){$z_2$}

\node[Nframe=n](n11)(10.25,-8.31){$x\in \Reals_{\ges 0}$}

\node[Nframe=n](n12)(10.25,-12.31){$\dot{x}=1$}

\drawedge[ELside=r,ELdist=0.89](n0,n9){$\frac{2}{3}$}
\end{picture}
}\end{center}\vspace{-0.3cm}\caption{An example \PDP\ with constant exit rate
$5$ and boundary measure $\mu\big((z_0,2),\{(z_1,2)\}\big):=1$\label{fig:PDP}}
\end{figure}

\begin{exa}
Figure~\ref{fig:PDP} depicts a $3$-location \PDP\ $\mc{Z}$ with $\mc{X} = x$,
where $\Inv(z_0) = x < 2$ and $\Inv(z_1) = \Inv(z_2) = x \in \Reals_{\ges 0}$.
Solving $\dot{x}=1$ yields the flow function $\phi(z_i, \eta(x), t) = \eta(x){+}t$
for $i=0,1,2$.
The state space of $\mc{Z}$ is $\mb{S}=\{(z_0,\eta)\mid \eta(x) < 2\} \cup
\{(z_1, \Reals_{\ges 0})\} \cup \{(z_2, \Reals_{\ges 0})\}$.
Let exit rate $\Lambda(\xi) = 5$ for any $\xi \in \mb{S}$.
For $\eta\models \Inv(z_0)$, let
$\mu\big((z_0,\eta),\{(z_1,\eta)\}\big):=\frac{1}{3}$,
$\mu\big((z_0,\eta),\{(z_2,\eta)\}\big):=\frac{2}{3}$ and the
boundary measure be given as $\mu\big((z_0,2),\{(z_1,2)\}\big):=1$.
The time for $\xi_0 = (z_0, 0)$ to hit the boundary is $\flat(\xi_0) = 2$.
For set of states $A = \{ (z_1,\Reals) \}$ and state $\xi_0$,
$(\mc{Q}\mathbf{1}_{A})(\xi_0 \oplus t) = \frac{1}{3}$ if $t{<}2$, and
$(\mc{Q}\mathbf{1}_{A})(\xi_0 \oplus t) = 1$ if $t{=}2$.
This yields for the transition probability from state $\xi_0$ to $A$ in
$\emb(\mc{Z})$ is:
$$
\hat\mu(\xi_0, A)
\ = \
\int_0^{2}\frac{1}{3}{\cdot} 5{\cdot} e^{-\int_0^t 5 \ d\tau} \  dt+1{\cdot}e^{-\int_0^2 5 \ d\tau}
\ = \
\frac{1}{3}+\frac{2}{3}e^{-10}.
$$
\end{exa}

\section{The Product of a CTMC and a DTA}\label{sec:DTA}

In this section, we will make the first steps towards the quantitative and qualitative
verification of CTMCs against linear real-time properties specified by \DTA.
The aim is to computing the probability of the set of paths in \CTMC\ $\mC$
accepted by a \DTA\ $\mA$, i.e., $\Pr(\mc{C} \models \mc{A})$.
We first prove that this question is well-defined, i.e., that this set of paths is
measurable.
The next step is to define the product of a CTMC $\mc{C}$ and a DTA $\mc{A}$.
As we will see, this is neither a CTMC nor a DTA, but a mixture of the two.
We define the semantics of such products and define a probability space on their
paths.
The central result of this section is that $\Pr(\mc{C} \models \mc{A})$ equals
the reachability probability in the product of $\mC$ and $\mA$, cf. Theorem
\ref{th:CTMC=MTA}.
In order to facilitate the effective computation of these reachability probabilities,
we adapt the region construction of timed automata to the product $\mc{C}
\otimes \mc{A}$, and show that this yields a PDP.
The analysis of these PDPs will be the subject of the next two sections.

To simplify the notations, we assume w.l.o.g.\ that a \CTMC\ has a single initial
state $s_0$, i.e., $\alpha(s_0) = 1$, and $\alpha(s) = 0$ for $s \neq s_0$.
The state labels of the CTMC will act as input symbols of the DTA.
Thus, the alphabet of DTA equals the powerset of the atomic propositions, i.e.,
$2^\AP$.
A timed path in a CTMC is accepted by a DTA $\mA$ if there exists a corresponding
accepting path in $\mA$.

\begin{defi}[CTMC paths accepted by a DTA]
Let \CTMC\ $\mC = (S, \AP, L, s_0, \P, E)$ and \DTA\ $\mA = (2^\AP, \mc{X}, Q,
q_0, Q_\mathbf{F},\rightarrow)$.
The \CTMC\ path $s_0\mv{t_0}s_1\mv{t_1}s_2\cdots $ is \emph{accepted
by $\mA$} if there exists a corresponding \DTA\ path
\[
q_0
\mv{L(s_0),t_0}
\underbrace{\widetilde{succ}\big(q_0,L(s_0),g_0\big)}_{= q_1}
\mv{L(s_1),t_1}
\underbrace{\widetilde{succ}\big(q_1,L(s_1),g_1\big)}_{= q_2}\cdots
\]
which is accepted by $\mA$, where $\eta_0 = \vec{0}$, $g_i$ is the (unique) guard
in $q_i$ such that $\eta_i{+}t_i \models g_i$ and $\eta_{i+1}=(\eta_i{+}t_i)[X_i:=0]$,
and $\eta_i$ is the clock evaluation when entering $q_i$, for all $i$.
\end{defi}

\subsection{Measurability}
The quantitative verification of \CTMC\ $\mC$ against \DTA\ $\mA$ amounts to
compute the probability of the set of paths in $\mC$ that is accepted by $\mA$.
Formally, let
\[
\Paths^{\mc{C}}(\mc{A}) \ = \
\{\; \rho\in\Paths^{\mc{C}} \mid \rho \mbox{ is }\mbox{accepted} \mbox{ by }
\DTA\ \mc{A}\;\}.
\]
We first prove its measurability:

\begin{thm}\label{lem:measurability}
For any \CTMC\ $\mc{C}$ and \DTA\ $\mc{A}$, $\Paths^{\mc{C}}(\mc{A})$ is measurable.
\end{thm}

\proof
It suffices to show that $\Paths^{\mc{C}}(\mc{A})$ can be written as a finite union or
intersection of measurable sets.
The proof is split in two parts: DTA with (1) reachability acceptance, and (2) Muller acceptance.
The proof of the first case is carried out by (1a) considering DTA that only contain strict
inequalities as guards, (1b) equalities, and (1c) non-strict inequalities.
(Note that constraint $x = K$ can be obtained by $x > K \wedge x \geq K$).
\begin{enumerate}[\hbox to8 pt{\hfill}]
\item\noindent{\hskip-12 pt\bf (1a):}\
Let \DTAr\ $\mA$ only contain strict inequalities as clock constraints.
As all accepting paths are finite, $\Paths^{\mc{C}}(\mc{A}) = \bigcup_{n\in \Nats}
\Paths^{\mc{C}}_n(\mc{A})$, where $\Paths^{\mc{C}}_n(\mc{A})$ is the set of paths
of length $n$ accepted by $\mc{A}$.
Let $\rho = s_0 \mv{t_0} s_1\cdots s_{n-1} \mv{t_{n-1}} s_n \in \Paths^{\mc{C}}_n(\mc{A})$.
Then there exists a corresponding path $\theta = q_0 \mv{L(s_0),t_0} q_1 \cdots q_{n-1}
\mv{L(s_{n-1}),t_{n-1}} q_n$ of $\mc{A}$ which is induced by the sequence:
\[
q_0 \mv{L(s_0), g_0, X_0} q_1 \cdots
q_{n-1} \mv{L(s_{n-1}), g_{n-1}, X_{n-1}} q_n,
\]
with $q_n \in Q_F$ such that there exist $\{\eta_i\}_{0\leqslant i< n}$
with 1) $\eta_0 = \vec{0}$; 2) $\eta_i{+}t_i \models g_{i}$; and 3) $\eta_{i+1} = (\eta_i{+}t_i)[X_i:=0]$, where $\eta_i$ is the clock valuation when entering $q_i$.

We prove the measurability of $\Paths^{\mc{C}}_n(\mc{A})$ by showing that for any path
\[\rho = s_0 \mv{t_0} \cdots\! \mv{t_{n-1}}s_n\in \Paths^{\mc{C}}_n(\mc{A}),\] there exists
a cylinder set $C(s_0,I_0,\ldots\!,I_{n-1},s_n)$ ($C_\rho$ for short) such that:
\begin{equation}\label{prfobli}
\rho \in C_\rho \quad \mbox{and} \quad C_\rho \subseteq \Paths^\mc{C}_n(\mc{A})
\quad \mbox{for} \quad |\rho|=n.
\end{equation}
This is proven in two steps:
\begin{enumerate}[a.]
\item
($\rho \in C_\rho$.)
Let $\rho = s_0 \mv{t_0} \cdots\! \mv{t_{n-1}}s_n\in \Paths^{\mc{C}}_n(\mc{A})$.
We define $C_\rho$ by considering intervals $I_i$ with rational bounds that are based on
$t_i$.
Let $I_i = [t_i^-, t_i^+]$ such that $t^-_i = t^+_i := t_i$ if $t_i \in \Rationals$, and
$t_i^-, t_i^+ \in \mb{Q}$ otherwise, such that:
$$
t_i^-\leqslant t_i\leqslant t_i^+, \quad \lfloor t_i^- \rfloor = \lfloor t_i \rfloor, \quad
\lceil t_i^+ \rceil = \lceil t_i \rceil, \quad \mbox{and} \quad
t_i^+-t_i^-< \dfrac{\Delta}{2\cdot n}
$$
where
$\Delta = \displaystyle \min_{0 \leqslant j< n,\ x\in \mc{X} }\Big\{ \, \{\eta_j(x){+}t_j\}, 1-\{\eta_j(x){+}t_j\} \ \big|\  \{\eta_j(x){+}t_j\}\neq 0 \, \Big\}$,
with $\{\cdot\}$ denoting the fractional part.
Since DTA $\mc{A}$ only contains strict inequalities, for any $i$ with $\eta_i{+}t_i \models
g_i$, it follows $\{ \eta_i(x){+}t_i \}\neq 0$.
\item
($C_\rho \subseteq \Paths^\mc{C}_n(\mc{A})$.)
Let $\rho' := s_0 \mv{t_0'} \cdots \!\mv{t_{n-1}'} s_n \in C_\rho$.
Let $\eta_0': = \vec{0}$ and $\eta'_{i+1} := (\eta'_i{+}t_i')[X_i:=0]$.
It remains to show that $\eta_i'{+}t_i' \models g_i$.
Observe that $\eta_0' = \eta_0$, and for any $i > 0$ and clock variable $x$,
$$
\qquad \qquad
\big|\eta'_i(x)-\eta_i(x)\big| \ \leqslant \
\sum_{j=0}^{i-1}\big|t_j'-t_j\big| \ \leqslant \
\sum_{j=0}^{i-1} t_j^+-t_j^-  \ \leqslant \
n\cdot (t_j^+-t_j^-) \ \leqslant \
\frac{\Delta}{2}.
$$
Given that guard $g_i$ only contains strict inequalities, it follows $\eta_i'{+}t_i' \models g_i$.
This can be seen as follows.
Let $g_i = x > K$ for some natural $K$.
As $|\eta_i'(x)-\eta_i(x)| \leqslant \frac{\Delta}{2}$ and $|t_i'-t_i|< \frac{\Delta}{2}$,
it follows $|(\eta_i'(x){+}t_i') - (\eta_i(x){+}t_i)|< \Delta$.
Note that $\eta_i(x){+}t_i > K$, and thus
$\eta_i(x){+}t_i- \{ \eta_i(x){+}t_i\} = \lceil \eta_i(x){+}t_i \rceil \geq K$.
Hence, $\eta_i(x){+}t_i -\Delta \geq K$ since, by definition, $\Delta\leqslant
\{\eta_i(x)+t_i\} $.
It follows that $\eta_i'(x) + t_i' > K$.
A similar argument applies to the case $x < K$ and extends to conjunctions of strict
inequalities.
Thus, $\eta_i'+t_i'\models g_{i}$, and $\rho' \in \Paths^\mc{C}_n(\mc{A})$.
\end{enumerate}
By (\ref{prfobli}) and the fact that $\Paths^\mc{C}_n(\mc{A})\subseteq \bigcup_{\rho\in
\Paths^\mc{C}_n(\mc{A})}C_\rho$, we have:
\[\Paths^\mc{C}_n(\mc{A}) =  \bigcup_{\rho\in \Paths^\mc{C}_n(\mc{A})}
C_\rho\quad\mbox{ and } \quad\Paths^\mc{C}(\mc{A}) = \bigcup_{n\in
\Nats}\bigcup_{\rho\in \Paths^\mc{C}_n(\mc{A})}C_\rho.
\]
As each interval in $C_\rho$ has rational bounds, $C_\rho$ is measurable.
It follows that $\Paths^\mc{C}(\mc{A})$ is a union of \emph{countably many}
cylinder sets, and hence is measurable.

\item\noindent{\hskip-12 pt\bf (1b):}\
Consider \DTAr\ $\mc{A}$ with equalities of the form $x = K$ for natural $K$.
Measurability is shown by induction on the number of equalities in $\mc{A}$.
The base case (only strict inequalities) has been shown above.
Now suppose there exists an edge $e = q \mv{a, g, X} q'$ in $\mc{A}$ where
$g$ contains the constraint $x=K$.
Let \DTAr\ $\mc{A}_{e}$ be obtained from $\mc{A}$ by deleting all the outgoing
edges from $q$ except $e$.
We then consider the \DTA\ $\bar{\mc{A}_e}$, $\mc{A}_{e}^{>}$, and
$\mc{A}_{e}^{<}$ where $\bar{\mc{A}_{e}}$ is obtained from
$\mc{A}_{e}$ by replacing $x=K$ by $true$; $\mc{A}_{e}^{>}$ is obtained
from $\mc{A}_{e}$ by replacing $x=K$ by $x>K$ and $ \mc{A}_{e}^{<}$ is
obtained from $\mc{A}_{e}$ by replacing $x=K$ by $x<K$.
Since $\mc{A}$ is deterministic, it follows that
$$
\Paths^\mc{C}(\mc{A}_{e}) = \Paths^\mc{C}(\bar{\mc{A}_{e}})\setminus
\big(\Paths^\mc{C}(\mc{A}_{e}^{>})\cup
\Paths^\mc{C}(\mc{A}_{e}^{<})\big). $$
By the induction hypothesis, the sets
$\Paths^\mc{C}(\bar{\mc{A}_{e}})$,
$\Paths^\mc{C}(\mc{A}_{e}^{>})$ and
$\Paths^\mc{C}(\mc{A}_{e}^{<})$ are measurable.
Hence, $\Paths^\mc{C}(\mc{A}_{e})$ is measurable.
Furthermore, as
$$\Paths^\mc{C}( \mc{A})=\bigcup_{e = q \mv{a, g, X} q'}\Paths^\mc{C}(\mc{A}_{e}),$$
where all guards $g$ of edge $e$ are equalities, it follows that $\Paths^\mc{C}( \mc{A})$
is measurable.

\item\noindent{\hskip-12 pt\bf (1c):}\
Let \DTAr\ $\mc{A}$ have clock constraints of the form $x \bowtie K$ where $\bowtie
\, \in \{\geq, \leq\}$.
We consider the \DTA\ $\mc{A}_=$ and $\mc{A}_{\overline{\bowtie}}$, where $\mc{A}_=$
is obtained from $\mc{A}$ by changing all constraints of the form $x \bowtie K$ by
$x = K$, and $\mc{A}_{\overline{\bowtie}}$ is obtained from $\mc{A}$ by changing any
constraint $x \bowtie K$ by $x \, \overline{\bowtie} \, K$, with $\overline{\geq} = \, >$
and $\overline{\leq} = \, <$ otherwise.
Clearly, $\Paths^\mc{C}(\mc{A}) = \Paths^\mc{C}(\mc{A}_{=}) \cup \Paths^\mc{C}(\mc{A}_{\overline{\bowtie}})$.
As it was shown before that $\Paths^\mc{C}(\mc{A}_{=})$ and $\Paths^\mc{C}(\mc{A}_{\overline{\bowtie}})$ are measurable, it follows that $\Paths^\mc{C}(\mc{A})$
is measurable.
\item\noindent{\hskip-12 pt\bf (2):}\
Let \DTAo $\mc{A}$ with $Q_\mF=\{F_1, \ldots ,F_k\}$.
$\Path^\mC (\mA) = \bigcap_{0 < i \les k}\Path^i$ where $\Path^i$ is the set of paths
in CTMC $\mC$ whose corresponding DTA paths are accepted by $F_i \in Q_\mF$, i.e., $\Path^i = \{\theta \in \Path^\mC(\mA) \mid \inf(\theta) = F_i\}$.
We have:
$$
\Path^i = \bigcap_{n\ges 0} \bigcup_{m\ges n} \bigcup_{s_0, \ldots, s_n,s_{n{+}1}
\ldots, s_m} C(s_0, I_0, \ldots, I_{n{-}1}, s_n, \ldots, I_{m{-}1}, s_m),
$$
where $\{s_{n+1}, \ldots ,s_m\}= L_{F_i}$ with $L_{F_i}$ the set of CTMC states whose
corresponding DTA states are $F_i$, and $C(s_0, I_0, \ldots, I_{n-1}, s_n, \ldots, I_{m-1},
s_m)$ is the cylinder set such that each timed path of the cylinder set of the form $s_0
\mv{t_0} \cdots \mv{t_{n-1}} s_n \cdots \mv{t_{m-1}} s_m$ is a prefix of an accepting
path of $\mA$.
It follows that $\Path^i$ is measurable.
Thus, $\Paths^\mc{C}(\mc{A})$ is measurable.\qed
\end{enumerate}


\subsection{The product of a CTMC and a DTA}\label{sec:big_product}
A central step in the verification of a CTMC $\mc{C}$ against a DTA $\mc{A}$ is
to construct its synchronous product $\mc{C} \otimes \mc{A}$.  The resulting
object is neither a CTMC nor a DTA, but a mixture of the two.  We first define this
model, called deterministic Markovian timed automata, and define a measurable
space over its paths.  In Section~\ref{sec:finite}, we consider
the computation of $\Pr(\mC \models \mA) = \Pr\left( \Paths^\mc{C}(\mc{A}) \right)$
which is based on this product.

\begin{defi}[\DMTA]\label{def:MTA} A \emph{deterministic Markovian
timed automaton} (\DMTA) is a tuple $\mc{M} = (Loc,\mc{X},\ell_0, Loc_\mathbf{F}, E,
\rightsquigarrow)$, where $Loc$ is a nonempty finite set of \emph{locations}; $\mc{X}$
is a finite set of \emph{clocks}; $\ell_0\in Loc$ is the \emph{initial location};
$Loc_\mathbf{F}$ is the \emph{acceptance condition} with $Loc_\mathbf{F} =
Loc_F\subseteq Loc$ the reachability condition and $Loc_\mathbf{F} = Loc_\mF
\subseteq 2^{Loc}$ the Muller condition; $E: Loc\to \Reals_{\ges 0}$ is the
\emph{exit rate function}; and $\rightsquigarrow \,\subseteq Loc \times \mCC
({\mc{X}})\times 2^\mX \times \Distr(Loc)$ is an \emph{edge relation} such that:
$$
\left( \updownsquigarrow{\ell}{g,X}{}{\zeta} \mbox{ and }
\updownsquigarrow{\ell}{g',X'}{}{\zeta'} \mbox{ with }
g\neq g' \right) \quad\mbox{implies}  \quad g\cap g'=\emptyset.
$$
\end{defi}

DMTA closely resemble DTA, but have in addition to DTA an exit rate function
that determines the random residence time in a location, and an edge relation
where the target of an edge is a probability distribution over the locations.
Concepts such as clock valuation, clock constraints and so forth are defined as
for \DTA.  We refer to $\updownsquigarrow{\ell}{g,X}{}{\zeta}$ for distribution
$\zeta\in\Distr(Loc)$ as an \emph{edge} and to $\updownmapsto{\ell}{g,X}{p}%
{\ell'}$ with $p = \zeta(\ell')$ as a \emph{transition} of this edge.  The intuition
is that when entering location $\ell$, the \DMTA\ chooses a residence time which
is governed by an exponential distribution with rate $E(\ell)$.  Thus, the
probability to leave $\ell$ within $t$ time units is $1-e^{-E(\ell)t}$.  Due to the
determinism of the edge relation, at most one edge, say $\updownsquigarrow%
{\ell}{g,X}{}{\zeta}$, is enabled.  The probability to jump to $\ell'$ via this edge
equals $\zeta(\ell')$.   Similar as for \DTA s, \DMTAr\ and \DMTAo\ are defined
and \DMTA\ refers to both classes.

\begin{defi}[DMTA paths]
An $($infinite$)$ \emph{symbolic path} of \DMTA\ $\mc{M}$ is of the form:
$$
\updownmapsto{\ell_0}{g_0,X_0}{p_0}{}\updownmapsto{\ell_1}{g_1,X_1}{p_1}{}\ell_2\cdots
\quad \mbox{where } \updownsquigarrow{\ell_i}{g_i,X_i}{}{\zeta_i} \mbox{ and }
p_i = \zeta_i(\ell_{i{+}1}), \mbox{ for all } i\in\Nats.
$$

A symbolic path induces \emph{infinite paths} of the form $\tau = \ell_0 \mv{t_0}
\ell_1\mv{t_1}\ell_2\cdots$ such that $\eta_0=\vec{0}$, $(\eta_i+t_i)\models g_{i}$, and
$\eta_{i+1}=(\eta_i+t_i)[X_i:=0]$ where $i\geqslant 0$ and
$\eta_i$ is the clock valuation of $\mc{X}$ in $\mc{M}$ when
\emph{entering} location $\ell_i$.
The path $\tau$ is \emph{accepted} by a \DMTAr\ if there exists $n\ges 0$,
such that $\tau[n]\in Loc_F$. It is \emph{accepted} by
\DMTAo\ if and only if $in\!f({\tau})\in Loc_\mF$.
\end{defi}

\subsection*{\DMTA\ semantics. }
Consider clock valuation $\eta$ in location $\ell$.
As the DMTA is deterministic, at most one guard is enabled in state $(\ell, \eta)$.
The \emph{one-jump} probability of taking the transition $\updownmapsto{\ell}{g,X}{p}{\ell'}$
within time interval $I$ starting at clock valuation $\eta$ in location $\ell$, denoted
$p_\eta(\ell,\ell',I)$, is defined as follows:
\begin{equation}
p_{\eta}(\ell,\ell',I) =
\int_I \ \underbrace{E(\ell)\cdot e^{-E(\ell)\tau}}_{\mbox{\begin{scriptsize}(i) density to leave
$\ell$ at $\tau$\end{scriptsize}}}
\cdot \quad
\underbrace{\mathbf{1}_{g}(\eta{+}\tau)}_{\textrm{(ii)}\eta{+}\tau\models g?}
\quad\cdot\quad
\underbrace{p}_{\mbox{\begin{scriptsize}(iii)
probabilistic jump\end{scriptsize}}}\quad
d\tau\label{eq:1_step_DMTA}\end{equation}
Note the resemblance with \eqref{eqn:ctmc}.
Actually, part (i) characterizes the delay $\tau$ at location $\ell$ which is exponentially
distributed with rate $E(\ell)$; (ii) is the \emph{characteristic function}, where
$\mathbf{1}_{g}(\eta{+}\tau)=1$ if and only if $\eta{+}\tau\models g$.
It compares the current valuation $\eta{+}\tau$ with guard $g$
and rules out those violating $g$.
Part (iii) indicates the probability of the transition under consideration.
Note that (i) and (iii) are features from \CTMC s while (ii) stems from \DTA.
The characteristic function $\mathbf{1}_{g}$ is Riemann integrable as it is bounded
and its support is an interval; therefore, $p_{\eta}(\ell,\ell',I)$ is well-defined.
The one-jump probability can be uniquely defined in this way because it relates to
a fixed clock evaluation $\eta$.

\bigskip

The above characterisation of the one-jump probability provides the basis for
defining the probability of a set of DMTA paths.
Let $C(\ell_0,I_0,\ldots\!,I_{n-1},\ell_n)$ be the cylinder set with $(\ell_0, \ldots\!, \ell_n)
\in {Loc}^{n+1}$ and $I_i \subseteq \Reals_{\ges 0}$.
It denotes a set of paths in DMTA $\mc{M}$ such that for any such path $\tau$, $\tau[i]
= \ell_i$ and $\tau\<i\> \in I_i$.
Let ${\Pr}^\mc{M}_{\eta_0}\left( C(\ell_{0},I_{0},\ldots\!,I_{n-1},\ell_n)\right)$ denote the
probability of $C(\ell_0,I_0,\ldots\!,I_{n-1},\ell_n)$ such that $\eta_0$ is the initial clock
valuation in location $\ell_0$.
Let ${\Pr}^\mc{M}_{\eta_0}\left( C(\ell_{0},I_{0},\ldots\!,I_{n-1},\ell_n)\right) =
\Pro^\mc{M}_0(\eta_0)$, where $\Pro^{\mc{M}}_{i}(\eta)$ is inductively defined as follows:
\begin{equation}\label{eq:semantics}
\Pro^{\mc{M}}_{i}(\eta) \ = \ \left\{
  \begin{array}{ll}
   1 & \mbox{ if } i = n \\[1ex]
   \displaystyle \int_{I_{i}}\
\underbrace{E(\ell_i){\cdot}e^{-E(\ell_{i})\tau} \cdot \mathbf{1}_{g_i}
(\eta+\tau) \cdot p_i}_{(\star)}\,\cdot\,\underbrace{\Pro^{\mc{M}}_{i+1}(\eta')}_{(\star\star)}\
d\tau & \mbox{ if } 0 \les i < n,
  \end{array} \right.
\end{equation}
where $\eta':=(\eta+\tau)[X_{i}:=0]$. Intuitively,
$\Pro^{\mc{M}}_i(\eta_i)$ is the probability of the suffix
cylinder set starting from $\ell_i$ and $\eta_i$ to $\ell_n$. It
is recursively defined by the product of the probability of
taking a transition from $\ell_i$ to $\ell_{i+1}$ within time interval
$I_i$ (cf.\,($\star$) and \eqref{eq:1_step_DMTA}) and the
probability of the suffix cylinder set from $\ell_{i+1}$ and
$\eta_{i+1}$ on (cf.\,($\star\star$)). For the same reason as
$p_{\eta}(\ell,\ell',I)$ is well-defined, $\Pro^{\mc{M}}_i(\eta)$ is well-defined.

\begin{exa}
The \DMTAr\ in Figure~\ref{fig:MTA_1c} has initial location $\ell_0$
with two outgoing edges, with guards $x<1$ and $1 < x < 2$.
We use the small black dots to indicate distributions.
Assume $t$ time units elapse in $\ell_0$.
If the current clock evaluation $\eta$ satisfies $\eta(x) < 1$, then the
upper edge is enabled and the probability to go to $\ell_1$ within time
$t$ is $p_{\vec{0}}(\ell_0,\ell_1,[0,t]) = (1-e^{-r_0t}){\cdot}1$, where
$E(\ell_0) = r_0$; no clock is reset.
It is similar when $1<\eta(x)<2$, except that $x$ will be reset (cf.\ the
lower edge emanating from location $\ell_0$).
If $\eta(x) \ges 2$, no outgoing edge is enabled, and the DMTA stays
in $\ell_0$ ad infinitum.
%
\end{exa}

\subsection{Product DMTA}\label{sec:product_DMTA}
\begin{figure*}[!ht]\begin{center}
 \subfigure[\DMTAr\
 $\mc{C}\otimes\mc{A}$]{\hspace{-0.6cm}\scalebox{0.8}{\begin{picture}(191,50)(0,-50)
\put(0,-50){}
\node[NLangle=0.0,Nmarks=i,Nw=20.0](n26)(21.35,-26.52){$\ell_0{=}\<s_0,q_0\>$}

\node[Nw=20.0](n31)(67.51,-26.52){$\ell_1{=}\<s_1,q_0\>$}

\node[Nfill=y,fillcolor=Black,Nw=2.0,Nh=2.0,Nmr=1.0](n36)(48,-26.52){}

\drawedge[ELpos=66,AHnb=0,curvedepth=4.23](n26,n36){ $x{<}1,\emptyset$}

\drawedge[ELpos=25,ELdist=0.89](n36,n31){1}

\node[Nw=20.0](n20)(126.63,-26.52){$\ell_2{=}\<s_2,q_0\>$}

\node[Nfill=y,fillcolor=Black,Nw=2.0,Nh=2.0,Nmr=1.0](n22)(48.0,-26.52){}

\drawedge[ELpos=57,AHnb=0,curvedepth=4.75](n22,n26){ $1{<}x{<}2{,}\{x\}$}

\node[Nfill=y,fillcolor=Black,Nw=2.0,Nh=2.0,Nmr=1.0](n23)(104.31,-26.52){}

\drawedge[ELdist=0.95,AHnb=0,curvedepth=4.73](n31,n23){ $x{<}1{,}\emptyset$}

\drawedge[ELside=r,ELpos=30](n23,n20){0.2}

\node[Nframe=n,NLangle=0.0](n156)(22.29,-19.82){$r_0$}

\node[Nframe=n,NLangle=0.0](n157)(68.13,-19.82){$r_1$}

\node[Nframe=n,NLangle=0.0](n158)(124.01,-33.49){$r_2$}

\node[Nw=20.0](n14)(126.76,-10.88){$\ell_4{=}\<s_3,q_0\>$}

\node[Nframe=n,NLangle=0.0](n15)(140.29,-12.0){$r_3$}

\node[Nmarks=r,Nw=20.9](n18)(174.26,-26.52){$\ell_3{=}\<s_2,q_1\>$}

\node[Nframe=n,NLangle=0.0](n19)(171.71,-33.49){$r_2$}

\node[Nfill=y,fillcolor=Black,Nw=2.0,Nh=2.0,Nmr=1.0](n22)(153.44,-26.52){}

\drawedge[ELside=r,ELdist=0.49,AHnb=0,curvedepth=-5.02](n31,n23){$1{<}x{<}2{,}\{x\}$}

\drawedge[ELpos=75,AHnb=0](n20,n22){$x{>}1{,}\emptyset$}

\drawedge[ELpos=23](n22,n18){1}

\drawedge[ELpos=34](n23,n14){0.3}

\drawedge[ELside=r,ELdist=1.06,curvedepth=-20.46](n23,n26){0.5}

\end{picture}\label{fig:MTA_1c}}
 }

\hspace{-1cm}\begin{minipage}{0.3\textwidth}
 \hspace{-0.2cm}\subfigure[\CTMC\
 $\mc{C}$]{\hspace{-0.8cm}\scalebox{0.8}{\begin{picture}(73,36)(0,-36)
\put(0,-36){}
\node[Nmarks=i](n0)(8.32,-16.3){$s_0$}

\node(n1)(32.51,-16.3){$s_1$}

\drawedge[curvedepth=3.0](n0,n1){1}

\drawedge[curvedepth=3.0](n1,n0){$0.5$}

\node(n2)(52.48,-10.19){$s_2$}

\node(n3)(52.48,-22.19){$s_3$}

\drawedge(n1,n2){0.2}

\drawedge[ELside=r,ELdist=1.31](n1,n3){0.3}

\drawloop[loopdiam=6.0,loopangle=0.0](n2){1}

\drawloop[loopdiam=6.0,loopangle=0.0](n3){1}

\node[Nframe=n](n4)(8.48,-10.16){$\{a\}$}

\node[Nframe=n](n5)(32.48,-10.16){$\{a\}$}

\node[Nframe=n](n6)(56.16,-4.16){$\{b\}$}

\node[Nframe=n](n7)(56.16,-28.16){$\{c\}$}

\node[Nframe=n](n8)(49.16,-28.16){$r_3$}

\node[Nframe=n](n9)(49.16,-4.16){$r_2$}

\node[Nframe=n](n10)(32.16,-22.16){$r_1$}

\node[Nframe=n](n11)(8.16,-22.16){$r_0$}

\end{picture}\label{fig:CTMC_1c}}
 }\\
\subfigure[\DTAr\
 $\mc{A}$]{\hspace{-0.8cm}\scalebox{0.8}{\begin{picture}(63,32)(0,-32)
\put(0,-32){}
\node[Nmarks=i](n24)(12.16,-12.32){$q_0$}

\node[Nmarks=r](n25)(48.16,-12.32){$q_1$}

\drawloop[loopdiam=6.0](n24){$\{a\},x<1,\emptyset$}

\drawloop[loopdiam=6.0,loopangle=270](n24){$\{a\},1<x<2,\{x\}$}

\drawedge(n24,n25){$\{b\},x>1,\emptyset$}

\end{picture}\label{fig:DTA_1c}}
 }\end{minipage}%
%
\begin{minipage}{0.4\textwidth}
 \subfigure[Reachable region graph of $\mc{C}\otimes \mc{A}$]{\vspace{-1cm}\scalebox{0.7}{\begin{picture}(128,84)(0,-84)
\put(0,-84){}
\node[Nmarks=i,Nw=22.74,Nmr=0.0](n57)(31.02,-14.49){$\ell_0,0{\les}
x{<}1$}

\node[Nw=22.74,Nmr=0.0](n60)(70.67,-14.49){$\ell_0,1{\les}
x{<}2$}

\node[Nw=22.74,Nmr=0.0](n61)(31.02,-34.43){$\ell_1,0{\les}
x{<}1$}

\node[NLangle=0.0,NLdist=0.45,Nw=22.74,Nmr=0.0](n62)(70.67,-34.43){$\ell_1,1{\les}
x{<}2$}

\drawedge[ELside=r,ELdist=1.13,curvedepth=-2.95](n57,n61){1}
\drawedge[ELpos=66](n60,n61){1}
\node[Nframe=n](n64)(21.78,-8.19){$v_0,r_0$}

\node[Nframe=n](n65)(79.78,-8.19){$v_1,r_0$}

\node[Nframe=n](n66)(21.78,-28.19){$v_2,r_1$}

\node[Nframe=n](n74)(79.6,-28.16){$v_3,r_1$}

\drawedge[ELside=r,curvedepth=-2.95](n61,n57){0.5}

\drawedge(n57,n60){$\delta$}

\drawedge[ELside=r,ELpos=20,ELdist=0.31](n62,n57){reset, 0.5}

\node[Nw=22.74,Nmr=0.0](n53)(31.02,-54.36){$\ell_2,0{\les}
x{<}1$}

\node[Nw=22.74,Nmr=0.0](n54)(70.67,-54.36){$\ell_2,1{\les} x{<}
2$}

\node[NLangle=0.0,NLdist=0.45,Nmarks=r,Nw=22.74,Nmr=0.0](n56)(70.67,-74.3){$\ell_3,1{\les}
x{<}2$}

\node[Nw=22.74,Nmr=0.0](n57)(110.49,-54.36){$\ell_2,x\ges
2$}

\node[Nmarks=r,Nw=22.74,Nmr=0.0](n58)(110.49,-74.3){$\ell_3,x\ges
2$}

\drawedge(n54,n56){$1$}

\node[Nframe=n](n59)(21.6,-48.32){$v_4,0$}

\node[Nframe=n](n60)(79.6,-48.32){$v_5,r_2$}

\node[Nframe=n](n162)(79.78,-68.19){$v_7,0$}

\drawedge(n53,n54){$\delta$}

\drawedge(n56,n58){$\delta$}

\drawedge(n57,n58){1}

\node[Nframe=n](n63)(117.6,-68.32){$v_8,0$}

\drawedge[ELpos=36,ELdist=0.3](n62,n53){reset,0.2}

\drawedge(n61,n53){0.2}

\drawedge(n54,n57){$\delta$}

\node[Nframe=n](n89)(117.78,-48.19){$v_6,r_2$}

\drawedge(n61,n62){$\delta$}

\end{picture}\label{fig:region_1c}}
 }\end{minipage}\vspace{0.5cm}\caption{Example product \DMTAr\ of \CTMC\ $\mc{C}$ and \DTAr\ $\mc{A}$}\label{fig:final}\end{center}
 \end{figure*}
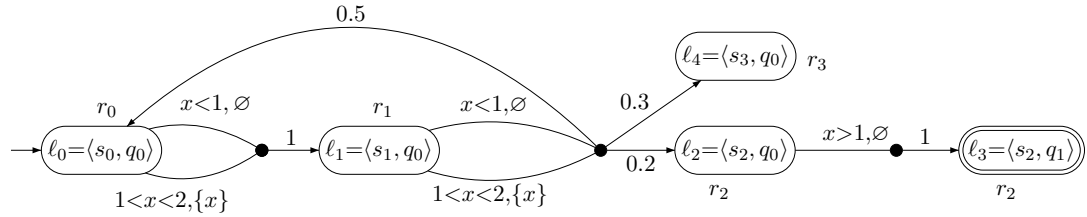
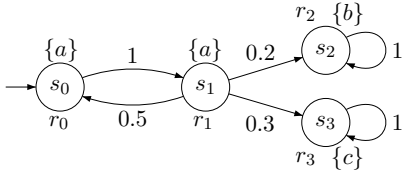
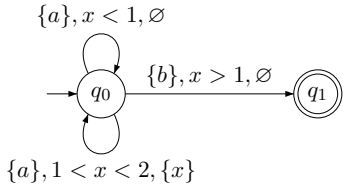
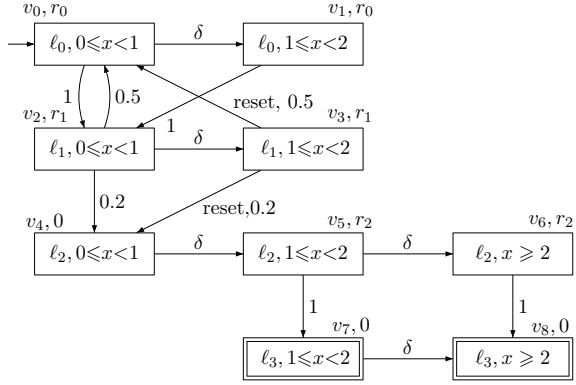

The product $\mc{C}\otimes\mc{A}$ for \CTMC\ $\mc{C}$ and \DTA\ $\mc{A}$,  is a \DMTA.

\begin{defi}[Product of \CTMC\ and \DTA]\label{def:product}
Let $\mc{C}=(S,\AP, L,s_0, \P,E)$ be a \CTMC\ and $\mc{A}=(2^\AP,
\mc{X},Q,q_0,Q_\mathbf{F},\rightarrow)$ be a \DTA.  Let
$\mc{C}\otimes \mc{A}= {(Loc, \mc{X}, \ell_0, Loc_\mathbf{F},
E,\rightsquigarrow)}$ be the product \DMTA, where $Loc =S\times Q$;
$\ell_0 =\<s_0,q_0\>$; $E(\<s,q\>) =E(s)$;  and
\begin{enumerate}[$\bullet$]
\item
$Loc_\mathbf{F}=Loc_F:=S\times Q_F$, if $Q_\mathbf{F}=Q_F$ $($reachability condition$)$
\item
$Loc_\mathbf{F}=Loc_\mF:=\bigcup_{F\in Q_\mF}S\times F$, if $Q_\mathbf{F}=Q_\mF$
$($Muller condition$)$
\end{enumerate}
and $\rightsquigarrow$ is defined as the smallest relation defined by the rule:
$$
\dfrac{\P(s,s')>0\ \wedge\ q\mv{L(s), g, X}q'}
    {\updownsquigarrow{\<s,q\>}{g,X}{}{\zeta}}\mbox{ such that } \zeta(\<s',q'\>)=\P(s,s').
$$
\end{defi}

The DMTA $\mc{C} \otimes \mc{A}$ is basically the synchronous product of CTMC
$\mc{C}$ and DTA $\mc{A}$ such that transition $s \to s'$ in $\mc{C}$ is matched
with the edge $q \mv{L(s), g, X} q'$, i.e., the set of atomic propositions of $s$ acts
as input symbol for the edge from location $q$ to $q'$ in $\mc{A}$.
The probability of the joint evolvement of $\mc{C}$ and $\mc{A}$ is given by $\P(s,s')$,
the discrete probability of $s \to s'$ in $\mc{C}$, whereas the residence time in the
location $\< s, q \>$ is given by $E(s)$, the exit rate of $s$ in $\mc{C}$.
It is easy to see from the construction that $\mc{C}\otimes\mc{A}$ is indeed a \DMTA.
The determinism of the \DTA\ $\mc{A}$ guarantees that the induced product is also
deterministic.
In $\mc{C} \otimes \mc{A}$, from each location there is at most one ``input symbol''
possible, viz.\ $L(s)$.
For the sake of convenience, input symbols can be omitted from $\mc{C} \otimes \mc{A}$.

\begin{exa}\label{ex:product_DMTAr}
Let \CTMC\ $\mc{C}$ and \DTAr\ $\mc{A}$ be given in Figure~\ref{fig:CTMC_1c}
and \ref{fig:DTA_1c}, respectively.
The product \DMTAr\ $\mc{C}{\otimes}\mc{A}$ is depicted in Figure~\ref{fig:MTA_1c}.
Since $Q_F=\{q_1\}$ in $\mc{A}$, the set of accepting locations in
\DMTAr\ is $Loc_F=\{\<s_2,q_1\>\}=\{\ell_3\}$.
\end{exa}

\begin{exa}\label{ex:product_DMTAo}
For the \CTMC\ $\mc{C}$ in Figure~\ref{fig:CTMC_Muller} and the
\DTAo\ $\mc{A}$ in Figure~\ref{fig:DTA_Muller} with
acceptance family $Q_\mF=\big\{\{q_1,q_2\},\{q_3,q_4\}\big\}$, the
product \DMTA$^\omega$ $\mc{C}\otimes\mc{A}$ is shown in
Figure~\ref{fig:DMTA_Muller}.
$Loc_\mF=\big\{\{\<s_i,q_1\>,\<s_j,q_2\>\},\{\<s_i',q_3\>,\<s_j',q_4\>\}\big\}$,
for any $s_i,s_i',s_j,s_j'\in S$, i.e.,
$Loc_\mF=\big\{\{\ell_1,\ell_2,\ell_3\},$ $\{\ell_4,\ell_5,\ell_6\}\big\}$.
\end{exa}
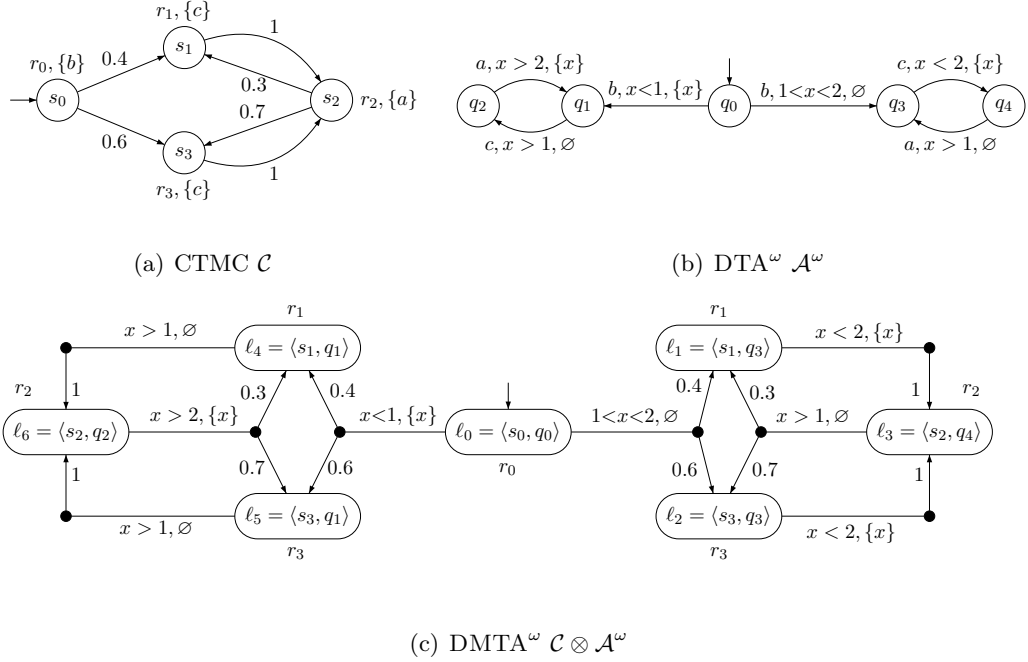
\begin{figure}[!ht]
 \centering
\subfigure[CTMC
$\mc{C}$]{\scalebox{0.7}{
%
%
%
%
%
%
%
%

\begin{picture}(84,48)(0,-48)
\put(0,-48){}
\node(n91)(40.07,-12.13){$s_1$}

\node(n92)(68.0,-22.13){$s_2$}

\node[Nmarks=i](n93)(16.0,-22.13){$s_0$}

\drawedge[curvedepth=7](n91,n92){$1$}

\drawedge[ELdist=0.23,curvedepth=-0.05](n92,n91){$0.3$}

\drawedge(n93,n91){$0.4$}

\node(n141)(40.0,-32.13){$s_3$}

\drawedge[ELside=r](n93,n141){$0.6$}

\drawedge[ELside=r](n92,n141){$0.7$}

\drawedge[curvedepth=-7,ELside=r](n141,n92){$1$}

\node[Nframe=n](n158)(16.0,-15.16){$r_0,\{b\}$}

\node[Nframe=n](n159)(40.0,-39.16){$r_3,\{c\}$}

\node[Nframe=n](n160)(79.0,-22.16){$r_2,\{a\}$}

\node[Nframe=n](n161)(40.0,-5.16){$r_1,\{c\}$}

\end{picture}\label{fig:CTMC_Muller}}
 }\subfigure[\DTAo\ $\mc{A}^\omega$]{\scalebox{0.7}{\begin{picture}(117,37)(0,-37)
\put(0,-37){}
\node[iangle=90.0,Nmarks=i](n119)(56.48,-12.21){$q_0$}

\node(n120)(88.45,-12.21){$q_3$}

\node(n121)(108.61,-12.21){$q_4$}

\node(n122)(28.64,-12.21){$q_1$}

\node(n123)(8.77,-12.21){$q_2$}

\drawedge(n119,n120){$b,1{<}x{<}2,\emptyset$}

\drawedge[ELside=r,ELdist=1.18](n119,n122){$b,x{<}1,\{x\}$}

\drawedge[curvedepth=5.0](n120,n121){$c,x<2,\{x\}$}

\drawedge[curvedepth=5.0](n121,n120){$a,x>1,\emptyset$}

\drawedge[curvedepth=5.0](n122,n123){$c,x>1,\emptyset$}

\drawedge[curvedepth=5.0](n123,n122){$a,x>2,\{x\}$}

\end{picture}\label{fig:DTA_Muller}}
 }

\subfigure[\DMTAo\
$\mc{C}\otimes\mc{A}^\omega$]{\scalebox{0.7}{\begin{picture}(193,59)(0,-59)
\put(0,-59){}
\node[iangle=90.0,Nmarks=i,Nw=23.88,Nh=8.63,Nmr=5.32](n138)(96,-24.0){$\ell_0=\<s_0,q_0\>$}

\node[Nw=23.88,Nh=8.63,Nmr=5.32](n139)(136.0,-8.0){$\ell_1=\<s_1,q_3\>$}

\node[Nw=23.88,Nh=8.63,Nmr=5.32](n140)(136.0,-40.0){$\ell_2=\<s_3,q_3\>$}

\node[Nw=23.88,Nh=8.63,Nmr=5.32](n141)(176,-24.0){$\ell_3=\<s_2,q_4\>$}

\node[Nw=23.88,Nh=8.63,Nmr=5.32](n142)(56,-8.0){$\ell_4=\<s_1,q_1\>$}

\node[Nw=23.88,Nh=8.63,Nmr=5.32](n143)(56,-40.0){$\ell_5=\<s_3,q_1\>$}

\node[Nw=23.88,Nh=8.63,Nmr=5.32](n144)(12,-24.0){$\ell_6=\<s_2,q_2\>$}

\node[Nfill=y,fillcolor=Black,Nw=2.0,Nh=2.0,Nmr=1.0](n150)(64,-24.0){}

\node[Nfill=y,fillcolor=Black,Nw=2.0,Nh=2.0,Nmr=1.0](n151)(132,-24.0){}

\drawedge[ELside=r,ELpos=65,ELdist=1.07,AHnb=0](n138,n150){$x{<}1,\{x\}$}

\drawedge[ELside=r,ELpos=33,ELdist=1.07,AHnb=0](n151,n138){$1{<}x{<}2,\emptyset$}

\drawedge[ELside=r,ELpos=38,ELdist=1.32](n150,n142){$0.4$}

\drawedge[ELpos=33,ELdist=0.4](n150,n143){$0.6$}

\drawedge[ELdist=1.08](n151,n139){$0.4$}

\drawedge[ELside=r,ELpos=38,ELdist=1.39](n151,n140){$0.6$}

\node[Nfill=y,fillcolor=Black,Nw=2.0,Nh=2.0,Nmr=1.0](n36)(144.0,-24.0){}

\node[Nfill=y,fillcolor=Black,Nw=2.0,Nh=2.0,Nmr=1.0](n37)(48.0,-24.0){}

\drawedge[ELside=r,ELpos=37,ELdist=0.92](n36,n139){$0.3$}

\drawedge[ELpos=34,ELdist=1.19](n36,n140){$0.7$}

\drawedge[ELpos=31,ELdist=1.18,AHnb=0](n36,n141){$x>1,\emptyset$}

\drawedge[ELpos=37](n37,n142){$0.3$}

\drawedge[ELside=r,ELpos=32,ELdist=0.8](n37,n143){$0.7$}

\drawedge[ELpos=68,ELdist=1.18,AHnb=0](n144,n37){$x>2,\{x\}$}

\node[Nframe=n](n538)(95.91,-31.0){$r_0$}

\node[Nframe=n](n539)(136.07,-1.0){$r_1$}

\node[Nframe=n](n540)(136.07,-47.0){$r_3$}

\node[Nframe=n](n541)(184.0,-16.0){$r_2$}

\node[Nframe=n](n542)(55.72,-1.0){$r_1$}

\node[Nframe=n](n543)(55.72,-47.0){$r_3$}

\node[Nframe=n](n544)(4.0,-16.0){$r_2$}

\node[Nfill=y,fillcolor=Black,Nw=2.0,Nh=2.0,Nmr=1.0](n190)(12.0,-8.0){}

\node[Nfill=y,fillcolor=Black,Nw=2.0,Nh=2.0,Nmr=1.0](n191)(12.0,-40.0){}

\node[Nfill=y,fillcolor=Black,Nw=2.0,Nh=2.0,Nmr=1.0](n192)(176.0,-8.0){}

\node[Nfill=y,fillcolor=Black,Nw=2.0,Nh=2.0,Nmr=1.0](n193)(176.0,-40.0){}

\drawedge[linewidth=0.12,ELside=r,ELpos=59,ELdist=1.54,AHnb=0](n142,n190){$x>1,\emptyset$}

\drawedge(n190,n144){$1$}

\drawedge[ELpos=61,AHnb=0](n143,n191){$x>1,\emptyset$}

\drawedge[ELside=r](n191,n144){$1$}

\drawedge[ELside=r,ELpos=63,ELdist=1.3,AHnb=0](n140,n193){$x<2,\{x\}$}

\drawedge(n193,n141){$1$}

\drawedge[ELpos=67,AHnb=0](n139,n192){$x<2,\{x\}$}

\drawedge[ELside=r,ELdist=1.65](n192,n141){$1$}

\end{picture}\label{fig:DMTA_Muller}}
 }\caption{Example product \DMTAo\ of \CTMC\ $\mc{C}$ and \DTAo\ $\mc{A}^\omega$}\label{fig:infinite_product}
\end{figure}

The set of accepted paths in DMTA $\mc{C}{\otimes}\mc{A}$ is defined by:
$$
\AccPaths^{\mc{C}\otimes\mc{A}} \ := \
\{\,{\tau\in\Paths^{\mc{C}\otimes\mc{A}}}\mid{\tau \mbox{
is accepted by }\mc{C}{\otimes}\mc{A}}\;\}.
$$
For $n$-ary tuple $J$, let $J{\downharpoonright_i}$ denote the $i$-th
entry in $J$, for $1 \leqslant i \leqslant n$.
For a $(\mc{C}{\otimes}\mc{A})$-path $\tau=\<s_0,q_0\> \mv{t_0} \<s_1,q_1\> \mv{t_1}
\cdots$, let $\tau{\downharpoonright_1} := s_0\mv{t_0}s_1\mv{t_1} \cdots$, and for
any set $\Pi$ of $(\mc{C}{\otimes}\mc{A})$-paths, let
$\Pi{\downharpoonright_1}=\bigcup_{\tau\in\Pi}\tau{\downharpoonright_1}$.
The following lemma asserts that there is a one-to-one relationship between
paths in CTMC $\mc{C}$ accepted by DTA $\mc{A}$ and accepting paths in
$\mc{C} \otimes \mc{A}$.

\begin{lem}\label{lemma:pathcorr}
For any \CTMC\ $\mc{C}$ and \DTA\ $\mc{A}$,
$\Paths^{\mc{C}}(\mc{A}) \ = \ \AccPaths^{\mc{C}\otimes\mc{A}}{\downharpoonright_1}.$
\end{lem}

\begin{proof}
We provide the proof for \DTAr\ $\mc{A}$; the proof for \DTAo\ $\mc{A}$ is similar.
\bigskip

\noindent $(\subseteq)$
Let $\rho \in \Paths^{\mc{C}}(\mc{A})$.
We prove that there exists a path $\tau\in \AccPaths^{\mc{C}\otimes\mc{A}}$ with
$\rho = \tau{\downharpoonright_1}$.
Assume w.l.o.g.\ that $\rho = s_0 \mv{t_0} s_1 \cdots s_{n-1} \mv{t_{n-1}} s_n \in
\Paths^{\mc{C}}(\mc{A})$, i.e., $s_n\in Q_F$, $\eta_0 \models \vec{0}$, and for
$0 \les i < n$, $\eta_i{+}t_i\models g_i$ and $\eta_{i+1} = (\eta_i{+}t_i)[X_i:=0]$,
where $\eta_i$ is the clock valuation in $\mc{A}$ when entering state $s_i$ in $\mc{C}$.
We construct a timed path $\theta \in \Paths^{\mc{A}}$ from $\rho$ such that
$\theta = q_0 \mv{L(s_0), t_0} q_1 \cdots q_{n-1} \mv{L(s_{n-1}), t_{n-1}} q_n$, where
the clock valuation on entering $s_i$ and $q_i$ coincides.
From $\rho$ and $\theta$, we can now construct the path
$$
\tau = \<s_0,q_0\> \mv{t_0} \<s_1,q_1\> \cdots \<s_{n-1},q_{n-1}\> \mv{t_{n-1}}
\<s_n,q_n\>,
$$
where $\<s_n,q_n\> \in Loc_F$.
It follows that $\tau \in \AccPaths^{\mc{C}\otimes\mc{A}}$ and $\rho = \tau{\downharpoonright_1}$.

\bigskip
\noindent $(\supseteq)$
Let $\tau \in \AccPaths^{\mc{C}\otimes\mc{A}}$.
We prove that $\tau{\downharpoonright_1}\in \Paths^{\mc{C}}(\mc{A})$.
Assume w.l.o.g.\ that
$$
\tau = \<s_0,q_0\> \mv{t_0} \cdots \mv{t_{n-1}} \<s_n,q_n\> \in
\AccPaths^{\mc{C}\otimes\mc{A}},
$$
with $\<s_n,q_n\> \in Loc_F$, $\eta_0\models\vec{0}$, and for $0 \les i < n$,
$\eta_i{+}t_i\models g_i$ and $\eta_{i+1}=(\eta_i{+}t_i)[X_i:=0]$, where
$\eta_i$ is the clock valuation when entering location $\<s_i,q_i\>$.
It then directly follows that $q_n\in Q_F$ and $\tau{\downharpoonright_1} \in
\Paths^{\mc{C}}(\mc{A})$, given the entering clock valuation $\eta_i$ of state
$s_i$.
\end{proof}

\begin{thm}\label{th:CTMC=MTA}
For any \CTMC\ $\mc{C}$ and \DTA\ $\mc{A}$,
$${\Pr}^{\mc{C}}\left(\Paths^{\mc{C}}(\mc{A})\right) \ = \
{\Pr}^{\mc{C}\otimes\mc{A}}_{\vec{0}}\left({\AccPaths}^{\mc{C}\otimes\mc{A}} \right).$$
\end{thm}
\begin{proof}
We provide the proof for \DTAr\ $\mc{A}$; the proof for \DTAo\ $\mc{A}$ goes along similar
lines as in the proof of Theorem~\ref{lem:measurability}.
\bigskip

According to Theorem~\ref{lem:measurability},
$\Paths^{\mc{C}}(\mc{A})$ can be rewritten as the combination of
cylinder sets of the form $C(s_0,I_0,\ldots,I_{n-1},s_n)$ which
are all accepted by \DTAr\ $\mc{A}$. Note that this means that
each path in the cylinder set is accepted by $\mc{A}$. By
Lemma~\ref{lemma:pathcorr}, namely by path lifting, we can
establish exactly the same combination of cylinder sets
$C(\ell_0,I_0,\ldots,I_{n-1},\ell_n)$ for $\AccPaths^{\mc{C}\otimes
\mc{A}}$, where
$s_i=\ell_i{\downharpoonright_1}$. It then suffices to show that
for each cylinder set $C(s_0,I_0,\ldots,I_{n-1},s_n)$ which is
accepted by $\mc{A}$, ${\Pr}^{\mc{C}}$ and
${\Pr}^{\mc{C}\otimes\mc{A}}$ yield the same probabilities.

For the measure $\Pr^{\mc{C}}$, according to Eq.\,(\ref{eqn:ctmc}) (cf.\
page \pageref{eqn:ctmc}),
\[{\Pr}^{\mc{C}}\big(C(s_0,I_0,\ldots,I_{n-1},s_n)\big)
= \prod_{0\leqslant  i< n}\int_{I_i}\P(s_i,s_{i+1})\cdot
E(s_i)\cdot e^{-E(s_i)\tau} d\tau. \]

The measure ${\Pr}^{\mc{C}\otimes\mc{A}}_{\vec{0}}$, according
to Section\,\ref{sec:big_product}, is given by
$\Pro^{\mc{C}\otimes\mc{A}}_0(\vec{0})$, where
$\Pro^{\mc{C}\otimes\mc{A}}_n(\eta)=1$ for any clock valuation
$\eta$ and for any $0\les i<n$:
\[
\Pro^{\mc{C}\otimes\mc{A}}_{i}(\eta_{i})=\int_{I_{i}}\mathbf{1}_{g_i}(\eta_i+\tau_i)
{\cdot}p_i{\cdot}E(\ell_i){\cdot}e^{-E(\ell_i)\tau_i}\cdot\Pro^{\mc{C}\otimes\mc{A}}_{i+1}(\eta_{i+1})\
d\tau_i,
\]
where $\eta_{i+1}=(\eta_i+\tau_i)[X_{i}:=0]$ and
$\mathbf{1}_{g_i}(\eta_i+\tau_i)=1$, if $\eta_i+\tau_i\models g_{i}$;
$0$, otherwise.

We will show, by induction, that
$\Pro^{\mc{C}\otimes\mc{A}}_{i}(\eta_{i})$ is a constant, i.e., is
independent of $\eta_i$, if the cylinder set
$C(\ell_0,I_0,\ldots,I_{n-1},\ell_n)$ is
accepted by $\mc{C}\otimes\mc{A}$.  %
First note that for this cylinder set
there must exist some sequence of transitions
\[
\updownmapsto{\ell_{0}}{g_{0},X_{0}}{p_{0}}{\ \ \ell_{1}} \cdots
\updownmapsto{\ell_{n-1}}{g_{n-1},X_{n-1}}{p_{n-1}}{\ \ \ell_{n}}
\]
with $\eta_0=\vec{0}$ and  $\forall t_i\in I_i$ with $0\leqslant
i<n$, $\eta_i+t_i\models g_i$ and
$\eta_{i+1}:=(\eta_i+t_i)[X_i:=0]$. Moreover, according to
Definition~\ref{def:product}, we have:
\begin{equation}p_i=\P(s_i, s_{i+1})\qquad\mbox{ and }\qquad E(\ell_i)=E(s_i).\label{eq:rate}\end{equation}
We apply a backward induction on $n$ down to 0. The base case is
trivial since $\Pro^{\mc{C}\otimes\mc{A}}_n(\eta_n)=1$. By the induction hypothesis,
$\Pro^{\mc{C}\otimes\mc{A}}_{i+1}(\eta_{i+1})$ is a constant.  For the
induction step, consider $i<n$. For any $\tau_i\in I_i$, since
$\eta_i+\tau_i\models g_i$, $\mathbf{1}_{g_i}(\eta_i+\tau_i)=1$, it
follows that
\begin{eqnarray*}
\Pro^{\mc{C}\otimes\mc{A}}_{i}(\eta_{i})&=&\int_{I_{i}}\mathbf{1}_{g_i}(\eta_i+\tau_i){\cdot}
p_i{\cdot}E(\ell_i){\cdot}e^{-E(\ell_i)\tau_i}\cdot\Pro^{\mc{C}\otimes\mc{A}}_{i+1}(\eta_{i+1})\
d\tau_i\\
&\stackrel{\textrm{I.H.}}{=}& \int_{I_{i}}p_i{\cdot}E(\ell_i){\cdot}e^{- E(\ell_i)\tau_i}d\tau_i\cdot\Pro^{\mc{C}\otimes\mc{A}}_{i+1}(\eta_{i+1})\\
&\stackrel{\textrm{Eq.}\eqref{eq:rate}}{=}&\int_{I_{i}}\P(s_i,s_{i+1}){\cdot}E(s_i){\cdot}e^{- E(s_i)\tau_i}d\tau_i\cdot\Pro^{\mc{C}\otimes\mc{A}}_{i+1}(\eta_{i+1}).\\
\end{eqnarray*}
Clearly, this is a constant. It is thus easy to see that
\[{\Pr}_{\vec{0}}^{\mc{C}\otimes
\mc{A}}\big(C(\ell_0,I_0,\ldots,I_{n-1},\ell_n)\big)\ :=\
\Pro^{\mc{C}\otimes \mc{A}}_{0}(\vec{0})= \prod_{0\leqslant i<
n}\int_{I_i}\P(s_i,s_{i+1}){\cdot}E(s_i){\cdot}e^{-E(s_i)\tau}
d\tau ,\]which completes the proof.
\end{proof}

\subsection{Region graph construction}\label{sec:region_DMTA}

Theorem~\ref{th:CTMC=MTA} asserts that the probability of CTMC $\mc{C}$
satisfying the DTA specification $\mc{A}$ equals the reachability probability
of some accepting location in $\mc{C} \otimes \mc{A}$.  The state space
of $\mc{C} \otimes \mc{A}$, however, is infinite.   As a next step towards obtaining
an effective procedure for computing reachability probabilities in $\mc{C} \otimes
\mc{A}$ we adopt the standard region construction of timed automata~\cite{AD94}
to \DMTA.   This yields a stochastic process, namely a PDP.  Here, we
consider the region construction for finite acceptance conditions, i.e, \DMTAr.  The
details for \DMTAo\ are slightly different (only the acceptance set differs) and are
provided in Section~\ref{sec:infinite}.

Let us briefly recall the concept of a region.
Formally, a region is an equivalence under $\cong$, an equivalence relation on
clock valuations.
A region is characterized by a specific form of a clock constraint.
Let $c_{x_i}$ be the largest constant with which $x_i \in \mX$ is compared in some
guard in the (DM)TA.
Clock evaluations $\eta, \eta'\in \mV(\mX)$ are \emph{clock-equivalent}, denoted
$\eta \cong \eta'$, if and only if either
\begin{enumerate}
\item
for any $x \in \mX$ it holds that $\eta(x) >c_x$ and $\eta'(x) > c_x$, or
\item
for any $x_i, x_j\in\mX$ with $\eta(x_i), \eta'(x_i) \les c_{x_i}$ and $\eta(x_j), \eta'(x_j)
\les c_{x_j}$ it holds:
	$$\lfloor \eta(x_i) \rfloor = \lfloor \eta'(x_i) \rfloor \quad \mbox{and} \quad
		             \{ \eta(x_i) \} \les \{ \eta'(x_i) \} \ \mbox{iff} \ { \eta(x_j) } \les { \eta'(x_j) },
    $$
    where $\lfloor d \rfloor$ and $\{ d \}$ are the integral and fractional part of $d \in \Reals$,
    respectively.
\end{enumerate}

This clock equivalence is coarser than the traditional definition by merging the ``boundary"
regions (those with point constraints like ``$x=0$") into the ``non-boundary'' regions
(those only with interval constraints like ``$0 < y <1$").
For instance, for $\mc{X} = \{ x_1, x_2 \}$, the boundary regions $(x_1 = 0, x_2 = 0)$,
$(0 < x_1 < 1, x_2 = 0)$ and $(x_1 = 0, 0 < x_2 < 1)$ are merged with the non-boundary
region $(0 < x_1 < 1, 0 < x_2 < 1)$ yielding $(0 \les x_1 < 1, 0 \les x_2 < 1)$.
The reason for this slight change will become clear later.

Let $\mRe(\mX)$ be the set of regions over the set $\mc{X}$ of clocks.
For $\Theta, \Theta' \in \mRe(\mc{X})$, $\Theta'$ is the \emph{successor region} of $\Theta$
if for all $\eta \models \Theta$ there exists $\delta \in \Reals_{>0}$ such that $\eta{+}\delta \models \Theta'$ and $\forall {\delta'<\delta}.\ \eta{+}\delta' \models \Theta\vee\Theta'$.
The region $\Theta$ \emph{satisfies} the guard $g$, denoted $\Theta\models g$, iff $\forall
\eta \models \Theta$.\ $\eta\models g$.
The \emph{reset operation} on region $\Theta$ is defined as $\Theta[X:=0] :=
\big\{\eta[X:=0]\mid \eta\models \Theta\big\}$.

\begin{defi}[Region graph of \DMTAr]\label{def:region}
The \emph{region graph} of \DMTAr\ $\mc{M} = (Loc, \mc{X}, \ell_0$, $Loc_F,
E, \rightsquigarrow)$ is $\mc{G}(\mc{M}) = (V ,v_0, V_F, \Lambda$, $\hookrightarrow)$,
where
\begin{enumerate}[$\bullet$]
\item $V = Loc\times \mRe(\mc{X})$ is a finite set of \emph{vertices} with
  \emph{initial vertex} $v_0 = (\ell_0,\vec{0})$;

\item $V_F = \big\{v \in V \mid v{\downharpoonright_1}\in Loc_F\big\}$ is
the set of \emph{accepting vertices};

\item
${\Lambda:V\to \Reals_{\ges 0}}$ is the \emph{exit rate function} where:
$$
\Lambda(v) \ = \ \left\{  \begin{array}{ll}
E(v{\downharpoonright_1}) & \mbox{ if } v \stackrel{p,X}{\hookrightarrow}
v' \mbox{ for some } v' \in V \\[1ex]
0 & \mbox{ otherwise}.

\end{array} \right.
$$

\item $\hookrightarrow\ \subseteq V\times\left(\left([0,1]\times
2^\mc{X}\right)\cup\{\delta\}\right)\times V$ is the
\emph{transition $($edge$)$ relation}, such that:
\begin{enumerate}[$\blacktriangleright$]
\item
$v\stackrel{\delta}{\hookrightarrow}v'$ 
if $v{\downharpoonright_1} = v'{\downharpoonright_1}$, and
$v'{\downharpoonright_2}$ is the successor region of $v{\downharpoonright_2}$;

\item
$v\stackrel{p,X}{\hookrightarrow}v'$ 
if 
$\updownmapsto{v{\downharpoonright_1}}{g,X}{p}{v'{\downharpoonright_1}}$ 
with $v{\downharpoonright_2} \models g$, and $v{\downharpoonright_2}[X:=0]= v'{\downharpoonright_2}$.
\end{enumerate}
\end{enumerate}
\end{defi}\medskip

\noindent Any vertex in the region graph is a pair consisting of a location and a region.
Edges of the form $v\stackrel{\delta}{\hookrightarrow}v'$ are called delay edges,
whereas those of the form $v\stackrel{p,X}{\hookrightarrow}v'$ are called Markovian
edges.   Note that Markovian edges emanating from a boundary region do \emph{not}
contribute to the reachability probability as the time to hit the boundary is always zero
(i.e., $\flat(v,\eta)=0$ in Eq.~\eqref{eq:Markovian}, page\,\pageref{eq:Markovian}).
Therefore, we can safely remove all the Markovian edges emanating from boundary
regions and combine each such boundary region with its unique non-boundary (direct)
successor.  In the sequel, by slight abuse of notation, we refer to this \emph{simplified region
graph} as $\mc{G}(\mc{M})$. Note that then
$v{\downharpoonright_2}[X:=0]\subseteq v'{\downharpoonright_2}$ in the last item of
Definition~\ref{def:region}.

\begin{rem}$[$Exit rates$]$
The exit rate $\Lambda(v)$ equals $0$ if only delay transitions emanate from $v$.
The probability to take the delay edge within time $t$ is $e^{-\Lambda(v)t} = 1$,
while the probability to take Markovian edges is $0$.
\end{rem}

\begin{exa}
For the \DMTAr\ $\mc{C}{\otimes}\mc{A}$ in Figure~\ref{fig:MTA_1c},
the reachable part (forward reachable from the initial vertex and
backward reachable from the accepting vertices) of the simplified
region graph $\mc{G}(\mc{C}{\otimes}\mc{A})$ is shown in
Figure~\ref{fig:region_1c}. Note that the exit rates on $v_4$ and $v_7$ are $0$,
as only a delay edge is enabled in these vertices.
\end{exa}

The following result asserts that the region graph obtained from a DMTA is
in fact a PDP.  This is an important observation, as verification
now reduces to analyzing this PDP.

\begin{lem}
The region graph of any \DMTA\ induces a PDP.
\end{lem}

\begin{proof}
Let \DMTAr\ $\mc{M} = (Loc, \mc{X}, \ell_0, Loc_{\bdF}, E, \rightsquigarrow)$
with region graph $\mc{G}(\mc{M}) = (V, v_0, V_{\bdF}, \Lambda,$ $\hookrightarrow)$.
Define $\mc{Z}(\mc{M}) = \left(V, \mc{X}, \Inv, \phi, \Lambda, \mu \right)$
where for any $v \in V$:
\begin{enumerate}[$\bullet$]
\item
$\Inv(v) := v{\downharpoonright_2}$ and the state space
$\mb{S} := \big\{(v,\eta)\mid v\in V, \eta\models\Inv(v)\big\}$;
\item
$\phi(v,\eta,t) :=\eta+t$;
\item
$\Lambda(v,\eta) :=\Lambda(v)$;
\item
if $v\stackrel{\delta}{\hookrightarrow}v'$ in $\mc{G}(\mc{M})$, then
$\mu((v,\eta),\{ (v',\eta) \}) := 1$, provided $\eta \models \partial \Inv(v)$;
\item
if $v\stackrel{p,X}{\hookrightarrow}v'$ in $\mc{G}(\mc{M})$, then
$\mu((v,\eta), \{ (v',\eta[X:=0])\}) := p$, provided $\eta \models \Inv(v)$.
\end{enumerate}
It follows directly that $\mc{Z}(\mc{M})$ is a
PDP.
\end{proof}
\noindent Note that the acceptance conditions play no role in the definition of a PDP,
thus this lemma applies to both \DMTAr\ and \DMTAo.

\section{Verifying CTMCs Against Finite DTA Specifications} \label{sec:finite}

The characterization of the region graph of $\mc{C} \otimes \mc{A}$ as a PDP paves
the way to the verification of CTMC $\mc{C}$ against \DTAr\ specification $\mc{A}$.
This section concentrates on the quantitative verification problem and deals with
single-clock DTA separately.

\subsection{Quantitative verification with arbitrarily many clocks}\label{sec:general_DTA}

The central issue in quantitative verification is to compute the probability of the set
of paths in $\mC$ accepted by $\mA$.   By Theorem~\ref{th:CTMC=MTA}, this is
equal to computing reachability probabilities in DTMA $\mC \otimes \mA$.   The
remaining question is how to determine these probabilities.  To that end, we show
that this amounts to determine reachability probabilities of untimed events in the
embedded PDP of $\mc{Z}(\mc{C} \otimes \mc{A})$ (cf.\ Theorem~\ref{th:MTA=%
DTMP} below).  These probabilities are characterized by a Volterra integral equation
system of second type.  As solving this integral equation system is typically hard, we
present an effective approximation algorithm.

\subsection*{Characterizing reachability probabilities.}
\label{sec:general_DTA_characterization}
We first consider determining unbounded reachability probabilities in the \PDP\
$\mc{Z} = \mc{Z}(\mc{C}\otimes\mc{A})$.
This is done by considering its embedded PDP, the DTMP $\emb(\mc{Z})$, as
for unbounded reachability probabilities, the timing aspects are not important.
Note that the set of locations of PDP $\mc{Z}$ and $\emb(\mc{Z})$ are equal.
Besides, the discrete probabilistic evolution of $\mc{Z}$ and $\emb(\mc{Z})$
coincide.  The main difference is that $\emb(\mc{Z})$ is time-abstract whereas
$\mc{Z}$ is not.

Let  initial state $(v_0,\vec{0})$ and $T \subseteq V$ be the set of goal locations.
For state $(v,\eta)$, let ${\Prob}^{\emb(\mc{Z})} \big((v,\eta), T \big)$,  ${\Prob}_v(\eta, T)$
for short, denote the probability to reach some state in $(T,\cdot)$ from state $(v,\eta)$ in $\emb(\mc{Z})$.
These probabilities are recursively defined as follows.
For vertex $v \in V$, we have: 
\begin{equation}\label{eq:basepdp}
\Prob_v(\eta,T) \ = \
\left\{ \begin{array}{ll}
1 & \mbox{if } v \in T \\[1ex]
\Prob_{v,\delta}(\eta,T) +
{\sum}_{v\stackrel{p,X}{\hookrightarrow}v'} \Prob_{v,v'}(\eta,T)
&\mbox{otherwise}
\end{array}\right.
\end{equation}
The case $v \in T$ is evident.  In case $v \not\in T$, then either a delay can
take place (first summand), or a Markovian edge is taken to vertex $v'$ (second
summand).

For a delay transition $v \stackrel{\delta}{\hookrightarrow} v'$ we have:
\begin{equation}\label{eq:delay}
{\Prob}_{v,\delta}(\eta, T) \ = \
   e^{-\Lambda(v){\cdot}\flat(v,\eta)} \cdot {\Prob}_{v'}\big(\eta{+}\flat(v,\eta), T \big),
\end{equation}
where $e^{{-}\Lambda(v){\cdot}\flat(v,\eta)}$ is the probability to stay in $v$ for at most
$\flat(v,\eta)$ time units.  Recall that $\flat(v,\eta)$ is the minimal time for state $(v,\eta)$
to hit the boundary $\partial \Inv(v)$.   Stated in other words, $e^{{-}\Lambda(v){\cdot}
\flat(v,\eta)}$ is the probability to reside in $v$ without violating the invariant.  The reachability
probability from the resulting state $\eta{+}\flat(v,\eta)$ is then given by the second
multiplicand in Eq.\,\eqref{eq:delay}.  This equation is based on Eq.\,\eqref{eq:embedded2}
by determining the multi-step reachability probability using a sequence of one-step
transition probabilities.

For the Markovian transition $v\stackrel{p,X}{\hookrightarrow}v'$, we have:
\begin{equation}\label{eq:Markovian}
{\Prob}_{v, v'} (\eta, T) \ = \
\int_0^{\flat(v,\eta)}p{\cdot}\Lambda(v) {\cdot} e^{{-}\Lambda(v){\cdot}\tau} \cdot
{\Prob}_{v'}\big((\eta+\tau)[X:=0], T\big)\ d\tau.
\end{equation}
Here, $\Lambda(v) {\cdot} e^{{-}\Lambda(v){\cdot}\tau}$ denotes the density to stay
for exactly $\tau$ time units in $v$.  As any delay up to $\flat(v,\eta)$ does not violate
the invariant, $\tau$ ranges over the dense interval $[0, \flat(v,\eta)]$.  The state after
first delaying $\tau$ time units and then taking the edge $v \stackrel{p,X}{\hookrightarrow}
v'$ is $(\eta+\tau)[X:=0]$.
Eq.\,\eqref{eq:Markovian} is derived from Eq.\,\eqref{eq:embedded}.

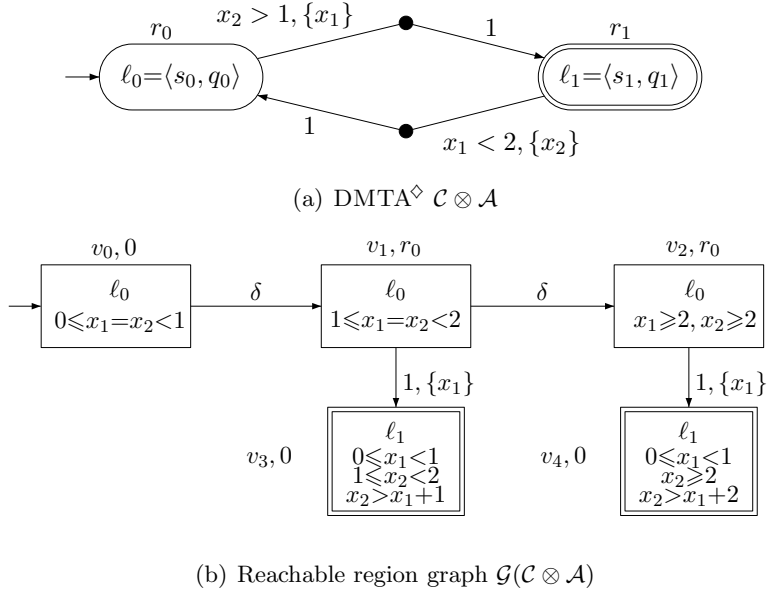
\begin{figure}[h]
 \centering
\subfigure[\DMTAr\
 $\mc{C}\otimes\mc{A}$]{\scalebox{0.9}{






%


\begin{picture}(106,31)(0,-31)
\put(0,-31){}
\node[NLangle=0.0,Nmarks=i,Nw=24.15,Nh=9.75,Nmr=4.83](n30)(22.78,-17){$\ell_0{=}\<s_0,q_0\>$}

\node[NLangle=0.0,Nmarks=r,Nw=24.15,Nh=9.75,Nmr=4.83](n32)(87.62,-17){$\ell_1{=}\<s_1,q_1\>$}

\node[Nframe=n](n16)(19.88,-10.23){$r_0$}

\node[Nframe=n](n17)(87.76,-10.23){$r_1$}

\node[Nfill=y,fillcolor=Black,Nw=2.0,Nh=2.0,Nmr=1.0](n10)(56.0,-9){}

\node[Nfill=y,fillcolor=Black,Nw=2.0,Nh=2.0,Nmr=1.0](n11)(56.0,-25){}

\drawedge[AHnb=0](n30,n10){$x_2>1,\{x_1\}$}

\drawedge[ELpos=38](n10,n32){$1$}

\drawedge[ELpos=57,AHnb=0](n32,n11){ $x_1<2,\{x_2\}$}

\drawedge[ELpos=41,ELdist=1.18](n11,n30){$1$}

\end{picture}\label{fig:MTA_r}}
 }
\subfigure[Reachable region graph
$\mc{G}(\mc{C}\otimes\mc{A})$]{\scalebox{0.85}{\begin{picture}(126,50)(0,-50)
\put(0,-50){}
\node[Nmarks=i,Nw=23.38,Nh=12.95,Nmr=0.0](n0)(20.0,-12.24){}

\node[Nframe=n](n9)(19.71,-3.3){$v_0, 0$}

\node[Nw=23.38,Nh=12.95,Nmr=0.0](n1)(109.62,-12.24){}

\node[Nframe=n](n11)(110.33,-3.3){$v_2, r_0$}

\node[Nw=23.38,Nh=12.95,Nmr=0.0](n2)(63.81,-12.24){}

\node[Nframe=n](n10)(63.52,-3.3){$v_1, r_0$}

\node[Nmarks=r,Nw=21.38,Nh=16.95,Nmr=0.0](n4)(63.81,-36.53){}

\node[Nframe=n](n12)(44.0,-36.0){$v_3, 0$}

\node[Nmarks=r,Nw=21.38,Nh=16.95,Nmr=0.0](n21)(109.62,-36.53){}

\drawedge(n1,n21){$1,\{x_1\}$}

\drawedge(n0,n2){$\delta$}

\drawedge(n2,n1){$\delta$}

\drawedge(n2,n4){$1,\{x_1\}$}

\node[Nframe=n](n13)(90.0,-36.0){$v_4, 0$}

\node[Nframe=n,NLangle=0.0,Nmr=0.0](n24)(20.65,-9.71){$\ell_0$}

\node[Nframe=n,NLangle=0.0,Nmr=0.0](n25)(20.65,-14.71){$0{\leqslant}x_1{=}x_2{<}1$}

\node[Nframe=n,NLangle=9.46,Nmr=0.0](n26)(63.81,-9.71){$\ell_0$}

\node[Nframe=n,NLangle=0.0,Nmr=0.0](n28)(63.75,-14.54){$1{\leqslant}x_1{=}x_2{<}2$}

\node[Nframe=n,NLangle=0.0,Nmr=0.0](n40)(110.62,-9.71){$\ell_0$}

\node[Nframe=n,NLangle=0.0,Nmr=0.0](n42)(110.62,-14.71){$x_1{\geqslant}2,x_2{\geqslant}2$}

\node[Nframe=n,NLangle=0.0,Nmr=0.0](n44)(63.81,-32.0){$\ell_1$}

\node[Nframe=n,NLangle=0.0,Nmr=0.0](n45)(63.81,-36.0){$0{\leqslant}x_1{<}1$}

\node[Nframe=n,NLangle=0.0,Nmr=0.0](n46)(63.81,-39.0){$1{\leqslant}x_2{<}2$}

\node[Nframe=n,NLangle=0.0,Nmr=0.0](n47)(63.81,-42.0){$x_2{>}x_1{+}1$}

\node[Nframe=n,NLangle=0.0,Nmr=0.0](n50)(109.62,-32.0){$\ell_1$}

\node[Nframe=n,NLangle=0.0,Nmr=0.0](n51)(109.62,-36.0){$0{\leqslant}x_1{<}1$}

\node[Nframe=n,NLangle=0.0,Nmr=0.0](n52)(109.62,-39.0){$x_2{\geqslant}2$}

\node[Nframe=n,NLangle=0.0,Nmr=0.0](n53)(109.62,-42.0){$x_2{>}x_1{+}2$}

\end{picture}\label{fig:reachable}}}
\caption{Reachable fragment of its region graph}\label{fig:region_construction}
\end{figure}

\begin{exa}\label{ex:integral}
Consider the \DMTAr\ in Figure~\ref{fig:MTA_r} and its region graph in Figure~\ref{fig:reachable}.
Let $T = V_F$ be the set of goal locations, i.e., the set of target states $\{ (v,\eta) \mid v \in V_F,
\eta \models \Inv(v) \}$.
The system of integral equations for $v_1$ in location $\ell_0$ is as follows.
For $1 \leqslant x_1 = x_2 < 2$:
$$\Prob_{v_1}(x_1,x_2)=\Prob_{v_1,\delta}(x_1,x_2)+\Prob_{v_1,v_3}(x_1,x_2),$$
where
$$\Prob_{v_1,\delta}(x_1,x_2)=e^{-(2-x_1)r_0}{\cdot}\Prob_{v_2}(2,2)$$
and
$$\Prob_{v_1,v_3}(x_1,x_2)=\int_0^{2-x_1}r_0{\cdot}e^{-r_0\tau}{\cdot}\Prob_{v_3}(0,x_2+\tau)\
d\tau
$$
where $\Prob_{v_3}(0,x_2+\tau)=1$.
The integral equations for vertices $v_2,v_4$ are similar.
\end{exa}

\begin{rem}
Clock valuations $\eta$ and $\eta'$ in region $\Theta$ may induce different
reachability probabilities.  This is due to the fact that $\eta$ and $\eta'$ may
have different periods of time to hit the boundary,  Thus, the probability for
$\eta$ and $\eta'$ to either delay or take a Markovian transition may differ.
This is in contrast with timed automata, as well as probabilistic extensions
thereof~\emph{\cite{KNSS02}}, where clock valuations in the same region are
not distinguished.
\end{rem}

Hence, reachability probabilities in the embedded PDP of $\mc{Z}(\mC \otimes \mA)$
are characterized by a system of \emph{Volterra integral equations} \eqref{eq:basepdp}.
One can read \eqref{eq:basepdp} either in the form $f(\xi)=\int_{Dom(\xi)}K(\xi,\xi')f(d\xi')$,
where $K$ is the kernel and $Dom(\xi)$ is the domain of integration depending on
the continuous state space $\mb{S}$; or in the operator form $f(\xi)=(\mc{J}\!f)(\xi)$, where
$\mc{J}$ is the integration operator.  Generally, \eqref{eq:basepdp} does \emph{not}
necessarily have a unique solution.  It turns out that the reachability probability ${\Prob}_{v_0}
(\vec{0})$ coincides with the least fixpoint of the operator $\mc{J}$ (denoted by ${\rm
lfp}\mc{J}$) i.e., ${\Prob}_{v_0}(\vec{0}) = ({\rm lfp}\mc{J})(v_0,\vec{0})$.

\begin{thm}\label{th:MTA=DTMP}
For any \CTMC\ $\mc{C}$ and \DTAr\ $\mc{A}$,
$$
{\Pr}^{\mc{C}\otimes \mc{A}}_{\vec{0}}\big(\AccPaths^{\mc{C}\otimes \mc{A}}\big)
\mbox{ is the least solution of } {\Prob}^\mc{D}_{v_0}(\vec{0},V_F),
$$
where DTMP $\mc{D} = \emb(\mc{Z}(\mc{C}\otimes \mc{A}))$.
\end{thm}
\begin{proof}
Let ${\Pr}^{\mc{C}\otimes \mc{A}}_{\vec{0}}\big(\AccPaths^{\mc{C}\otimes \mc{A}}\big)$
be the least solution of the system of integral equations:
\begin{equation*}
\Pr(\ell,\eta) = \left\{ \begin{array}{ll}
1 & \mbox{ if } \ell\in Loc_F \\[1ex]
\displaystyle \int_0^{\infty} E(\ell){\cdot}e^{-E(\ell)\tau} \cdot\!\!\!\!\!
\sum_{\updownmapsto{\ell}{g, X}{p}{\ell'}}
\mathbf{1}_{g}(\eta{+}\tau){\cdot}p{\cdot}\Pr(\ell',(\eta{+}\tau)[X:=0])\ d\tau\!\!\!\!
& \mbox{ otherwise, }
\end{array} \right.
\end{equation*}
Informally, $\Pr(\ell,\eta)$ is the probability to reach the set of locations $Loc_F$ from
location $\ell$ and clock valuation $\eta$.
The above integral can be simplifed as follows.
W.l.o.g. assume clock constraints to be of the form $x \trianglelefteq c$, where $c \in
\Nats$ and $\trianglelefteq \, \in\{\le,<,\ge,>\}$.
Then we have:
\begin{equation*}
\Pr(\ell,\eta) =
\int_{t_1}^{t_2} E(\ell){\cdot}e^{{-}E(\ell)\tau} \cdot
\sum_{\updownmapsto{\ell} {g,X}{p}{\ell'}}p\cdot\Pr(\ell',(\eta{+}\tau)[X:=0])\ d\tau,
\end{equation*}
where $t_1, t_2\in\Rationals_{\geqslant 0}\cup\{\infty\}$ and $\eta{+}\tau \models g$
for any $t_1 < \tau < t_2$.

If $\ell\in Loc_F$, the theorem follows directly.  In the remainder of the proof, assume
$\ell\notin Loc_F$.  Our proof is based on showing that for any $\ell\notin Loc_F$ and
clock valuation $\eta$,
\begin{equation}\label{eq:unbpres}
\Pr(\ell,\eta) \ = \ {\Prob}_{v_0}(\eta,V_F),
\end{equation}
where $v_0$ is the initial vertex in the region graph $\mc{Z}(\mc{C} \otimes
\mc{A})$ with $v_0{\downharpoonright_1} = \ell$, and $V_F = \{ v \in V \mid
v{\downharpoonright_1} \in Loc_F \}$.
This is done as follows.
For natural $n$, let ${\Pr}^n(\ell,\eta)$ be the probability to reach $Loc_F$ in $n$
steps in $\mc{C} \otimes \mc{A}$.
For $n=0$, we have ${\Pr}^n(\ell,\eta) = 1$ if $\ell \in Loc_F$ and 0, otherwise.
For $n > 0$, we define inductively:
\begin{equation*}
{\Pr}^{n}(\ell,\eta) =
\int_{t_1}^{t_2} E(\ell){\cdot}e^{-E(\ell)\tau} \cdot
\sum_{\updownmapsto{\ell}{g,X}{p}{\ell'}} p \cdot {\Pr}^{n{-}1}(\ell',\eta')\ d\tau.
\end{equation*}
Similarly, let $\Prob_v^{n}(\eta, V_F)$ be the probability to reach the set of goal states
$V_F$ in $n > 0$ steps:
\begin{eqnarray}
\Prob_v^n(\eta,V_F)&=&
\left\{\begin{array}{ll}\Prob_{v,\delta}^n(\eta,V_F)+{\Prob}_{v}
^{s,n}(\eta,V_F), &\quad
\mbox{if } v\notin V_F \\[1ex]
1,& \quad \mbox{otherwise}
\end{array}\right.\\
{\Prob}_{v}^{s,n}(\eta,V_F)&=&
\int_0^{\flat(v,\eta)}\hspace{-0.6cm}\Lambda(v) {\cdot}
e^{-\Lambda(v)\tau}{\cdot}\sum_{v\stackrel{p,
X}{\hookrightarrow}v'}p{\cdot}{\Prob}_{v'}^{n-1}\big((\eta{+}\tau)[X{:=}0],V_F
\big)\ d\tau,\\
{\Prob}_{v,\delta}^n(\eta,V_F)&=&e^{-\Lambda(v)\flat(v,
\eta)}\cdot{\Prob}_{v'}^n\big(\eta+\flat(v,\eta),V_F\big).\label{eq:prob_n_v_delta}
\end{eqnarray}
In the sequel, we show that for any $n\in\Nats$, it holds:
\begin{equation}\label{eq:unbpresind}
{\Pr}^{n}(\ell,\eta) = \Prob_{v_0}^{n}(\eta,V_F).
\end{equation}
The theorem then follows from the fact that $\displaystyle
\lim_{n\to\infty}{\Pr}^{n}(\ell,\eta)={\Pr}(\ell,\eta)$ and, similarly,
$\displaystyle \lim_{n\to\infty}\Prob_v^{n}(\eta,V_F)=\Prob_v(\eta,V_F)$.

\begin{figure}[!ht]
\centering \scalebox{0.85}{

\begin{picture}(172,38)(0,-38)
\node[Nmarks=i,Nw=20.38,Nh=10.95,Nmr=0.0](n0)(13.0,-8.37){}

\node[Nframe=n,NLangle=0.0](n26)(35.84,-8.37){$\cdots$}

\node[Nw=28.38,Nh=10.95,Nmr=0.0](n20)(65.81,-8.37){}

\node[Nw=20.38,Nh=10.95,Nmr=0.0](n28)(101.78,-8.37){}

\node[Nframe=n,NLangle=0.0](n34)(124.62,-8.37){$\cdots$}

\node[Nw=22.38,Nh=10.95,Nmr=0.0](n31)(151.59,-8.37){}

\node[Nframe=n,NLangle=0.0,Nmr=0.0](n24)(12.03,-6.16){$v_0{=}(\ell{,}
\Theta_0)$}

\node[Nframe=n,NLangle=0.0,Nmr=0.0](n25)(12.97,-10.57){\begin{small}$\flat(v_{0}
{,}\hat\eta_{0}){\leqslant}
1$\end{small}}

\node[Nframe=n,NLangle=0.0,Nmr=0.0](n21)(65.81,-6.16){$v_{m-1}{=}(\ell{,}
\Theta_{m-1})$}

\node[Nframe=n,NLangle=0.0,Nmr=0.0](n22)(65.81,-10.57){\begin{small}$\flat(v_{m{
-}1}{,}\hat\eta_{m{-}1}){=}
1$\end{small}}

\drawedge[ELpos=65,ELdist=1.18](n0,n26){$\delta$}

\drawedge[ELpos=30,ELdist=1.18](n26,n20){$\delta$}

\node[Nframe=n,NLangle=0.0,Nmr=0.0](n29)(101.78,-6.16){$v_m{=}(\ell{,}
\Theta_m)$}

\node[Nframe=n,NLangle=0.0,Nmr=0.0](n30)(101.78,-10.57){\begin{small}$\flat(v_{m
}{,}\hat\eta_{m}){=}
1$\end{small}}

\node[Nframe=n,NLangle=0.0,Nmr=0.0](n32)(151.59,-6.16){$v_{k}{=}(\ell{,}
\Theta_{k})$}

\node[Nframe=n,NLangle=0.0,Nmr=0.0](n33)(151.59,-10.57){\begin{small}$\flat(v_{k
}{,}\hat\eta_{k}){=}
1$\end{small}}

\drawedge[ELpos=65,ELdist=1.18](n28,n34){$\delta$}

\drawedge[ELpos=35,ELdist=1.18](n34,n31){$\delta$}

\drawedge[ELpos=57,ELdist=2.36](n20,n28){$\delta$}

\node[Nw=28.38,Nh=10.95,Nmr=0.0](n76)(101.78,-28.16){}

\node[Nframe=n,NLangle=0.0,Nmr=0.0](n77)(101.78,-25.95){$v_{m}'{=}{(}\ell'{,}
\Theta_m{)}$}

\node[Nframe=n,NLangle=0.0,Nmr=0.0](n78)(101.78,-30.36){\begin{small}$\flat(v_{m
}'{,}\hat\eta_{m}'){\leqslant}
1$\end{small}}

\drawedge(n28,n76){$p,X$}

\node[Nw=28.38,Nh=10.95,Nmr=0.0](n86)(151.59,-28.16){}

\node[Nframe=n,NLangle=0.0,Nmr=0.0](n87)(151.59,-25.95){$v_{k}'{=}{(}\ell'{,}
\Theta_{k}{)}$}

\node[Nframe=n,NLangle=0.0,Nmr=0.0](n88)(151.59,-30.36){\begin{small}$\flat(v_{k
}'{,}\hat\eta_{k}'){\leqslant}1$\end{small}}

\drawedge(n31,n86){$p,X$}

\node[Nframe=n,Nmr=0.0](n44)(76.0,-24.0){}

\drawedge(n28,n44){}

\node[Nframe=n,NLangle=0.0,Nmr=0.0](n46)(96.0,-16.0){$\cdots$}

\node[Nframe=n,NLangle=0.0,Nmr=0.0](n47)(148.0,-16.0){$\cdots$}

\node[Nframe=n,Nmr=0.0](n49)(128.0,-24.0){}

\drawedge(n31,n49){}

\end{picture}}
\caption{The sub-region graph $\mc{Z}(\mc{C} \otimes \mc{A})$ for the
transition from $\ell$ to $\ell'$.}\label{fig:subregion}
\end{figure}
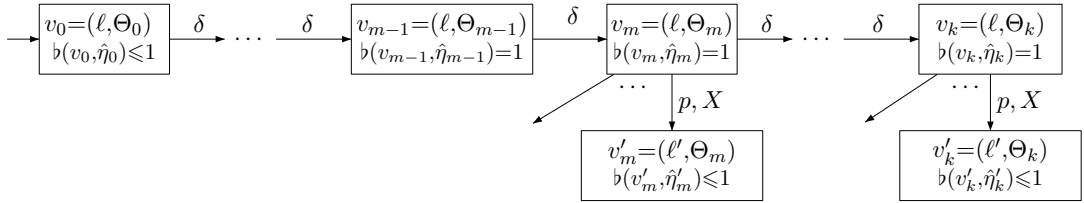
The proof of ${\Pr}^{n}(\ell,\eta) = \Prob_{v_0}^{n}(\eta,V_F)$ is by induction
on $n$.
\begin{enumerate}[(1)]
\item
(Base case.) For $n=0$, ${\Pr}^{0}(\ell,\eta) = 0 = \Prob_{v_0}^{0}(\eta,V_F)$ if
$\ell \notin Loc_F$, and 1 otherwise.
\item
(Induction step.)
Consider $n{+}1$.
Let edge $\ell \stackrel{g,X}{\rightsquigarrow} \zeta$ in $\mc{C} \otimes \mc{A}$.
Assume the fragment of the region graph $\mc{Z}(\mc{C} \otimes \mc{A})$ that
corresponds to this edge with $\zeta(\ell, \ell') > 0$ is as shown in Fig.\ \ref{fig:subregion}.
Location $\ell$ induces the vertices $\{ v_i = (\ell,\Theta_i) \mid 0 \leqslant i \leqslant k \}$.
Intuitively speaking, the transition from location $\ell$ to $\ell'$ is enabled in
region $\Theta_i$ for $m \leqslant i \leqslant k$, whereas only a delay can take
place in all regions $\Theta_i$ with $i < m$ (while staying in location $\ell$).

Let $\hat\eta_i$ be the clock valuation when entering vertex $v_{i}$, i.e., $\hat\eta_0
= \eta$ and $\hat\eta_i = \hat\eta_{i-1} + \flat(v_{i-1},\hat\eta_{i-1})$ for $0 < i \leqslant
k$.
It is assumed that $\hat\eta_i \not\models g$, where $g$ is the guard of the edge
at hand, for $i < m$ and $i > k$.
Accordingly,
$$
t_1=\sum_{i=0}^{m-1}\flat(v_i,{\hat\eta}_i) \qquad\mbox{and} \qquad
t_2=\sum_{i=0}^{k}\flat(v_i,{\hat\eta}_i)
$$
are the lower and upper bound, respectively, of the interval during which guard $g$ holds.

For convenience, let $p_{v}^{n}(\eta) := \Prob_{v,\delta}^n(\eta,V_F) +
{\Prob}_{v}^{s,n}(\eta,V_F)$.
Given the fact that only a delay transition can be taken before time $t_1$, it holds that
\begin{eqnarray*}
p_{v_0}^{n+1}(\eta) & = &
e^{-t_1\Lambda(v_0)}\cdot p_{v_m}^{n+1}(\hat\eta_m), \mbox{ where} \nonumber \\[1ex]
p_{v_m}^{n+1}(\hat\eta_m) & = &
\Prob^{n+1}_{v_m,\delta}(\hat\eta_m,V_F)+\Prob^{s,n+1}_{v_m}(\hat\eta_m,V_F).
\end{eqnarray*}

\noindent
We now derive:
\begin{eqnarray*}
& & e^{-t_1\Lambda(v_0)}{\cdot}\Prob^{s,n+1}_{v_m}(\hat\eta_m,V_F) \nonumber \\[1ex]
& = &e^{-t_1\Lambda(v_0)}{\cdot}\int_0^{\flat(v_m,\hat\eta_m)}
\!\!\!\Lambda(v_m){\cdot}e^{-\Lambda(v_m)\tau}{\cdot}\!\!\!\sum_{v_m\stackrel{
p,X}{\hookrightarrow}v_m'}p{\cdot}{\Prob}^{n}_{v_{m}'}
\big((\hat\eta_m{+}\tau)[X:=0],V_F \big)d\tau\nonumber\\
&=& \int_{t_1}^{t_1+\flat(v_m,\hat\eta_m)}
\!\!\!\Lambda(v_m){\cdot}e^{-\Lambda(v_m)\tau}{\cdot}\!\!\!\sum_{v_m\stackrel{
p,X}{\hookrightarrow}v_m'}p{\cdot}{\Prob}^{n}_{v_{m}'}
\big((\hat\eta_m{+}\tau{-}t_1)[X:=0],V_F \big)d\tau.
\end{eqnarray*}
Now consider:
$$
p_{v_0}^{n+1}(\eta) =
e^{-t_1\Lambda(v_0)}{\cdot}\Prob^{n+1}_{v_m,\delta}(\hat\eta_m,V_F) +
e^{-t_1\Lambda(v_0)}{\cdot}\Prob^{s,n+1}_{v_m}(\hat\eta_m,V_F).
$$
Using the definition of $\Prob^{n+1}_{v_m,\delta}(\hat\eta_m,V_F)$ (see Eq.\,\eqref{eq:prob_n_v_delta}),
together with the result derived above, yields the following sum of integrals:
\begin{eqnarray*}
\quad \quad p_{v_0}^{n+1}(\eta) & \!\! = \!\!
&\sum_{i=0}^{k-m}\int_{t_1+\sum_{j=0}^{i-1}
\flat(v_{m+j},\hat\eta_{m+j})}^{t_1+\sum_{j=0}^{i}\flat(v_{m+j},\hat\eta_{
m+j})}\!\!\!\Lambda(v_{m+i}){\cdot}e^{-\Lambda(v_{m+i})\tau}
\!\!\!\\\nonumber\\
&\cdot&
\underbrace{\sum_{v_{m+i}\stackrel{
p,X}{\hookrightarrow}v_{m+i}'}p{\cdot}{\Prob}^{n}_{v_{m+i}'}
\big((\hat\eta_{m+i}{+}\tau{-}t_1{-}\sum_{j=0}^{i-1}
\flat(v_{m+j},\hat\eta_{m+j}))[X:=0],V_F \big)}_{= F^n(\tau)} \, d\tau.
\end{eqnarray*}
Using $F^{n}(t)$ we obtain:
\begin{equation}\label{eq:genxxx}
p_{v_0}^{n+1}(\eta)=\int_{t_1}^{t_2} \Lambda(v_0){\cdot}e^{
-\Lambda(v_0)\tau}{\cdot}F^{n}(\tau) \, d\tau.
\end{equation}
Notice that
\begin{equation*}
\hat\eta_{m+i}=\eta+ \underbrace{\sum_{j=0}^{m-1}\flat(v_j,\hat\eta_j)}_{= \, t_1} +\sum_{j=0}^{i-1} \flat(v_{m+j},\hat\eta_{m+j}).
\end{equation*}
Therefore, for any
$t\in[t_1+\sum_{j=0}^{i-1}\flat(v_{m+j},\hat\eta_{m+j}),t_1+\sum_{j=0}^{i}
\flat(v_{m+j},\hat\eta_{m+j})]$, $i\leqslant k-m$ we obtain
\begin{equation*}
\hat\eta_{m+i}+t-t_1-\sum_{j=0}^{i-1}\flat(v_{m+j},\hat\eta_{m+j})=\eta+t.
\end{equation*}
From the induction hypothesis (for $n$), it follows that $\Pr^{n}(\ell,\eta) =
\Prob^{n}_{v_0}(\eta,V_F)$ with $v_0{\downharpoonright_1}=\ell$.
Therefore, for any $t\in[t_1+\sum_{j=0}^{i-1}\flat(v_{m+j},\hat\eta_{m+j}),t_1+\sum_{j=0}^{i}
\flat(v_{m+j},\hat\eta_{m+j})]$  and
$v_{m+i}'{\downharpoonright_1}=\ell'$, $i\leqslant k-m$, we get
\begin{eqnarray*}
\quad \quad \quad
F^{n}(t)
& = &
\sum_{v_{m+i}\stackrel{p,X}{\hookrightarrow}v_{m+i}'}p{\cdot}{\Prob}^{n}_{v_{m+i}'}
\big((\hat\eta_{m+i}{+}t{-}t_1{-}\sum_{j=0}^{i-1} \flat(v_{m+j},\hat\eta_{m+j}))[X:=0],V_F \big)\\
&=&
\sum_{v_{m+i}\stackrel{p,X}{\hookrightarrow}v_{m+i}'}p{\cdot}{\Prob}^{n}_{v_{m+i}'}
\big((\eta{+}t))[X:=0],V_F \big)\\
&=&
\sum_{v_{m+i}\stackrel{p,X}{\hookrightarrow}v_{m+i}'}p{\cdot}{\Pr}^{n}(\ell',(\eta{+}t))[X:=0])\\
&=&\sum_{\updownmapsto{\ell}
{g,X}{p}{\ell'}}p{\cdot}{\Pr}^{n}(\ell',(\eta+t))[X:=0]).
\end{eqnarray*}
\noindent
Substituting this result into equation \eqref{eq:genxxx} results in
\begin{equation*}
p_{v_0}^{n+1}(\eta)=\int_{t_1}^{t_2} \Lambda(\ell){\cdot}e^{
-\Lambda(\ell)\tau}{\cdot}\sum_{\updownmapsto{\ell}
{g,X}{p}{\ell'}}p{\cdot}{\Pr}^{n}(\ell',(\eta{+}\tau))[X:=0])d\tau.
\end{equation*}
As for $v_0\notin V_F$,
$\Prob_{v_0}^{n+1}(\eta,V_F)=p_{v_0}^{n+1}(\eta)$ we get that
$\Prob_{v_0}^{n+1}(\eta,V_F)={\Pr}^{n+1}(\ell,\eta)$.
\end{enumerate}
\end{proof}
Note that, similar to the computation of reachability probabilities in DTMCs~\cite{HJ94},
the goal states in $T \subseteq \mathbb{S}$ as well as all states that cannot reach $T$
can be made absorbing, i.e., all outgoing edges can be removed, without affecting the
reachability probabilities.  This may yield a substantial state-space reduction.

\subsection*{Approximating reachability probabilities.}\label{sec:PDE}
The results so far assert that $\Pr(\mc{C} \models \mc{A})$ coincides with reachability
probabilities in an embedded PDP that is obtained via a region construction applied
on the product $\mc{C}\otimes \mc{A}$.
The previous result shows that such reachability probabilities are characterized by
Volterra equations of the second type~\cite{AWW95}.
Such integral equation systems can be solved using techniques explained in standard
textbooks, such as~\cite{Cord91}.
An alternative option ---inspired by a formulation of bounded reachability probabilities
in arbitrary PDPs~\cite{Dav93}--- is to \emph{approximate} the probability $\Pr \big(\Paths^{\mc{C}}(\mc{A})\big)$ by a system of partial differential equations (\PDE s,
for short).
The intuition is to consider paths that are accepted within some time bound $t_{\!f}$.
Let DTA $\mc{A}[t_{\!f}]$ be obtained by adding a single fresh clock $z$, say, to DTA
$\mc{A}$ which is never reset, and strengthening all guards of incoming edges into
$q \in Q_F$ by adding the conjunct $z \leqslant t_{\!f}$.
Obviously, $\Paths^{\mc{C}}(\mc{A}[t_{\!f}]) \subseteq \Paths^{\mc{C}}(\mc{A})$.
Note that $\displaystyle \lim_{t_{\!f}\to\infty}\Pr (\Paths^{\mc{C}}(\mc{A}[t_{\!f}])) =
\Pr (\Paths^{\mc{C}}(\mc{A}))$.

Given \CTMC\ $\mc{C}$, \DTAr\ $\mc{A}$, time bound $t_f$ and \PDP\ $\mc{Z}(\mc{C}\otimes\mc{A})=$ $(V,\mc{X},\Inv,\phi,\Lambda, \mu)$, we have:
$$
{\Pr}^{\mc{C}}\left(\Paths^{\mc{C}}(\mc{A}[t_{\!f}])\right) \ = \
\sum_{\bar v\in V_F}\int_{Inv(\bar v)}\hbar_{v_0}^{\bar v}(t_f,\vec{0},d\eta),
$$
where $\hbar_{v_0}^{\bar v}(t_f,\vec{0},\bar\eta)$ is the probability to reach the state
$(\bar v,\bar\eta)$, with $\bar v\in V_F$ and $\bar\eta\models Inv(\bar v)$ at time $t_f$
from state $(v_0,\vec{0})$.
The transition probability function $\hbar_{v_0}^{\bar v}(t_f,\vec{0},\bar\eta)$ is described
by the following equations:
\begin{enumerate}[$\bullet$]
\item
for $v\in V\setminus V_F$, $\bar v\in V_F$ with $\eta\models \Inv(v)$,
$\bar\eta\models \Inv(v_f)$ and $y\in(0,t_{\!f})$:
\begin{equation}\label{eq:davispde}
\frac{\partial\hbar_{v}^{\bar v}(y,\eta,\bar\eta)}{\partial
y}+\sum_{i=1}^{|\mc{X}|}
\frac{\partial\hbar_{v}^{\bar v}(y,\eta,\bar\eta)}{\partial\eta^{(i)}}+
\Lambda(v){\cdot}\!\!\sum_{v\stackrel{p,X}{\hookrightarrow}v'}p{\cdot}
\big(\hbar_{v'}^{\bar v}(y,\eta[X:=0],\bar\eta)-\hbar_{v}^{\bar
v}(y,\eta,\bar\eta)\big) \ = \ 0,
\end{equation}
where $\eta^{(i)}$ is the $i$'th clock variable.
\item
$\hbar_{v}^{\bar v}(0,\eta,\bar\eta)=1$, when $v=\bar v$ and $\eta=\bar\eta$,
$\hbar_{v}^{\bar v}(0,\eta,\bar\eta)=0$, otherwise.
\item the boundary conditions are:
for $v,\bar v\in V$, $\eta\models\partial \Inv(v),\bar\eta\models\partial
\Inv(\bar v)$ and transition $v\stackrel{\delta}{\hookrightarrow}v'$
we have $\hbar_v^{\bar v}(y,\eta,\bar\eta)=\hbar_{v'}^{\bar
v}(y,\eta,\bar\eta)$.
\end{enumerate}
Equation~\eqref{eq:davispde} is obtained by simplifying a corresponding characterisation in
Davis~\cite{Dav93}, where the author defines the function $\hbar_{v}^{\bar v}(\cdot)$ as an
expectation.
In our setting, $\hbar_{v_0}^{\bar v}(t_f,\vec{0},\bar\eta) = \mb{E}[\mathbf{1}(X_{t_f})\vert
X_0=\xi]$, where $X_\tau$ is the underlying stochastic process of the \PDP\ $\mc{Z}$ with
the state space $\mb{S}$, $\xi = (v,\vec{0})$ and $\mathbf{1}(X_{t_f})$ is the characteristic
function such that $\mathbf{1}(X_{t_f}) =1$ if and only if $X_{t_f}=(\bar v,\bar\eta)$.
The PDE~\eqref{eq:davispde} is a special case of \cite{Dav93} as the flow function in $\mc{Z}$
is linear and the probabilistic jumps to the continuous part of the state space $\mb{S}$ are
non-uniform.

\subsection{Single-clock \texorpdfstring{\DTAr}{DTAdiamond} specifications}\label{sec:singleDTA}

For \emph{single-clock} \DTAr\ specifications, we can simplify the system of Volterra integral
equations (of second type) obtained in the previous section.
As we will show in this subsection, the probability that a CTMC satisfies a single-clock DTA is
given by a system of linear equations whose coefficients are a solution of a system of \ODE s
that can be solved efficiently.
The key observation is that the region graph corresponding to $\mC \otimes \mA$ can be
naturally divided into a number of subgraphs, each of which is a CTMC.

Let $\mA$ be a single-clock DTA with finite acceptance criterion, and $\{c_0, \ldots, c_m\}$
be the set of natural numbers that appear in the clock constraints of $\mA$.
Assume $0=c_0< c_1<\cdots< c_m$, and let $\Delta c_i = c_{i{+}1} - c_i$ for $0 \les i <m$.
Note that for single-clock \DTA, the regions in the region graph of $\mC \otimes \mA$ can
be partitioned by the following intervals: $[c_0,c_1), [c_1, c_2), \ldots, [c_m,\infty)$.
Using this observation, we partition the region graph $\mc{Z}(\mc{C} \otimes \mc{A})$ as
follows.

\begin{figure}\begin{center}
 \scalebox{0.7}{\begin{picture}(158,100)(0,-100)
\put(0,-100){}
\node[Nmarks=i,Nw=22.74,Nmr=0.0](n57)(31.02,-14.49){$\ell_0,0{\les}
x{<}1$}

\node[Nw=22.74,Nmr=0.0](n60)(82.48,-14.49){$\ell_0,1{\les}
x{<}2$}

\node[Nw=22.74,Nmr=0.0](n61)(31.02,-34.43){$\ell_1,0{\les}
x{<}1$}

\node[NLangle=0.0,NLdist=0.45,Nw=22.74,Nmr=0.0](n62)(82.48,-34.43){$\ell_1,1{\les}
x{<}2$}

\drawedge[ELside=r,ELdist=1.13,curvedepth=-3.63](n57,n61){1}

\drawedge[ELpos=66](n60,n61){reset,1}

\node[Nframe=n](n64)(21.78,-8.19){$v_0,r_0$}

\node[Nframe=n](n65)(91.59,-8.19){$v_1,r_0$}

\node[Nframe=n](n66)(21.78,-28.19){$v_2,r_1$}

\node[Nframe=n](n74)(91.41,-28.16){$v_3,r_1$}

\drawedge[ELside=r,curvedepth=-3.73](n61,n57){0.5}

\drawedge(n57,n60){$\delta$}

\drawedge[ELside=r,ELpos=20,ELdist=0.31](n62,n57){reset, 0.5}

\node[Nw=22.74,Nmr=0.0](n53)(31.02,-54.36){$\ell_2,0{\les}
x{<}1$}

\node[Nw=22.74,Nmr=0.0](n54)(82.48,-54.36){$\ell_2,1{\les} x{<}
2$}

\node[NLangle=0.0,NLdist=0.45,Nmarks=r,Nw=22.74,Nmr=0.0](n56)(82.48,-74.3){$\ell_3,1{\les}
x{<}2$}

\node[Nw=22.74,Nmr=0.0](n57)(134.11,-54.36){$\ell_2,x\ges
2$}

\node[Nmarks=r,Nw=22.74,Nmr=0.0](n58)(134.11,-74.3){$\ell_3,x\ges
2$}

\drawedge(n54,n56){$1$}

\node[Nframe=n](n59)(21.6,-48.32){$v_4,r_2$}

\node[Nframe=n](n60)(91.41,-48.32){$v_5,r_2$}

\node[Nframe=n](n162)(91.59,-68.19){$v_7,r_2$}

\drawedge(n53,n54){$\delta$}

\drawedge(n56,n58){$\delta$}

\drawedge(n57,n58){1}

\node[Nframe=n](n63)(141.22,-68.32){$v_8,r_2$}

\drawedge[ELpos=36,ELdist=0.31](n62,n53){reset,0.2}

\drawedge(n61,n53){0.2}

\drawedge(n54,n57){$\delta$}

\node[Nframe=n](n89)(141.4,-48.19){$v_6,r_2$}

\node[dash={2.0 2.0 2.0 3.0}{0.0},Nw=41.05,Nh=85.06,Nmr=20.53](n112)(82.99,-44.75){}

\node[dash={2.0 2.0 2.0 3.0}{0.0},Nw=36.03,Nh=46.37,Nmr=18.02](n113)(134.83,-62.61){}

\node[dash={2.0 2.0 2.0 3.0}{0.0},Nw=36.62,Nh=65.57,Nmr=18.31](n114)(29.98,-34.41){}

\node[Nframe=n](n115)(28.0,-92.0){$\mG_0$}

\node[Nframe=n](n116)(84.0,-92.0){$\mG_1$}

\node[Nframe=n](n117)(136.0,-92.0){$\mG_2$}

\drawedge[ELside=r](n61,n62){$\delta$}

\end{picture}}\end{center}\vspace{-0.6cm}{\caption{Partitioning the region graph of Figure~\ref{fig:region_1c}\label{fig:oval}}}
\end{figure}
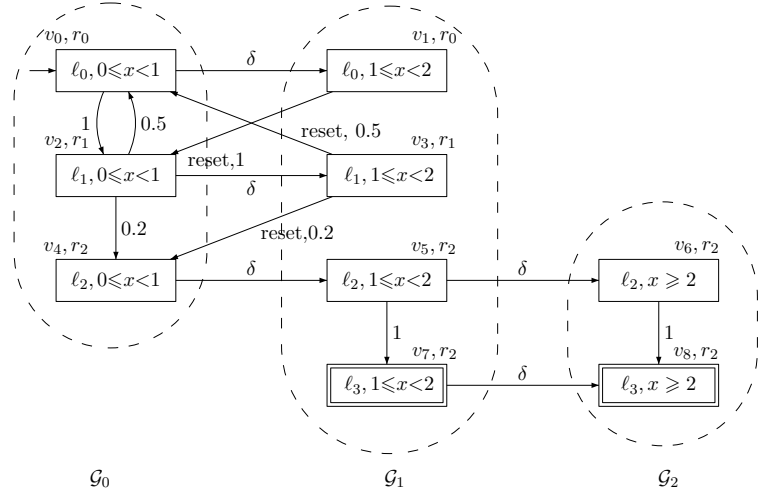

\begin{defi}[Partitioning of region graph]
Let $\mc{G}(\mC \otimes \mA) = (V, v_0, V_F, \Lambda, \hookrightarrow)$,
or $\mc{G}$ for short, for single-clock \DTAr\ $\mA$.
The \emph{partitioning} of $\mc{G}$ is defined as the collection of subgraphs
$\mc{G}_i = (V_i, {V_{F_i}}, \Lambda_i, \hookrightarrow_i)$, for $0 \les
i \les m$ where:
\begin{enumerate}[$\bullet$]
\item $V_i = \{ \, (\ell,\Theta) \in V \mid \Theta\subseteq [c_i,c_{i{+}1}) \, \}$
\item $V_{F_i} = V_i \, \cap \, V_F$,
\item $\Lambda_i(v) = \Lambda(v)$ if $v\in V_i$, and 0 otherwise, and
\item $\hookrightarrow \ = \displaystyle \bigcup_{0\les i\les m } M_i\cup F_i\cup
 B_i$, where
\begin{enumerate}[$-$]
\item
$M_i$ is the set of Markovian edges (without reset) between vertices in $V_i$,
\item
$F_i$ is the set of delay edges between $V_i$ and $V_{i{+}1}$,
\item
$B_i$ is the set of Markovian edges (with reset) from $V_i$ to $V_0$.
\end{enumerate}
\end{enumerate}
\end{defi}
Since the initial vertex of $\mG_0$ is $v_0$ and the initial vertices of $\mG_i$ for
$0< i \les m$ are implicitly given by the edges in $F_{i{-}1}$, we omit them.  Note
that the subgraph $\mG_m$ involves only infinite regions and has no outgoing
delay transitions.
\begin{exa}
Consider the region graph in Figure~\ref{fig:MTA_1c} (page~\pageref{fig:MTA_1c}).
The partitioning of this region graph is depicted in Figure~\ref{fig:oval}.
The edges in $M_i$, $F_i$ and $B_i$ are labeled with probabilities, $\delta$ (delay),
and ``reset'' with probabilities, respectively.
Observe that if $v = (\ell, [c_i,c_{i+1})) \in V_F$, then $v' = (\ell, [c_j,c_{j+1})) \in V_F$
for $i < j \les m$.
(In this example, this applies to $v = v_7$ and $v' = v_8$.)
This is true since $V_F = \{(\ell,\mathsf{true})\mid \ell\in Loc_F\}$.
Thus, from any final vertex in $V_i$ with $i < m$, there is a delay transition to the
next region (if any).
\end{exa}

Assume $|V_i| = k_i$.
We now define for each type of edge ($M, B$, or $F$) a matrix ($\bdM, \bdB$, and $\bdF$, respectively).
Let $x \in \mathbb{R}$ with $x \in [0, \Delta c_i]$.
Then:
\begin{enumerate}[$\bullet$]
\item
$\bdD_i(x) \in \Reals^{k_i\times k_i}$ is the delay probability matrix, where for any
$0\les j\les k_i$, $\bdD_i(x)[j,j]=e^{-E(v_{i}^j){\cdot}x}$ and the off-diagonal elements
are zero.
\item
$\bdM_i(x) = \bdD_i(x){\cdot}\bdE_i{\cdot} \P_i \in \Reals^{k_i\times k_i}$ is the
probability density matrix for $M_i$-edges, where $\P_i$ and $\bdE_i$ are the
transition probability matrix and exit rate matrix respectively, for vertices in $V_i$.
\item
$\bdB_i(x) \in \Reals^{k_i\times k_0}$ is the probability density matrix for the
$B_i$-edges, where $\bdB_i(x)[j,k]$ indicates the probability density function
to take a $B_i$-edge from $v^j \in V_i$ to $v^k \in V_0$.
\item
$\bdF_i \in \Reals^{k_i\times k_{i+1}}$ is the incidence matrix for $F_i$-edges,
i.e., $\bdF_i[j,k]= 1$ if and only if there is a delay transition between $v^j \in V_i$
and $v^k \in V_{i{+}1}$.
\end{enumerate}\medskip

\noindent Due to the fact that in any subgraph $\mG_i$ there are only Markovian jumps
without resets, and no delay transitions, the subgraph $(V_i,\Lambda_i,M_i)$, i.e.,
$\mc{G}$ restricted to Markovian jumps (without resets) forms a \CTMC\ $\mC_i$,
say.  To take the effect of Markovian jumps with resets into account, we define for
each $\mG_i$ the \emph{augmented} \CTMC\ $\mC_i^a$ with state space $V_i \cup
V_0$, where all $V_0$-vertices are absorbing, i.e., do not have any outgoing edges.
The edges connecting $V_i$ to $V_0$ are kept.  The augmented \CTMC\ is used to
calculate the probability to start from a vertex in $\mG_i$ and take a reset edge within
a certain period of time.

\begin{figure}
\begin{center}
\hspace{-0.7cm}
\subfigure[$\mc{C}_0$]{\scalebox{0.7}{\begin{large}

\begin{picture}(28,68)(0,-68)
\put(0,-68){}
\node[iangle=90,Nmarks=i](n0)(12.03,-12.0){$v_0$}

\node(n1)(12.03,-31.0){$v_2$}

\node(n2)(12.03,-50.0){$v_4$}

\drawedge[curvedepth=-4,ELside=r](n0,n1){1}

\drawedge[curvedepth=-4,ELside=r](n1,n0){0.5}

\drawedge(n1,n2){0.2}

\node[Nframe=n](n3)(20.03,-12.0){$r_0$}

\node[Nframe=n](n4)(20.03,-36.0){$r_1$}

\node[Nframe=n](n5)(20.03,-60.0){$r_2$}

\drawloop[loopangle=270.0](n2){1}

\end{picture}
\end{large}}}\hspace{-0.3cm}
\subfigure[$\mc{C}_1$]{\scalebox{0.7}{\begin{large}
%
%
%
%
%
%
%
%
%
%
%
%
%
%

\begin{picture}(37,58)(0,-58)
\put(0,-58){}
\node(n2)(20.6,-12.15){$v_5$}

\node[Nframe=n](n5)(28.6,-12.15){$r_2$}

\node(n6)(20.6,-36.15){$v_7$}

\node[Nframe=n](n7)(28.57,-36.15){$r_2$}

\drawedge(n2,n6){1}

\drawloop[loopangle=270.0](n6){1}

\end{picture}
\end{large}}}\hspace{-0.5cm}
\subfigure[$\mc{C}_1^a$]{\scalebox{0.7}{\begin{large}

%
%
%
%
%
%
%
%
%
%
%
%
%
%
%
%
%
%
%
%
%
%
%

\begin{picture}(81,69)(0,-69)
\put(0,-69){}
\node[iangle=90.0,Nmarks=i](n0)(40.0,-15.84){$v_1$}

\node(n1)(40.0,-47.84){$v_3$}

\node(n2)(64.16,-15.84){$v_5$}

\node[Nframe=n](n3)(48.0,-15.84){$r_0$}

\node[Nframe=n](n4)(48.0,-49.84){$r_1$}

\node[Nframe=n](n5)(72.16,-15.84){$r_2$}

\node(n6)(64.16,-47.84){$v_7$}

\node[Nframe=n](n7)(72.16,-47.84){$r_2$}

\drawedge(n2,n6){1}

\drawloop[loopangle=270.0](n6){1}

\node(n16)(20.0,-15.84){$v_0$}

\node(n17)(20.0,-47.84){$v_4$}

\node(n18)(20.0,-31.84){$v_2$}

\drawedge[ELside=r](n1,n16){0.5}

\drawedge(n1,n17){0.2}

\node[Nframe=n](n19)(12.0,-15.84){$r_0$}

\node[Nframe=n](n20)(20.0,-38.84){$r_1$}

\node[Nframe=n](n21)(12.0,-47.84){$r_2$}

\drawedge[ELside=r](n0,n18){1}
\drawloop[loopangle=90.0](n16){1}
\drawloop[loopangle=180.0](n18){1}
\drawloop[loopangle=270.0](n17){1}
\end{picture}
\end{large}}}\hspace{-0.6cm}
\subfigure[$\mc{C}_2$]{\scalebox{0.7}{\begin{large}
\begin{picture}(36,63)(0,-63)
\put(0,-63){}
\node[iangle=90.0,Nmarks=i](n22)(20.32,-12.16){$v_6$}

\node[Nmarks=r](n23)(20.32,-44.16){$v_8$}

\node[Nframe=n](n24)(28.32,-12.16){$r_2$}

\node[Nframe=n](n25)(28.32,-44.16){$r_2$}

\drawedge(n22,n23){1}

\drawloop[loopangle=270.0](n23){1}

\end{picture}
\end{large}}}
\end{center}
\caption{CTMCs corresponding to the (augmented) subgraphs\label{fig:CTMC_in_region}}
\end{figure}
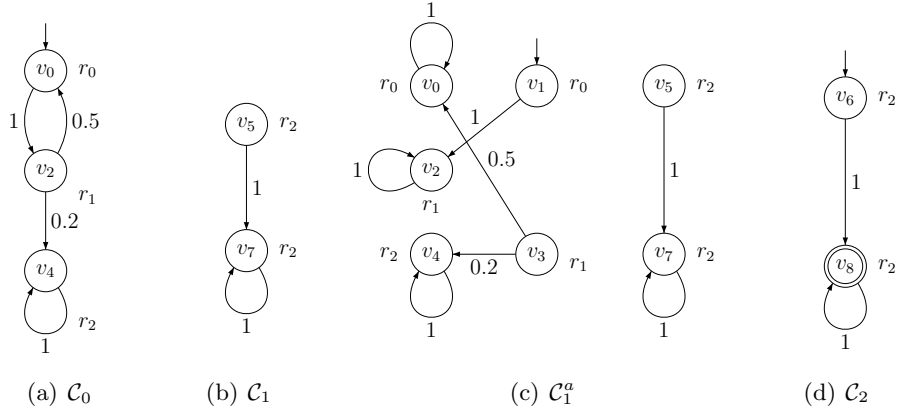

\begin{exa}
Consider the partitioned region graph in Figure~\ref{fig:oval}.
The matrices for $\mG_0$ are:
\[
\bdM_0(x)=\left(
\begin{array}{ccc}
0& 1{\cdot} r_0{\cdot}e^{-r_0x} & 0\\
0.5{\cdot}r_1{\cdot} e^{-r_1x}& 0 &0.2{\cdot}r_1{\cdot} e^{-r_1x} \\
0 &0& 0
\end{array} \right)
\quad
\bdF_0 =\left(
\begin{array}{cccc}
1& 0 & 0 & 0\\
0& 1 & 0 & 0\\
0& 0 & 1 & 0
\end{array} \right)
\]
The matrices for $\mG_1$ and its augmented version are given by:
\[
\bdM_1(x)=\left(
\begin{array}{cccc}
0& 0 & 0 & 0\\
0 & 0& 0&0\\
0 & 0 & 0 & r_2{\cdot} e^{-r_2x} \\
0 &0& 0 &0
\end{array} \right)
\quad 
\bdB_1(x) =\left(
\begin{array}{ccc}
0 & r_0{\cdot}e^{-r_0x} &0\\
0.5{\cdot}r_1{\cdot}e^{-r_1x} & 0 &0.2{\cdot}r_1{\cdot}e^{-r_1x}\\
0 & 0 &0\\
0 & 0 &0
\end{array} \right)
\]
\[
\bdF_1 =\left(
\begin{array}{cc}
0 & 0\\
0 & 0 \\
1 & 0 \\
0 & 1
\end{array} \right)
\quad \bdM_1^a(x)=\left(
\begin{array}{ccccccc}
0& 0 & 0 & 0 & 0&  r_0{\cdot}e^{-r_0x}&0\\
0 & 0& 0&0 & 0.5 {\cdot} r_1{\cdot}e^{-r_1x} & 0 & 0.2 {\cdot} r_1{\cdot}e^{-r_1x}\\
0 & 0 & 0 & r_2{\cdot} e^{-r_2x} & 0 & 0 & 0\\
0 &0& 0 &0 & 0 & 0 & 0 \\
0 &0& 0 &0 & 0 & 0 & 0 \\
0 &0& 0 &0 & 0 & 0 & 0 \\
0 &0& 0 &0 & 0 & 0 & 0
\end{array} \right)
\]
The corresponding \CTMC s and their augmented version are depicted in Figure~\ref{fig:CTMC_in_region}.
\end{exa}

For \CTMC\ $\mc{C}$ with $k$ states and rate matrix $\bdE \cdot \P$, let:
\begin{equation}
\bdPi(x) =\int_{0}^{x}\bdM(\tau)\bdPi(x-\tau)d\tau+\bdD(x).
\end{equation}
Intuitively, $\bdPi(x)[j,m]$ indicates the probability to move from vertex $j$
to $m$ at time $x$.
The following proposition states the close relationship between $\bdPi(x)$
and the transient probability vector of $\mc{C}$.
Let $\vec{\wp}(t)$ be the transient probability vector where $\wp_s(t)$ is the
probability to be in state $s$ at time $t$ given the initial distribution $\alpha$.

\begin{prop}\label{prop:pi2tran}
Given a \CTMC\ $\mc{C}$ with initial distribution $\alpha$, rate
matrix $\bdE \cdot \P$ and $\bdPi(t)$, $\vec{\wp}(t) $ satisfies
the following two equations:
\begin{eqnarray}
\vec{\wp}(t)&=&\alpha\cdot\bdPi(t),\\
\frac{d\vec{\wp}(t)}{dt}&=&\vec{\wp}(t)\cdot\Q,\label{eq:trandistr}
\end{eqnarray}
where $\Q=\bdE{\cdot}\P-\bdE$ is the infinitesimal generator.
\end{prop}

Equation~\eqref{eq:trandistr} is the well-known forward Chapman-Kolmogorov equation.
According to this proposition, solving the integral equation for $\bdPi(t)$ boils down to
solving the system of \ODE s \eqref{eq:trandistr} given some initial distribution vector
$\alpha$.
This can be done using standard means for CTMCs such as uniformization.

Now let the probability vector $\vec{U}_i(x) = {[u_i^{1}(x), \dots, u_i^{k_i}(x)]}^{\top}
\in\Reals^{k_i \times 1}$ where $u_i^{j}(x)$ is the probability to move from vertex
$v_{i}^j \in V_i$ to some vertex in $V_F$ (in $\mG$) at time $x$.
Based on the equations \eqref{eq:basepdp}-\eqref{eq:Markovian}, we provide a set of
integral equations for $\vec{U}_i(x)$ which later on is reduced to a system of linear
equations.
Distinguish two cases:

\medskip
\noindent\textbf{Case $0\les i<m$: } for
\begin{equation}\label{eq:delay_m}
\vec{U}_i(x)= \int_{0}^{\Delta
c_i-x}\bdM_i(\tau)\vec{U}_i(x+\tau)d\tau+\int_{0}^{\Delta
c_i-x}\bdB_i(\tau)d\tau\cdot\vec{U}_0(0) +\bdD_i(\Delta
c_i-x)\cdot \bdF_i\vec{U}_{i+1}(0),
\end{equation}
where $x\in[0,\Delta c_i]$.  Let us explain this equation. The last summand is
obtained from \eqref{eq:delay} where $\bdD_i(\Delta c_i{-}x)$ is the probability
to delay until the ``end'' of region $i$, and $\bdF_i\vec{U}_{i{+}1}(0)$ denotes
the probability to continue in $\mG_{i+1}$ (at relative time $0$).
Similarly, the first and second summands are obtained from \eqref{eq:Markovian};
the former reflects the case where clock $x$ is not reset, while the latter considers
the reset of $x$ (thus, implying a return to $\mG_0$).

\medskip
\noindent\textbf{Case $i=m$: }
\begin{equation}
\vec{U}_m(x)= \int_{0}^{\infty}\hat{\bdM}_m(\tau)\vec{U}_m(x{+}\tau)d\tau+ \vec{1}_F+\int_{0}^{\infty}\bdB_m(\tau)d\tau\cdot\vec{U}_0(0)\label{eq:last_m} %
\end{equation}
where for $x \in [c_m, \infty)$, $\hat{\bdM}_m(\tau)[v, \cdot] = \bdM_m(\tau)[v,
\cdot]$ for $v\notin V_F$, $0$ otherwise, and $\vec{1}_F$ is the characteristic
vector for $V_F$.
Note that $\vec{1}_F$ stems from the second clause of \eqref{eq:basepdp},
and $\hat{\bdM}_m$ is obtained by setting the corresponding elements of
$\bdM_m$ to 0.

\begin{exa}
The matrices for $\mG_2$ are given as:
\[\hat{\bdM}_2(x)=\left( \begin{array}{cc}  0&  r_2{\cdot}e^{-r_2x}\\ 0&0 \end{array}
\right) \quad \hat{\P}_2=\left( \begin{array}{cc} 0& 1\\ 0&0
\end{array} \right)\]
\end{exa}

For augmented CTMC $\mC^a_i$, let
\[\bdPi_i^a(x)=\left(\begin{array}{c|c}  \bdPi_i(x) & \bar\bdPi_i^a(x)\\
\hline \mathbf{0}&\mathbf{I} \end{array}\right),
\]
where $\mathbf{0}\in\Reals^{k_0\times k_i}$ is the zero matrix and
$\mathbf{I}\in\Reals^{k_0\times k_0}$ is the identity matrix.
Matrix $\bdPi_i$ indicates the transient probabilities for the CTMC $\mc{C}_i$.
Intuitively speaking, $\bar{\bdPi}_i^a$ contains the probabilities starting from
$V_i$ and ending in $V_0$.

\begin{thm}\label{thm:integraltotransient}
For subgraph $\mG_i$ (with $k_i$ vertices) of $\mG$, it holds that:
$$
\vec{U}_i(0) \ = \
\left\{
\begin{array}{ll}
\bdPi_i(\Delta c_i)\cdot \bdF_i \cdot \vec{U}_{i+1}(0)+\bar \bdPi_i^{a}(\Delta c_i)\cdot \vec{U}_{0}(0)
& \mbox{ if } i < m
\\[1ex]
\hat{\P}_m\cdot\vec{U}_m(0)+\vec{1}_F + \hat \bdB_m\cdot \vec{U}_0(0) & \mbox{ if } i = m
\end{array} \right.
$$
where $\hat{\P}_i(v,v') = \P_i(v,v')$ if $v \not\in V_F$; $0$ otherwise and
$\hat \bdB_m = \int_{0}^{\infty} \bdB_m(\tau) \, d\tau$.
\end{thm}

\proof
Distinguish two cases: $i < m$ and $i = m$.
\begin{enumerate}[(1)]
\item ($i <m $.)
Consider the augmented \CTMC\ $\mc{C}_i^a$ with $k_i^a=k_i+k_0$ states.
From equation~\eqref{eq:delay_m}, and the fact that $\mc{C}_i^a$ contains
reset edges of $\mc{C}_i$, we have:
\[
\vec{U}_i^a(x)= \int_{0}^{\Delta c_i{-}x} \bdM_i^a(\tau){\cdot}\vec{U}_i^a(x{+}\tau) \, d\tau \, + \,
\bdD_i^a(\Delta c_i{-}x)\cdot \bdF_i^a \cdot \vec{\hat U}_i(0)
\]
where $\vec{U}_i^a(x) = \left(\begin{array}{c} \vec{U}_i(x)\\ \hline \vec{U}_i'(x) \\
\end{array}\right) \in\Reals^{k_i^a \times 1}$, $\vec{U}_i'(x)\in\Reals^{k_0\times1}$ is
the vector representing the reachability probabilities for the augmented states in $\mc{G}_i$,
$\bdF_i^a=\left(\begin{array}{c|c}{\bdF'_i}&{\bdB'_i}\end{array}
\right)\in\Reals^{k_i^a\times(k_{i+1}+k_0)}$ such that
$\bdF'_i=\left(\begin{array}{c}
\bdF_i\\ \hline \mathbf{0}\\\end{array}\right)\in\Reals^{k_i^a\times k_{i+1}}$
is the incidence matrix for delay edges and $\bdB'_i=\left(\begin{array}{c}
\mathbf{0}\\ \hline \mathbf{I}\\\end{array}\right)\in\Reals^{k_i^a\times k_0}$, and finally
$\vec{\hat U}_i(0)=\left(\begin{array}{c}
\vec{U}_{i+1}(0)\\ \hline
\vec{U}_{0}(0)\\\end{array}\right)\in\Reals^{(k_{i+1}+k_0)\times 1}$.
\noindent
The proof of the theorem for $i < m$ proceeds in two steps.
\begin{enumerate}[(a)]
\item[(a)]
We first show that:
$$
\begin{array}{rcl}
\vec{U}_i^a(x) & = & \bdPi_i^a(\Delta c_i{-}x)\cdot\bdF_i^a\cdot\vec{\hat U}_i(0) \mbox{ where } \\[1ex]
\bdPi_i^a(x) & = & \displaystyle \int_{0}^{x}\bdM_i^a(\tau)\cdot \bdPi_i^a(x{-}\tau) \, d\tau + \bdD_i^a(x).
\end{array}
$$
Consider the following system of inductively defined integral equations.
Let $c_{i,x}=\Delta c_i-x$.
\begin{eqnarray*}
\vec{U}_i^{a,(0)}(x)&=& \vec{0}\\
\vec{U}_i^{a,(j+1)}(x)&=&\int_{0}^{c_{i,x}}\bdM_i^a(\tau)\cdot\vec{U}_i^{a,(j)}(x{+}
\tau) \, d\tau+\bdD_i^a(c_{i,x})\cdot\bdF_i^a\cdot\vec{\hat U}_i(0).
\end{eqnarray*}
and
\begin{eqnarray*}
\bdPi_i^{a,(0)}(c_{i,x})&=& \mathbf{0}\\
\bdPi_i^{a,(j+1)}(c_{i,x})&=&\int_{0}^{c_{i,x}}\bdM_i^a(\tau)\cdot\bdPi_i^{a,(j)}(c_{
i,x}{-}\tau) \, d\tau+\bdD_i^a(c_{i,x}).
\end{eqnarray*}
Clearly, $\displaystyle \bdPi_i^a(c_{i,x})=\lim_{j\to\infty}\bdPi_i^{a,(j+1)}(c_{i,x})$ and
$\displaystyle \vec{U}_i^a(x)=\lim_{j\to\infty}\vec{U}_i^{a,(j+1)}(x)$.

\noindent
By induction on $j$, we prove the following relation:
\begin{equation}\nonumber
\vec{U}_i^{a,(j)}(x)=\bdPi_i^{a,(j)}(c_{i,x})\cdot\bdF_i^a\cdot\vec{\hat
U}_i(0).
\end{equation}
\begin{enumerate}[(i)]
\item (Base case.) $\vec{U}_i^{a,(0)}(x)=\vec{0}$ and $\bdPi_i^{a,(0)}(c_{i,x})={\mathbf{0}}$.
\item (Induction step.)
By exploiting the induction hypothesis (in the second step), we derive:
\begin{eqnarray*}
\quad\qquad \qquad\qquad\vec{U}_i^{a,(j+1)}(x) & =&
\int_{0}^{c_{i,x}}\bdM_i^a(\tau)\vec{U}_i^{a,(j)}(x+\tau)d\tau+\bdD_i^a(c_{i,x}
)\cdot\bdF_i^a\vec{\hat U}_i(0)\\
&=&\int_{0}^{c_{i,x}}\bdM_i^a(\tau)\bdPi_i^{a,(j)}(c_{i,x}{-}
\tau)\cdot\bdF_i^a\vec{\hat U}_i(0)d\tau+\bdD_i^a(c_{i,x})\cdot\bdF_i^a\vec{\hat
U}_i(0)\\
& =& \Bigg(\int_{0}^{c_{i,x}}\bdM_i^a(\tau)\bdPi_i^{a,(j)}(c_{i,x}-\tau)d\tau
+\bdD_i^a(c_{i,x})\Bigg)\cdot \bdF_i^a\vec{\hat
U}_i(0)\\
&  =&\bdPi_i^{a,(j+1)}(c_{i,x})\cdot\bdF_i\vec{\hat U}_i(0).
\end{eqnarray*}
\end{enumerate}
\item[(b)]
$\bdPi_i^a(\Delta c_i)\cdot\bdF_i^a\vec{\hat
U}_i(0)=\left(\begin{array}{c}
\bdPi_i(\Delta c_i)\bdF_i\vec{U}_{i+1}(0)+\bar \bdPi_i^a(\Delta
c_i)\vec{U}_{0}(0)\\ \hline
\vec{U}_{0}(0)\\\end{array}\right)=\left(\begin{array}{c}
\vec{U}_i(0)\\ \hline \vec{U}_i'(0)\\\end{array}\right)$.
Let $x=0$ and we obtain \begin{equation}\nonumber
\vec{U}_i^a(0)=\bdPi_i^a(c_{i,0})\cdot\bdF_i^a\vec{\hat U}_i(0).
\end{equation}

We can also write the above relation for $x=0$ as:
\begin{small}
\begin{eqnarray*}\nonumber
\left(\begin{array}{c}
\vec{U}_i(0)\\ \hline \vec{U}_i'(0)\\\end{array}\right)&=&\bdPi_i^a(\Delta c_i)\left(\begin{array}{c|c}{\bdF'_i}&{\bdB'_i}\end{array}\right)\left(\begin{array}{c}
\vec{U}_{i+1}(0)\\ \hline \vec{U}_{0}(0)\\\end{array}\right)\\
&=&\left(\begin{array}{c|c}\bdPi_i(\Delta c_i)& \bar\bdPi_i^a(\Delta c_i)\\ \hline \mathbf{0}&\mathbf{I} \end{array}\right)\left(\begin{array}{c|c}\bdF_i& \mathbf{0}\\ \hline \mathbf{0}&\mathbf{I} \end{array}\right)\left(\begin{array}{c}
\vec{U}_{i+1}(0)\\ \hline \vec{U}_{0}(0)\\\end{array}\right)\\
&=&\left(\begin{array}{c|c}\bdPi_i(\Delta c_i)\bdF_i& \bar\bdPi_i^a(\Delta c_i)\\ \hline \mathbf{0}&\mathbf{I} \end{array}\right)\left(\begin{array}{c}
\vec{U}_{i+1}(0)\\ \hline \vec{U}_{0}(0)\\\end{array}\right)\\
&=&\left(\begin{array}{c}
\bdPi_i(\Delta c_i)\bdF_i\vec{U}_{i+1}(0)+\bar \bdPi_i^a(\Delta c_i)\vec{U}_{0}(0)\\ \hline \vec{U}_{0}(0)\\\end{array}\right).
\end{eqnarray*}\end{small}
As a result we can represent $\vec{U}_i(0)$ in the following matrix form
\begin{equation}\nonumber
\vec{U}_i(0)=\bdPi_i(\Delta c_i)\bdF_i\vec{U}_{i+1}(0)+\bar \bdPi_i^a(\Delta c_i)\vec{U}_{0}(0)
\end{equation}
by noting that $\bdPi_i$ is formed by the first $k_i$ rows and
columns of matrix $\bdPi_i^a$ and $\bar \bdPi_i^a$ is formed by
the first $k_i$ rows and the last $k_i^a-k_i$ columns of
$\bdPi_i^a$.
\end{enumerate}
\item ($i=m$.)
The proof of this case follows almost immediately from equation~\eqref{eq:last_m}.
As any region in $\mc{G}_m$ is unbounded, delay transitions do not exist.
As $\vec{U}_m(x{+}\tau)$ does not depend on $x$, the integral  $\int_{0}^{\infty}
\hat{\bdM}_m(\tau)\vec{U}_m(x{+}\tau) \, d\tau$ reduces to
$\int_{0}^{\infty}\hat{\bdM}_m(\tau) \, d\tau \cdot\vec{U}_m(0)$.
In addition, $\int_{0}^{\infty}\hat{\bdM}_m(\tau) \, d\tau$ boils down to $\hat{\P}_m$
and $\int_{0}^{\infty}\bdB_m(\tau) \, d\tau$ to $\hat \bdB_m$.\qed
\end{enumerate}

Since the coefficients of the linear equations are all known,
solving the system of linear equations yields $\vec{U}_0(0)$,
which contains the probability $\Prob_{v_0}(0)$ of reaching $V_F$
from initial vertex $v_0$.

Theorem~\ref{thm:integraltotransient} is based on the equations~\eqref{eq:delay_m}
(for $i < m$), and \eqref{eq:last_m} (for $i{=}m$).
The term $\bdPi_i(\Delta c_i) \cdot \bdF_i \cdot \vec{U}_{i+1}(0)$ stands for the delay
transitions, where $\bdF_i$ specifies how the delay transitions are connected between
the sub-graphs $\mG_i$ and $\mG_{i+1}$.
The term $\bar\bdPi_i^a(\Delta c_i)\cdot \vec{U}_{0}(0)$ stands for Markovian transitions
with reset.
The term $\bar\bdPi_i^a(\Delta c_i)$ in the augmented \CTMC\ $\mC_i^a$ specifies
the probabilities to first take transitions inside $\mG_i$ followed by a one-step
Markovian transition back to $\mG_0$.

\begin{rem}
The approach in this section is focused on single-clock DTA (with finite acceptance
criteria).
For two-clock \DTAr\ the approach fails.
In case of a single clock $x$, any reset (of $x$) from $\mG_i$ yields a state in
$\mG_0(0)$, and any delay (of $x$) yields some state in $\mG_{i+1}(0)$.
However, in the setting of two clocks, after a reset generally only one clock has a fixed
value while the value of the other one is not determined.
\end{rem}

\begin{lem}
For CTMC $\mC$ and single-clock \DTAr\ $\mA$, computing
${\Pr}^\mC \big( \Paths^\mC(\mA) \big)$ can be done in time
$\mathcal{O}(m^2{\cdot}{\vert S\vert}{\cdot}{\vert Loc\vert}
{\cdot}\lambda{\cdot}\Delta c+m^3{\cdot}{\vert S\vert}^3{\cdot}{\vert Loc\vert}^3)$,
where $m$ is
the number of constants appearing in $\mA$,  $\vert S\vert$ is
the number of states in $\mC$, $\vert Loc\vert$ is the number
of locations in $\mA$, $\lambda$ is the maximal exit rate in
$\mC$ and $\Delta c = \max_{0 \les i<m}\{\Delta c_{i}\}$.
\end{lem}

\begin{proof}
The DMTA $\mc{C} \otimes \mc{A}$ has at most $| S | {\cdot} | Loc |$ locations.
The number of vertices in the PDP $\mc{Z}(\mc{C} \otimes \mc{A})$ is at most
$m {\cdot} | S | {\cdot} | Loc |$, as there are $m$ possible regions.
CTMC $\mc{G}_i$ and its annotated version $\mc{G}^a_i$ thus have at most
$\mathcal{O} \left( m {\cdot} | S | {\cdot} | Loc | \right)$ states.
Calculating the transient distribution $\bdPi_i(\Delta c_i)$ on CTMC $\mc{G}_i$
for any state in $\mc{G}_i$ takes at most $\mathcal{O} \left( m {\cdot} | S | {\cdot}
| Loc | {\cdot} \lambda {\cdot} \Delta c \right)$ where $\lambda$ is the maximal
exit rate in $\mc{G}_i$ (and thus in $\mC$) and $\Delta c = \max_{0 \les i<m}
\{\Delta c_{i}\}$ is the maximal width of a region.
Given that this computation needs to be performed for any subgraph yields the
first summand in the time complexity.
Subsequently, according to Theorem~\ref{thm:integraltotransient}, a system of
linear equations has to be solved with at most
$\mathcal{O}\left( m {\cdot} | S | {\cdot} | Loc | \right)$ variables.
This takes at most $\mathcal{O}\left( m^3 {\cdot} | S |^3 {\cdot} | Loc |^3 \right)$
operations.
\end{proof}

\section{Verifying CTMCs Against Muller DTA Specifications}\label{sec:infinite}

Finally, we deal with the verification of CTMCs against DTA with Muller acceptance
conditions.
The procedure is very similar to the one for DTA with finite acceptance conditions.
Let $\mA$ be a \DTAo, and $\mC$ a CTMC.
The region graph of the product $\mC \otimes \mA$ is defined as before (cf.\
Def.~\ref{def:region}, page~\pageref{def:region}), except that the accepting set
$V_F$ is defined using bottom (or: terminal) SCCs (BSCCs for short).
A strongly connected component (SCC) is terminal if it cannot be left once entered.

\begin{defi}[Region graph of \DMTAo]\label{def:regionagain}
The \emph{region graph} of \DMTAo\ $\mc{M} = (Loc, \mc{X}, \ell_0$,
$Loc_{\mF}, E, \rightsquigarrow)$ is $\mc{G}(\mc{M}) = (V ,v_0, V_F,
\Lambda$, $\hookrightarrow)$, where $V$, $v_0$, $\Lambda$ and
$\hookrightarrow$ are defined as in Def.~\ref{def:region} (page
\pageref{def:region}), and $V_F = \big\{v \in B \mid B\in a\mB \big\}$
where $a\mB$ is the set of accepting \BSCC s in $\mc{G}(\mc{M})$.
BSCC $B \subseteq V$ is accepting if there exists $L_F\in Loc_{\mF}$
such that for any $v \in B$, $v{\downharpoonright}_1\in L_F$.
\end{defi}

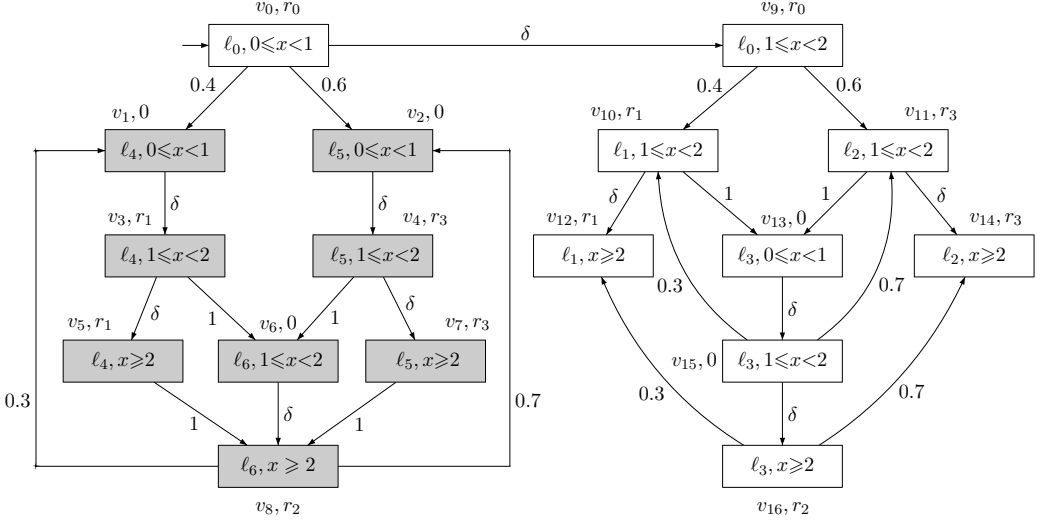
\begin{figure}
\begin{center}
\scalebox{0.7}{
\begin{picture}(208,107)(0,-107)
\put(0,-107){}
\node[Nmarks=i,Nw=22.74,Nmr=0.0](n57)(58.19,-8.19){$\ell_0,0{\les} x{<}1$}

\node[Nfill=y,fillgray=0.8,Nw=22.74,Nmr=0.0](n60)(38.5,-48.19){$\ell_4,1{\les} x{<}2$}

\node[Nfill=y,fillgray=0.8,Nw=22.74,Nmr=0.0](n61)(38.5,-28.19){$\ell_4,0{\les} x{<}1$}

\node[Nfill=y,fillgray=0.8,NLangle=0.0,NLdist=0.45,Nw=22.74,Nmr=0.0](n62)(77.97,-48.19){$\ell_5,1{\les} x{<}2$}

\node[Nframe=n](n64)(60.0,-1.19){$v_0,r_0$}

\node[Nframe=n](n65)(31.97,-21.19){$v_1,0$}

\node[Nframe=n](n66)(87.97,-21.19){$v_2,0$}

\node[Nframe=n](n74)(31.97,-41.19){$v_3,r_1$}

\node[Nfill=y,fillgray=0.8,Nw=22.74,Nmr=0.0](n53)(77.97,-28.19){$\ell_5,0{\les} x{<}1$}

\node[Nfill=y,fillgray=0.8,Nw=22.74,Nmr=0.0](n54)(30.53,-68.19){$\ell_4,x{\ges} 2$}

\node[Nfill=y,fillgray=0.8,NLangle=0.0,NLdist=0.45,Nw=22.74,Nmr=0.0](n56)(88.0,-68.19){$\ell_5,x{\ges}2$}

\node[Nfill=y,fillgray=0.8,Nw=22.74,Nmr=0.0](n157)(60.0,-68.19){$\ell_6,1{\les}x{<} 2$}

\node[Nfill=y,fillgray=0.8,Nw=22.74,Nmr=0.0](n58)(60.0,-88.19){$\ell_6,x\ges 2$}

\node[Nw=22.74,Nmr=0.0](n272)(155.83,-8.19){$\ell_0,1{\les} x{<}2$}

\node[Nw=22.74,Nmr=0.0](n274)(132.15,-28.19){$\ell_1,1{\les} x{<}2$}

\node[Nw=22.74,Nmr=0.0](n275)(175.86,-28.19){$\ell_2,1{\les} x{<}2$}

\node[Nw=22.74,Nmr=0.0](n276)(155.77,-48.19){$\ell_3,0{\les} x{<}1$}

\node[Nw=22.74,Nmr=0.0](n277)(119.86,-48.19){$\ell_1, x{\ges}2$}

\node[Nw=22.74,Nmr=0.0](n278)(155.77,-68.19){$\ell_3,1{\les} x{<}2$}

\node[Nw=22.74,Nmr=0.0](n279)(155.77,-88.19){$\ell_3,x{\ges}2$}

\node[Nw=22.74,Nmr=0.0](n280)(192.18,-48.19){$\ell_2, x{\ges}2$}

\node[Nframe=n](n59)(87.97,-41.19){$v_4,r_3$}

\node[Nframe=n](n160)(24.29,-61.19){$v_5,r_1$}

\node[Nframe=n](n162)(96.0,-61.19){$v_7,r_3$}

\node[Nframe=n](n263)(60.0,-96.19){$v_8,r_2$}

\node[Nframe=n](n389)(60.0,-61.19){$v_6,0$}

\drawedge(n57,n53){$0.6$}

\drawedge[ELside=r,ELdist=1.03](n57,n61){$0.4$}

\drawedge(n61,n60){$\delta$}

\drawedge(n53,n62){$\delta$}

\drawedge[ELside=r](n60,n157){$1$}

\drawedge(n60,n54){$\delta$}

\drawedge(n62,n157){$1$}

\drawedge(n62,n56){$\delta$}

\drawedge[ELside=r,ELdist=0.56](n54,n58){$1$}

\drawedge(n157,n58){$\delta$}

\drawedge(n56,n58){$1$}

\node[Nw=0.0,Nh=0.0,Nmr=0.0](n268)(14.0,-88.19){}

\node[Nw=0.0,Nh=0.0,Nmr=0.0](n269)(14.0,-28.19){}

\node[Nw=0.0,Nh=0.0,Nmr=0.0](n270)(104.0,-88.19){}

\node[Nw=0.0,Nh=0.0,Nmr=0.0](n271)(104.0,-28.19){}

\drawedge[AHnb=0](n58,n268){ }

\drawedge[ELpos=21,AHnb=0](n268,n269){$0.3$}

\drawedge(n269,n61){}

\drawedge[AHnb=0](n58,n270){ }

\drawedge[ELside=r,ELpos=21,AHnb=0](n270,n271){$0.7$}

\drawedge(n271,n53){}

\drawedge(n57,n272){$\delta$}

\drawedge[ELside=r,ELdist=0.68](n272,n274){$0.4$}

\drawedge(n272,n275){$0.6$}

\drawedge(n274,n276){$1$}

\drawedge(n276,n278){$\delta$}

\drawedge(n278,n279){$\delta$}

\drawedge[ELside=r,ELdist=1.25](n274,n277){$\delta$}

\drawedge(n275,n280){$\delta$}

\drawedge[curvedepth=7.36](n278,n274){$0.3$}

\drawedge[ELside=r,curvedepth=-8.23](n278,n275){$0.7$}

\drawedge[curvedepth=5.59](n279,n277){$0.3$}

\drawedge[ELside=r,ELdist=0.9,curvedepth=-5.06](n279,n280){$0.7$}

\node[Nframe=n](n312)(155.97,-1.19){$v_9,r_0$}

\node[Nframe=n](n313)(124.35,-21.19){$v_{10},r_1$}

\node[Nframe=n](n314)(184.0,-21.19){$v_{11},r_3$}

\node[Nframe=n](n315)(116.0,-41.19){$v_{12},r_1$}

\node[Nframe=n](n316)(156.0,-41.19){$v_{13},0$}

\node[Nframe=n](n317)(196.32,-41.19){$v_{14},r_3$}

\node[Nframe=n](n318)(139.0,-68.19){$v_{15},0$}

\node[Nframe=n](n319)(155.97,-96.19){$v_{16},r_2$}

\drawedge[ELside=r](n275,n276){$1$}

\end{picture}}
\end{center}\vspace{-0.8cm}\caption{Region graph of the product
\DMTAo\ in Figure~\ref{fig:DMTA_Muller}\label{fig:infinite_region}}
\end{figure}

\begin{exa}
Consider the \DMTA$^\omega$ in Figure~\ref{fig:DMTA_Muller} with
$Loc_\mF = \{ L_{F_1}, L_{F_2} \}$ with $L_{F_1} = \{\ell_1,\ell_2$, $\ell_3\}$, and
$L_{F_2} = \{\ell_4,\ell_5,\ell_6\}$.
Its region graph is depicted in Figure~\ref{fig:infinite_region}.
There is one accepting \BSCC, whose vertices are colored gray,
corresponding to the set $L_{F_2}$.
There is no \BSCC\ corresponding to $L_{F_1}$, due to the presence of
the sink vertices $v_{12}$ and $v_{14}$.
These vertices are reachable from locations $\ell_1$ and $\ell_2$ if
$x \ges 2$.
\end{exa}

Two remarks are in order.
A first observation is that the probability to stay in an accepting \BSCC\ is one,
considering both the delay and Markovian transitions.  That is to say, there are
no outgoing transitions from which some probability can ``leak away''.
In addition, any pair of accepting BSCCs is disjoint, which allows the addition of,
e.g., their reachability probabilities.

\begin{thm}\label{th:inf_2_fin}
For any \CTMC\ $\mC$, \DTAo\ $\mA$,
${\Pr}^{\mC} \big( \Paths^{\mC}(\mA) \big)$ is the least solution of
$\Prob^{\mD}_{v_0}(\vec{0},U)$,
where DTMP $\mD = \emb(\mc{Z}(\mC \otimes \mA)) \mbox{ and } U = \bigcup_{B \in a\mB} B.$
\end{thm}

\begin{proof}
To start off, observe that ${\Pr}^{\mC} \big( \Paths^{\mC}(\mA) \big)$ is measurable,
cf.\ Theorem~\ref{lem:measurability} (page~\pageref{lem:measurability}).
The proof follows from Theorem~\ref{th:CTMC=MTA} and the following observations.
%
For any DTMP expanded with a finite set of locations---like for
finite DTMCs--- almost surely the states that are visited
infinitely often along a path constitute a BSCC. It thus follows
that the probability for visiting a set of states infinitely often
equals the reachability probability of some BSCC in the DTMP
$\emb(\mc{Z}(\mC \otimes \mA))$.
The result now follows from Theorem~\ref{th:MTA=DTMP}.
\end{proof}

\begin{exa}
Consider the region graph in Figure~\ref{fig:infinite_region}.  The only \BSCC\
is indicated by the gray shaded states.  To determine ${\Pr}^{\mC}\big(
\Paths^{\mC}(\mA) \big)$, it suffices to consider the reachability probability for
$T = \{ v_1, v_2 \}$.
For the delay transition $v_0 \stackrel{\delta}{\hookrightarrow} v_9$, we
have
$$
\Prob_{v_0,\delta}(0) \ = \ e^{-r_0{\cdot}1}{\cdot}\Prob_{v_9}(1) \ = \
  e^{-r_0{\cdot}1}{\cdot}0 \ = \ 0.
$$
For the Markovian transition $v_0\stackrel{0.4,\{x\}}{\hookrightarrow}v_1$,
$$
\Prob_{v_0,v_1}(0)
\ = \
\int_0^10.4{\cdot} r_0{\cdot}e^{-r_0{\cdot}\tau}{\cdot}\Prob_{v_1}(\tau) \, d\tau
\ = \
\int_0^10.4{\cdot} r_0{\cdot}e^{-r_0{\cdot}\tau} \, d\tau.
$$
A similar reasoning applies to $v_0\stackrel{0.6,\{x\}}{\hookrightarrow}v_2$.
Gathering the results we obtain:
$$
{\Pr}^\mC \big( \Paths^\mC(\mA) \big)
\ = \
\int_0^1(0.4+0.6){\cdot} r_0{\cdot}e^{-r_0{\cdot}\tau} \, d\tau
\ = \
\int_0^1 r_0{\cdot}e^{-r_0{\cdot}\tau}d\tau=1-e^{-r_0}.
$$
\end{exa}

\subsection*{Verifying qualitative specifications}\label{sec:qualitative}
Until now we have investigated the quantitative verification problem, which is to determine
the value of $\Pr(\mc{C} \models \mc{A})$.
The qualitative verification problem, on the other hand, is to determine whether the
probability that $\mc{C}$ satisfies $\mc{A}$ exceeds zero, or, dually, equals one.
For stochastic processes such as finite CTMCs and finite DTMCs, qualitative verification
problems are known to be decidable by means of a simple graph analysis.

\begin{prop}
 For any \CTMC\ $\mc{C}$ and \DTA\ $\mc{A}$,
\begin{enumerate}[(1)]
\item
$\Pr^\mC\!\big(\Paths^\mc{C} (\mc{A})\big) > 0
\quad \mbox{iff} \quad
\mc{Z}(\mc{C} \otimes \mc{A}) \models \exists \Ever V_F$,
\item
$\Pr^\mC\!\big(\Paths^\mc{C} (\mc{A})\big) = 1
\quad \mbox{iff} \quad
\mc{Z}(\mc{C} \otimes \mc{A}) \models \forall \left(\left(\exists\Ever V_F\right)\W V_F\right)$,
\end{enumerate}
where $V_F{=}\{v{\in} V\mid v{\downharpoonright_1}{\in} Loc_F\}$ for \DTAr,
$V_F{=}\big\{v {\in} B \mid B{\in} \, a\mB \big\}$ for \DTAo, and $\W$ denotes the
weak until operator.
\end{prop}

\begin{proof}
Similar to the case for discrete-time Markov chains~\cite[Chapter 10]{BaKa08}.
\end{proof}

From the above theorem, it follows that the qualitative properties can be verified using a
standard graph-based CTL model checking algorithm, i.e., by just considering the
underlying finite digraph of the PDP $\mc{Z}(\mc{C} \otimes \mc{A})$ ---basically the
region graph of $\mC\otimes\mA$--- while ignoring the transition probabilities.


\section{Conclusion}\label{sec:concl}
This paper addressed the quantitative (and qualitative) verification of a finite \CTMC\
$\mC$ against a linear real-time specification given as a deterministic timed automaton
(DTA).
We studied DTA with finite and Muller acceptance criteria.
The key result (for finite acceptance) is that the probability of $\mC \models \mA$ equals
the reachability probability in the embedded discrete-time Markov process of a \PDP.
This PDP is obtained via a standard region construction.
Reachability probabilities in the thus obtained PDPs are characterized by a system of
Volterra integral equations of the second type and are shown to be approximated by a
system of  PDEs.
For Muller acceptance criteria, the probability of $\mC \models \mA$ equals the reachability
probability of the accepting terminal SCCs in the embedded \PDP.
These results apply to DTA with arbitrarily (but finitely) many clocks.
For single-clock DTA with finite acceptance, $\Pr(\mC\models \mA)$ is obtained by
solving a system of linear equations whose coefficients are solutions of a system of ODEs.
As the coefficients are in fact transient probabilities in CTMCs, this result implies that
standard algorithms for CTMC analysis suffice to verify single-clock DTA specifications.

An interesting future research direction is the verification against non-deterministic timed
automata (NTA).
NTA are strictly more expressive than DTA, and thus would allow more linear real-time
specification.
Following the approach in this paper requires a nondeterministic variant of PDP.
Another challenging open problem is to consider real-time linear temporal logics as
specifications such as metric temporal logic (MTL)~\cite{Koy90} or variants thereof.

\section*{Acknowledgement}
\begin{small}
We thank Jeremy Sproston (University of Turin) for fruitful and insightful discussions and
Beno\^it Barbot (ENS Cachan) for pointing out some flaws in an earlier version of this
paper.
We are grateful to the reviewers for providing many useful suggestions on improving the
presentation of the paper.
\end{small}

%
%
\bibliographystyle{abbrv}
\bibliography{../bib/myBib,../bib/mybibProceedings,../bib/mybibArticles,../bib/mybibInroceedings,../bib/mybibInbooks}
%
%

\end{document}